\documentclass{aa}
\usepackage{graphicx}
\usepackage{txfonts,rotate,psfig,subfigure,graphicx,supertabular,lscape,times}
\begin{document}
\title{A Multi-epoch VLBI Survey of the Kinematics of \\ Caltech-Jodrell Bank Flat-Spectrum Sources \thanks{Table C.1. and Fig. C.1. are available in electronic form at the CDS via anonymous ftp to cdsarc.u-strasbg.fr (130.79.128.5) or via http://cdsweb.u-strasbg.fr/cgi-bin/qcat?J/A+A/}}
\subtitle{Part II: Analysis of the Kinematics}
\author{S. Britzen\inst{1,2,3}
\and R.C. Vermeulen\inst{3}
\and R.M. Campbell\inst{4}
\and G.B. Taylor\inst{5,6}
\and T.J. Pearson\inst{7} \and A.C.S. Readhead\inst{7} \and W. Xu\inst{7}
\and I.W. Browne\inst{8} \and D.R. Henstock\inst{8} \and P. Wilkinson\inst{8}}
\offprints{Silke Britzen, sbritzen@mpifr-bonn.mpg.de}
\institute{present address: Max-Planck-Institut f\"ur Radioastronomie, Auf dem H\"ugel 69, D-53121 Bonn, Germany \and Landessternwarte, K\"onigstuhl, D-69117 Heidelberg, Germany \and ASTRON, Netherlands Foundation for Research in Astronomy, P.O. Box 2, NL-7990 AA Dwingeloo, The Netherlands \and Joint Institute for VLBI in Europe, Oude Hoogeveensedijk 4, NL-7991 PD Dwingeloo, The Netherlands \and Kavli Institute of particle Astrophysics and Cosmology, Menlo Park, CA 94025, USA \and National Radio Astronomy Observatory, P.O. Box O, Socorro, NM 87801, USA \and California Institute of Technology, Department of Astronomy, 105-24, Pasadena, CA 91125, USA \and University of Manchester, Nuffield Radio Astronomy Laboratories, Jodrell Bank, Macclesfield, Cheshire SK11 9 DL, England UK}
\date{Received; accepted}
\abstract
{This is the second of a series of papers presenting VLBI observations of the 293 Caltech-Jodrell Bank Flat-Spectrum (hereafter CJF) sources and their analysis.}{To obtain a consistent motion dataset large enough to allow the systematic properties of the population to be studied.}{We present the detailed kinematic analysis of the complete flux-density limited CJF survey. We computed 2-D kinematic models based on the optimal model-fitting parameters of multi-epoch VLBA observations. This allows us to calculate not only radial, but also orthogonal motions, and thus to study curvature and acceleration. Statistical tests of the motions measured and their reliability have been performed. A correlation analysis between the derived apparent motions, luminosities, spectral indices, and core dominance and the resulting consequences is described.}{With at least one velocity in each of 237 sources, this sample is much larger than any available before and allows a meaningful statistical investigation of apparent motions and any possible correlations with other parameters in AGN jets. The main results to emerge are as follows:
- In general motions are not consistent with a single uniform velocity applicable to all components along a jet.
- We find a slight trend towards a positive outward  acceleration and also adduce some evidence for greater acceleration in the inner-most regions.
- We find a lack of fast components at physical distances less than a few pc from the reference feature.     
- Only $\sim$4\% of the components from galaxies and $<$2\% of those from quasars undergo large bends i.e. within  $15^\circ$ of $\pm90^\circ$.
- The distribution of radial velocities shows a broad distribution of velocities (apparent velocities up to 30 $c$). 15\% of the best sampled jet components exhibit "low" velocities which may need to be explained in a different manner to the "fast" motions.
- Some negative or "backwards" superluminal motions are seen and in 15 cases (6\%) these are definitely significant.
- We find a strong correlation between the 5 GHz luminosity and the apparent velocity.
- The CJF galaxies, on average, show slower apparent jet component velocities than the quasars.             
- The mean velocity in the VLBA 2cm survey (Kellermann et al. 2004) is substantially higher than in the CJF survey -- the ratio could be roughly a factor of 1.5--2. This supports the observed trend of increasing apparent velocity with increasing observing frequency.}{This AGN survey provides the basis for any statistical analysis of jet and jet-component properties.}
\keywords{Techniques: interferometric -- Surveys -- Galaxies:active -- Radio continuum:Galaxies}
\maketitle

\section{Introduction}
\label{sec:intro}

Extragalactic radio jet sources have been intensively studied for several
decades now, for a number of important reasons. These include their association
with supermassive accreting black holes, which are recognized to play a
substantial role in the formation and evolution of galaxies. Radio jets can
help to reveal their energetics, and the way they interact with their
environment. The formation, acceleration, and propagation of radio jets also
challenges our understanding of the laws of physics under extreme
circumstances, such as (ultra-)relativistic speeds, high magnetic field
strengths, and, possibly, the occurrence on galactic scales of Poynting flux
jets, and/or, of non-baryonic matter. In addition, radio sources can be bright
enough to be visible at very large redshifts, and thus they allow population
studies to be carried out over a large fraction of cosmic time.

The parsec-scale properties of jets have been observed with all available VLBI
networks at all available wavelengths (e.g., the review of Zensus 1997, and
references therein). Immediately alongside morphological studies have come
studies of the motions of features seen within these jets. Those motions often
appear to be faster-than-light, or superluminal. This phenomenon occurs as a
result of light travel time compression when features move with a speed close
to $c$ at a small angle to the line-of-sight; a useful introduction to the
phenomenon can be found in Pearson \& Zensus (1987). Detailed studies and 
intensive monitoring campaigns of individual objects have proven to be important 
to understand jet component motions in a handful of selected sources
(e.g., 3C120: G$\acute{\rm o}$mez et al. 2001; 1803+784: Britzen et al.
2005a\&b; 0735+178: Agudo et al. 2006). It has emerged that different sources can 
reveal quite different motions; see for example the compilations of Barthel et al. (1988), Zensus \& Pearson (1990), Ghisellini et al.\ (1993), Vermeulen \& Cohen (1994), Jorstad et al.\
(2001), Kellermann et al.\ (2004), Cohen et al. (2007), and Piner et al. (2007). This indicates a clear need to study large, well-defined samples that can be subjected to
statistical analysis.

In the 1980s and 1990s it was thought that a careful study of the statistics of superluminal motions as a function of redshift could have a major impact on cosmology (e.g., Cohen et al. 1988; Vermeulen \& Cohen 1994) and that this would be the major driver for studies of large samples of sources.  
The attention has now turned to the study of the astrophysics of the jets themselves.   

The CJF survey (Taylor et al.\ 1996a) on which we report here provides a large, well-defined, complete flux-density limited
flat-spectrum sample that has been treated fully homogeneously as regards
observational strategy, data reduction, and data analysis. Images at 5 GHz have been obtained for nearly all objects from global VLBI or VLBA data at least at three observing epochs from 1990. A detailed description of the observations and data is presented in Britzen et al. 2007a (hereafter Paper~I). 
With at least one velocity in each of 237 sources, this sample is much larger than any available
heretofore. We have not just derived a single speed per object, but have been
able to obtain motions for several individual features in many of the sources.
We have also studied curvature and accelerations, rather than obtaining only
total speeds, or radial motions.

Preliminary results from some subsets of the CJF data have already been
published in Vermeulen (1995), Britzen et al. (1999, 2001b), and Britzen
(2002). Polarization information of a sub-sample of CJF-sources has been
published in Pollack et al. (2003). Detailed information on the VLBI
observations, data reduction, analysis, imaging,  model fitting, and component
identification is given in Paper~I; the database is
available online at the CDS via anonymous ftp to cdsarc.u-strasbg.fr (130.79.128.5) or via http://cdsweb.u-strasbg.fr/cgi-bin/qcat?J/A+A/. In addition, the data can be downloaded from a CJF-archive at the MPIfR in Bonn (http://www.mpifr-bonn.mpg.de/staff/sbritzen/cjf.html). The present, second paper, deals with the derivation of
motions from the component positions. The motion statistics are analyzed in
depth, and correlations with some other parameters, in particular the observed
luminosity, are discussed. In a third paper (Britzen et al.\ 2007b, Paper~III),
 the correlation between soft X-ray and VLBI radio properties has been discussed.

In this paper we first give an overview, in \S~\ref{sec:sample}, of the sources
available for the motion analysis. There is a full list of CJF source
identifications and redshifts, which have not previously appeared
systematically; 
we give particulars concerning new redshifts in Appendix~\ref{app:z}.
In \S~\ref{sec:modelf} we review the cross-epoch
identification of components, and discuss the calculation of proper motions and
their uncertainties.  Appendix~\ref{app:fitmeth} delves into more details about
the kinematic model fitting. 
\S~\ref{sec:beta} discusses the conversion to apparent velocities and
some further consistency analysis.
Appendix~\ref{app:veltab} presents the full table of the
kinematic results, together with pertinent source and component data.
We begin the analysis by scrutinizing in
\S~\ref{sec:within} velocity differences between components within individual
jets, as well as the related topics of acceleration and bending. Then, in
\S~\ref{sec:vel-dis} we present and analyze the velocity statistics of the full
sample.  \S~\ref{sec:main-cor} investigates correlations with other radio parameters, and in particular with the
observed radio luminosity, while
\S~\ref{sec:comp} compares our results with other apparent velocity
datasets. We give our conclusions in \S~\ref{sec:conclusions}.

\section{The sample}
\label{sec:sample}

\subsection{Identifications and redshifts}
\label{sub:id-red}

The CJF sample contains 293 radio sources. The properties of their
optical hosts have not yet been tabulated for all sources in previous
papers. We present the relevant data here (Table~\ref{table:proper}) as
it is required to analyze and interpret the apparent jet component motions.\\

We have labeled the hosts as either Q(SO), G(alaxy), or B(L Lac
Object), using one consistent classification from the literature, as
given by V\'{e}ron-Cetty \& V\'{e}ron (2003), whenever this was
available. To classify the remainder of the sources, we have inspected
optical images, initially the POSS, supplemented later with Palomar 60"
and 200" images obtained by some of us (Taylor \&\ Vermeulen,
unpublished). There are 8 optical hosts which have
never been detected, and are therefore U(nclassified); these probably
include galaxies at substantial redshifts as well as some highly
reddened quasars. 
The method of classifying is subject to
some uncertainty; in particular, faint extended emission, and therefore
classification as G, might have been missed in some cases. However,
these ``interlopers'' into the Q class are expected to be, at most, a
minor contaminant of that large class.\\ 

We have taken redshifts in preference from our own work (Vermeulen \&
Taylor 1995; Vermeulen et al. 1996; Henstock et al. 1995), and we have
otherwise used NED to check for the existence of a known redshift
measurement, which we have traced back to the original reference
whenever possible for verification and maximal accuracy.  For 19
sources, our listed redshift (or, in one case, the absence of a
redshift) constitutes an update to previously available information
(see Table~\ref{table:proper}, column 3 for values). Details for these 19
objects are given in Appendix~\ref{app:z}. 

\subsection{Sources omitted from the motion analysis}
\label{sub:dropped}

Not all of the CJF sources could be included in the kinematic analysis.
Table \ref{table:overview} gives a detailed overview of the number of sources
and jet components that have been used for the kinematic analysis, and
those that had to be removed from the analysis for various reasons. The
first column lists the various reasons for source elimination, and source
name(s) that fell victim to each criterion. For each separate reason,
column 2 lists the number of sources affected, column 3 provides the
cumulative number of rejected sources, and column 4 lists the total
number of sources that remain available for subsequent kinematic
analysis.\\

The first part of Table \ref{table:overview} summarizes the loss of sources
during the component model-fit stage, described in Paper~I. We have
found usable information for 266 of the 293 CJF sources on the proper
motions of 779 jet components. Of the 9
sources which could not be used because they are unresolved or have a
single component in all epochs, as many as 6 are quasars at $z>1$. 
Such redshifts are rather common for quasars in the sample. 
The 13 sources in which the components cannot be uniquely identified across
epochs do not stand out from the sample as a whole in their redshifts
or identifications.\\

\begin{figure}[htb]
\begin{center}
\hspace*{0.15cm}\psfig{figure=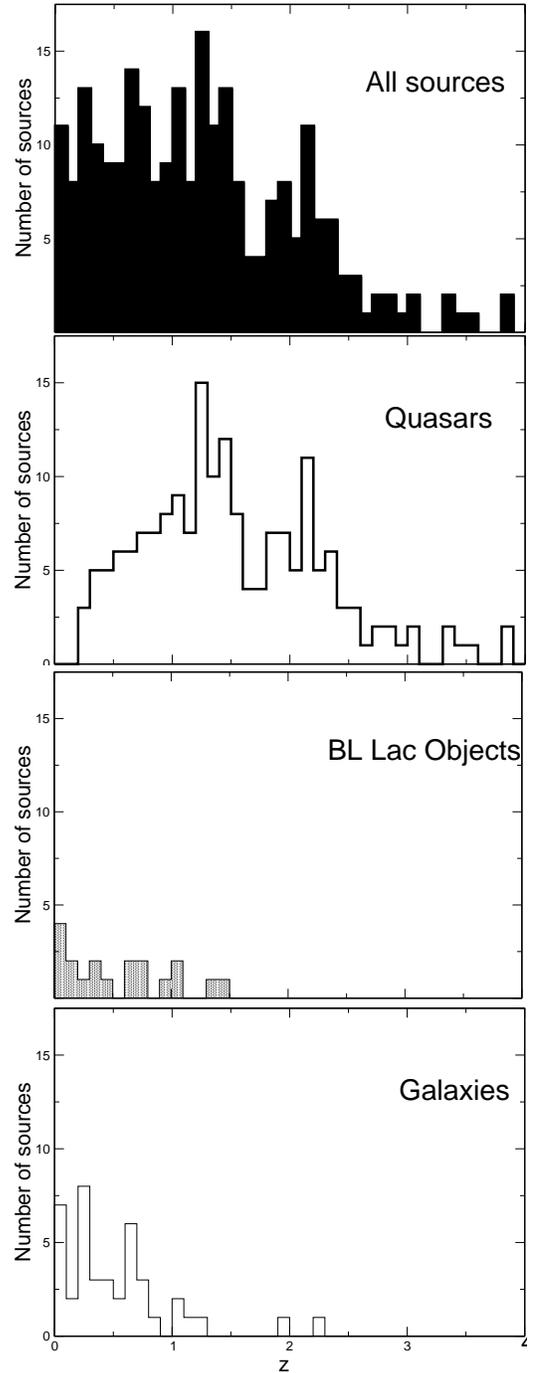,width=7.0cm}
\caption[]{The redshift distribution of the sources included in the
apparent-velocity analysis (see Table~\ref{table:overview}). The top panel
shows all 237 sources; the lower three panels show separate
distributions by source class.}
\label{fig:redshift}
\end{center}
\end{figure}

The second part of Table \ref{table:overview} summarizes those sources lost
from the apparent-velocity analysis (eliminating sources without reliable
redshifts), and breaks down the number of components in all sources
remaining in both the proper-motion and apparent-velocity analysis.
The ``quality" of a component is discussed in
\S~\ref{sec:modelf}. In Fig.~\ref{fig:redshift} we show the
distribution in redshift for the 237 sources which finally remain. This sample comprises
177 quasars, 19 BL Lac objects, and 41 galaxies.  \\

\begin{table*}[htb]                                                                                     
\caption{A brief summary of the numbers of sources and components
taken along in the subsequent proper-motion and apparent-velocity
analysis, following through the various stages of elimination.}
\label{table:overview}
\renewcommand{\arraystretch}{1.2}
\begin{tabular}{|l|lll|}
\hline
(1)	&         (2)  & (3) & (4)\\  
\hline
From the complete CJF to the sources with monitored components.& This & Cum.&Cum.\\
                  &  categ.& elim.& left\\
     \hline
{\bf Paper~I}&&&\\
\hline
CJF sources: &                                        293 &    0 &293 \\
Paper~I: Wrong coordinates used for observations: &   1  &   1& 292\\
0344+405  &&&\\
Paper~I: Too faint or too extended to image:   &      4  &   5 &288\\
0256+424, 0424+670, 0945+664, 1545+497   &&&\\
Paper~I: Unresolved or single component in all epochs: & 9  &  14& 279\\
0621+446 (no $z$, Q), 0636+680 ($z$=3.18, Q), 1254+571 ($z$=0.04217, G) &&&\\
1308+471 (no $z$, VisS by NED), 1342+663 ($z$=1.351, Q)  &&&\\
1417+385 ($z$=1.832, Q), 1638+398 ($z$=1.66, Q), 1851+488 ($z$=1.25, Q), 2005+642 ($z$=1.574, Q)&&&\\
Paper~I: components cannot be identified across epochs: & 13 &  27 & 266\\
0205+722 ($z$=0.895, G): Component identification not sufficiently unambiguous&&&\\
0218+357 ($z$=0.68466, G, BL): Gravitationally lensed&&&\\
0402+379 ($z$=0.055, G): Core identification not unambiguous&&&\\
0615+820 ($z$=0.71, Q): Components not securely identifiable&&&\\
0650+371 ($z$=1.982, Q): Drastic flux-density variability&&&\\
0650+453 ($z$=0.933, Q): Large position angle changes across epochs&&&\\
0718+793 (no $z$): Components not securely identifiable&&&\\
0954+556 ($z$=0.895504, Q): Not sufficient epochs due to faintness of source&&&\\
1144+402 ($z$=1.088, Q): No reliable component identification found&&&\\
1206+415 (no $z$, Q)  Components not securely identifiable&&&\\
1531+722 ($z$=0.899, G)  Blend between core and first jet component&&&\\
1800+440 ($z$=0.663, Q)  Resolution of epochs too different&&&\\
1839+389 ($z$=3.095, Q)  Jet component too faint&&&\\
	\hline
	{\bf This Paper}&&&\\
	\hline
	{\it Proper motion:}&&&\\
Sources with proper motion measurement(s): 266&&&\\
Number of jet components (moving or stationary), all quality classes: 779 &&&\\
Number of jet components, Q1: 305 (Q: 187, G: 81, B: 26, U: 11)&&&\\
Number of jet components, Q2 \& Q3: 474 (Q: 299, G: 101, B: 63, U: 11)&&&\\
{\it Further kinematic analysis:}&&&\\
Sources without redshifts: 29 &29&56&237\\
0018+729, 0102+480, 0346+800, 0633+596, 0700+470, 0702+612, 0716+714, 0740+768,&&&\\
0749+540, 0843+575, 0925+504, 0927+352, 1107+607, 1205+544, 1221+809, 1246+586,&&&\\
1250+532, 1312+533, 1322+835, 1333+589, 1357+769, 1656+482, 1746+470, 1747+433,&&&\\
1828+399, 1926+611, 2010+723, 2023+760, 2138+389&&&\\
Sources with redshifts: 237 (Q: 177, G: 41, B: 19)&&&\\
Number of jet components (moving or stationary), all quality classes: 699 &&&\\
Number of jet components, Q1: 272 (Q: 186, G: 70, B: 16)&&&\\
Number of jet components, Q2 \& Q3: 427 (Q: 295, G: 93, B: 39)&&&\\

\hline
\end{tabular}
\end{table*}

\section{Component identification and proper motion fitting}
\label{sec:modelf}

For all epochs of all sources we first obtained Clean images. We then
fitted circular Gaussian model components directly to the observed
complex visibilities using the Levenberg-Marquardt non-linear least
squares minimization technique (program ``modelfit" within {\it
Difmap}, Shepherd 1997). This
database is presented in Paper~I.\\

The proper-motion analysis is based on a careful identification of the
components across the epochs and multiple checks of the resulting
motions parameterized in rectangular coordinates in the {\it x-y} plane. We do
realize that ``components'' are unlikely to be sharply defined moving
physical entities; rather, they are likely to represent those regions
in jets which produce a surfeit of synchrotron emission as a result of
passing shocks and the details of the magnetic field configuration.
Nevertheless, with observing epochs spanning a number of years it is
possible in most of the sources to trace with confidence the evolution
of such emitting regions, including their propagation along the jets,
and this is what, for brevity, is traditionally called component
motion. When identifying components across epochs in jets with multiple
features, we followed the assumption that those emission regions having
the smallest deviations in flux-density, core separation, position
angle and size between adjacent epochs most likely can be identified
with the same component.\\
\begin{figure*}[htb]
\hspace*{-0.5cm}\subfigure[]{\includegraphics[clip,width=6.0cm,angle=-90]{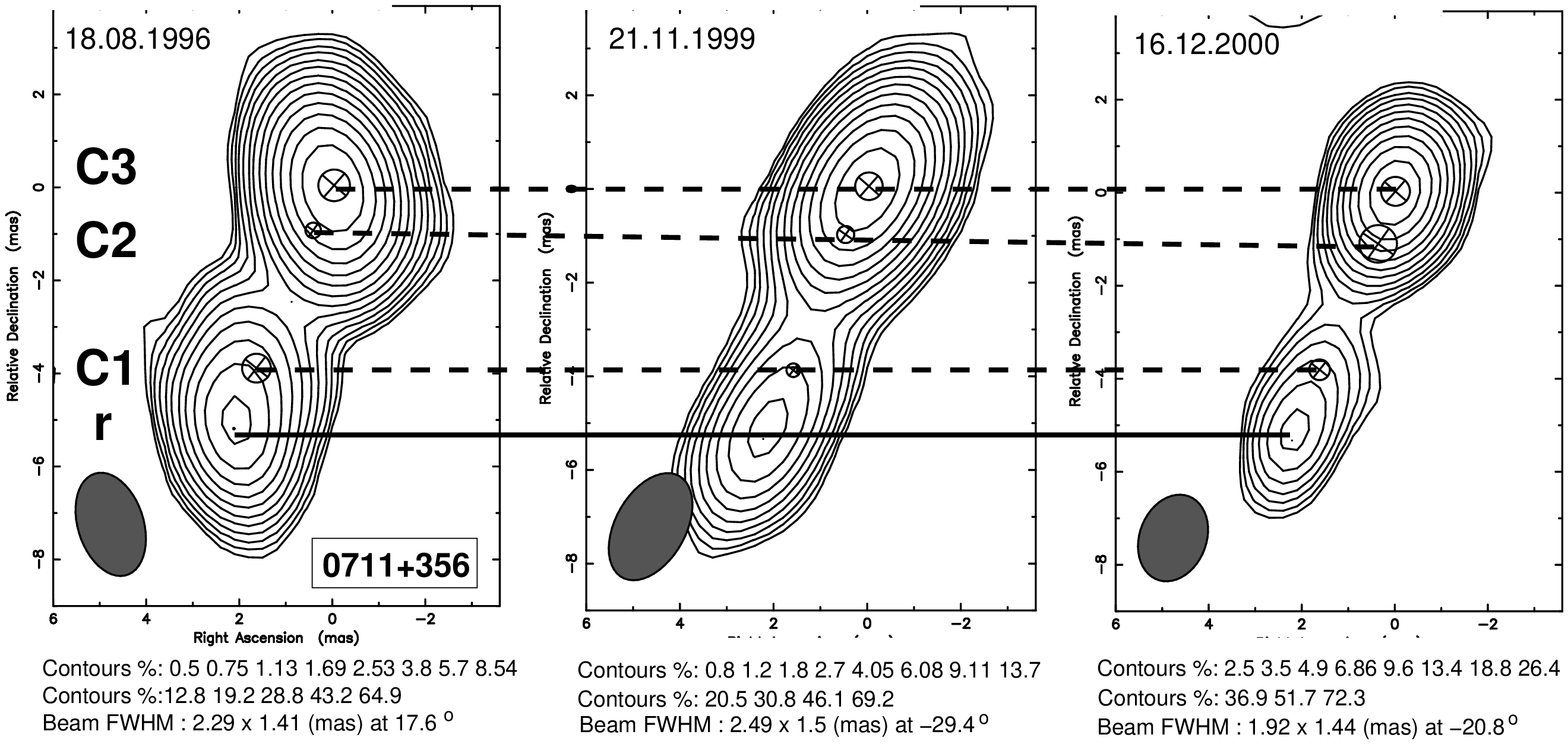}}\\
\hspace*{-0.5cm}\subfigure[]{\includegraphics[clip,width=7.0cm,angle=-90]{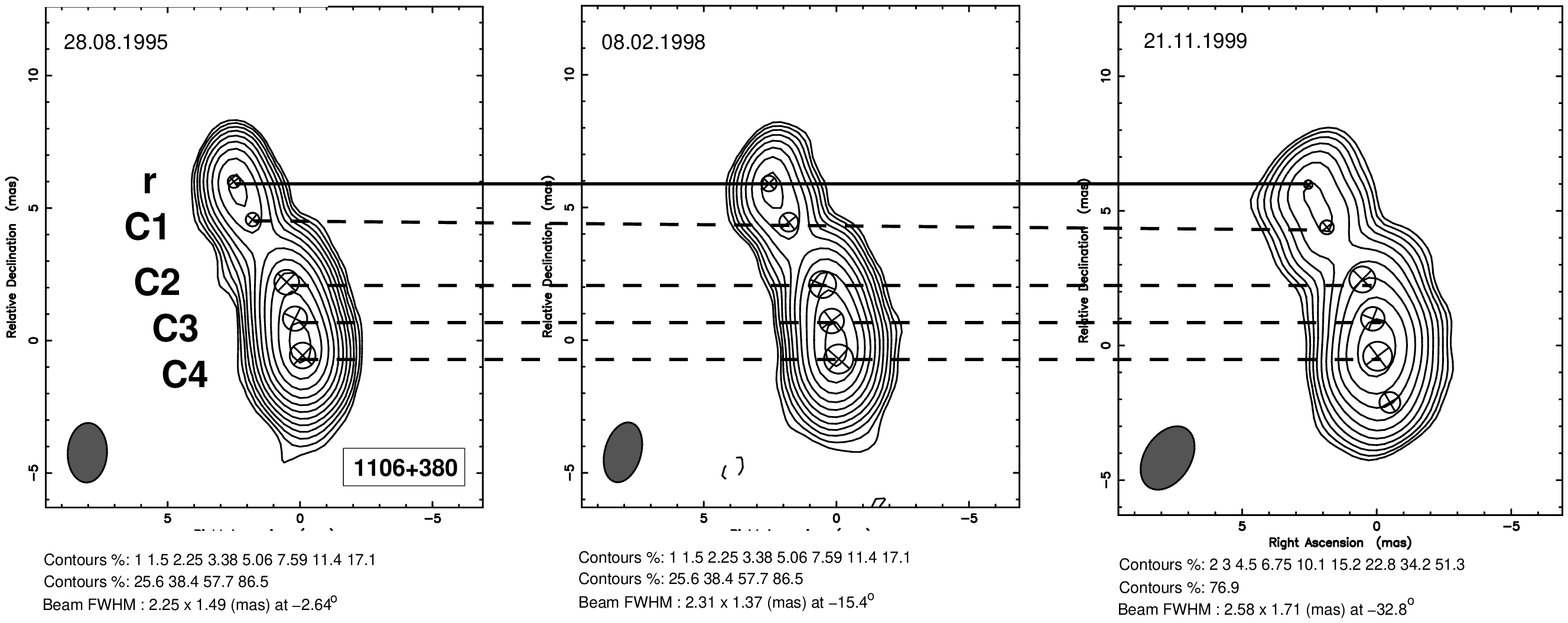}}\\
\hspace*{-1.0cm}\subfigure[]{\includegraphics[clip,width=6.0cm,angle=-90]{fig2c.save.ps1}}
\caption[]{Images are shown, built up by restoring the model-fitted components, convolved with the clean beam, into the residual image, which was made by Fourier transforming the visibility data after first subtracting the model-fitted components in the uv-plane. Over-plotted we show symbols to represent the model components. (a) shows the quasar 0711+356 (all jet components of Q1), (b) the galaxy 1106+380 (all jet components of Q1), and (c) the quasar 2255+416 (C1: Q1; C2,3,4: Q2; C5,6,7: Q3). The components C4, C5, and C6 in 2255+416 split in the last epoch. The identified jet components that can be traced across the
epochs are connected by dashed lines.}
\label{fig:maps1}
\end{figure*}
\clearpage
\pagebreak

It turned out, however, that not all jet components yield proper motion
values of equal significance. The uncertainties of, and correlations
among, the circular Gaussian parameters determined in the jet-component
model-fitting were propagated through the proper-motion analysis
(Table~\ref{table:proper}), but there are probably other sources of
(systematic) errors as well. A quality classification 1, 2, or 3 (high to low) was
assigned to every component, based on a careful assessment of the
component model fits, the images, and the motion in the {\it x-y} plane. Single, bright, jet
components that are clearly separate in all epochs were given a quality
class 1. More diffuse, extended, or fainter
components, and to those in close proximity to others (to the
resolution of the images) were given a quality class 2.  
Any component not uniquely identifiable at
all epochs received quality class 3, which also includes a number of
components that appeared to merge or split, rather than having a
single brightness peak at all epochs. Such phenomena are to be expected
if components represent a complex underlying shock geometry, and in the
more extreme cases have prevented us from deriving a useful motion at
all. We will denote the quality classification of
a component with the syntax Q followed by a numeral 1--3 (a ``Q" without
a trailing numeral can be used to refer to quasars as a class of sources).\\

In Fig.~\ref{fig:maps1} we show examples for jet component identification,
quality-class assignment, and motion in three sources: 0711+356 ($z$=1.620,
Q), 1106+380 ($z$=2.29, G), and 2255+416 ($z$=2.150, Q). The jet
components that can be followed across the epochs are connected by a
dashed line.  The component labels comprise the letter ``C" plus a
number increasing with ordinal separation from the core (although not
illustrated here, the prefix ``CC" denotes a counter-jet component).
The quality classification of all components in these sources are Q1, 
except for some in 2255+416:  components C2--C4 are Q2 and C5--C7  are Q3.
\\

The position of the reference point (r) is marked by a solid black
line. We usually call the most compact feature the reference point.
In a few cases, however,
we have chosen a feature at one extreme end of the source, even if at
some epochs it appeared somewhat extended, because that could then be
ascribed to a flare in progress leading to the formation (``ejection'')
of a new feature in the jet. In some obviously two-sided sources we
chose the most plausible central, core, component to be the reference. 
In others, however, the core was too faint, or unclear, and 
we then used a feature that appears to be the hot spot at the 
end of one of the jets. It is generally assumed that the ``core" feature 
at the base of radio jets is stationary, but it has only rarely been possibly to investigate this directly with phase referencing (e.g., Bartel et al.\ 1986), and has in fact been called into question in some cases (e.g., Guirado et al.\ 1998). The analysis in this paper assumes that the core is stationary to within the errors (but see also \S~\ref{sub:inward}).

\begin{figure}[htb]
\begin{center}
\includegraphics[clip,width=8.5cm]{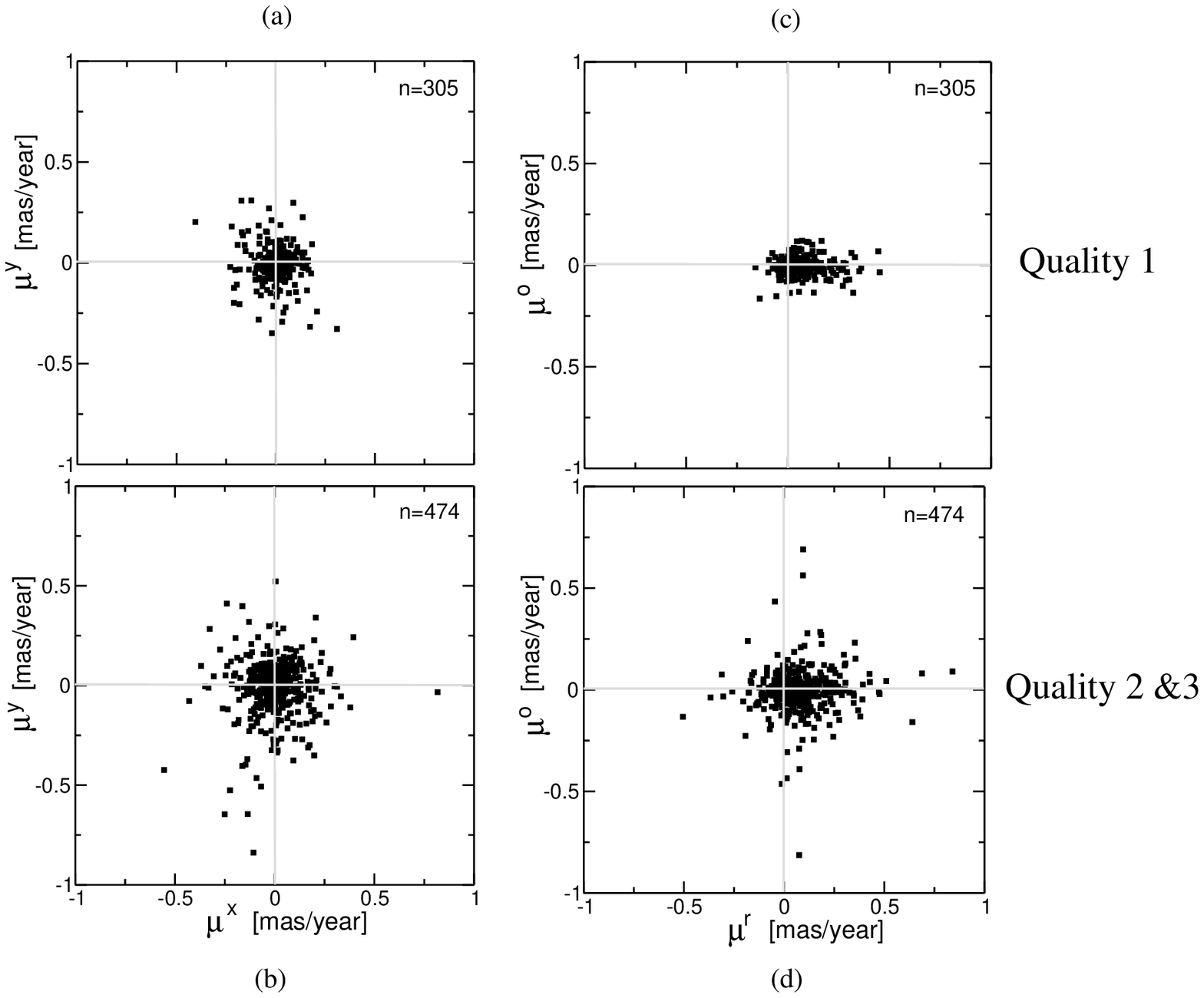}
\end{center}
\caption{The $\mu^x$ and $\mu^y$ components of the proper
motions are shown (a) for Q1 only, and (b) for Q2 \& Q3.
The radial and orthogonal proper motion pairs $(\mu^{r},
\mu^{o})$ are also shown (c) for Q1 only, and (d) for Q2 \& Q3. ``n'' is the number
of values (components) displayed in this figure.}
\label{fig:muxy}
\end{figure}

\subsection{Proper motion calculations}
\label{sub:mucalc}

We calculated the proper motions based on the component identifications
and model-fit parameters presented in Paper~I.  However, effects based on
the dynamic range of the observations and source-intrinsic peculiarities
complicated this calculation.  In particular, jet components could be seen
to merge or split from one epoch to another.  This can result from
comparing observations with different dynamic range or angular resolution
arising from the use of different telescope arrays at different epochs (see
Table~1 in Paper~I).  
Furthermore, there were some sources in which all 
jet components could not be traced through all the epochs.
We investigated several
different ways of handling these sorts of complications in the
estimation of the kinematic parameters for individual jet components.
In Appendix~\ref{app:fitmeth} we provide some details of these procedures but it is appropriate to note a few generic points here.

The kinematic model, for each of the numbered C/CC components, comprises the position $(X,Y)$ at the
reference epoch relative to the reference point, the proper
motion $(\mu^x, \mu^y)$, the associated uncertainties $(\sigma_{\mu^x}, \sigma_{\mu^y})$ and the
correlation matrix.  The reference epoch is the mean of the
observing epochs for that source -- even
if particular components weren't detected in some epochs.
These kinematic models make no allowance for accelerations or other
non-linear motions.  The uncertainties in the kinematic-model parameters are scaled to a reduced chi-square of unity. This procedure helps to allow comparisons among models computed using alternate (relative) weighting schemes for the uncertainties in the component positions ({\it cf} Appendix~\ref{app:fitmeth}).

We also convert the motions for each component from an
$(x,y)$ to a $(radial, orthogonal)$ representation -- both are Cartesian frames,
but the latter is shifted to each component's
$(X,Y)$ and rotated such that $\hat r$ points away from the
component's reference point ({\it i.e.,} the units of $\mu^{o}$
remain mas/yr). The computation of the uncertainties in $(\mu^r,
\mu^{o})$ follow from standard error-propagation from the original
parameters, including correlation information.  The total proper motion
($\mu^{\rm tot}$) is the quadrature sum of the two proper-motion
components (in either representation).

We present the table of kinematic model parameters in Appendix~\ref{app:veltab},
along with corresponding plots. The complete table of model parameters and plots are available in electronic form at the CDS via anonymous ftp to cdsarc.u-strasbg.fr (130.79.128.5) or via http://cdsweb.u-strasbg.fr/cgi-bin/qcat?J/A+A/  

\subsection{Proper motion consistency analysis}
\label{sub:consis}

\begin{figure*}[htb]
\begin{center}
\subfigure[]{\rotate[r]{\psfig{figure=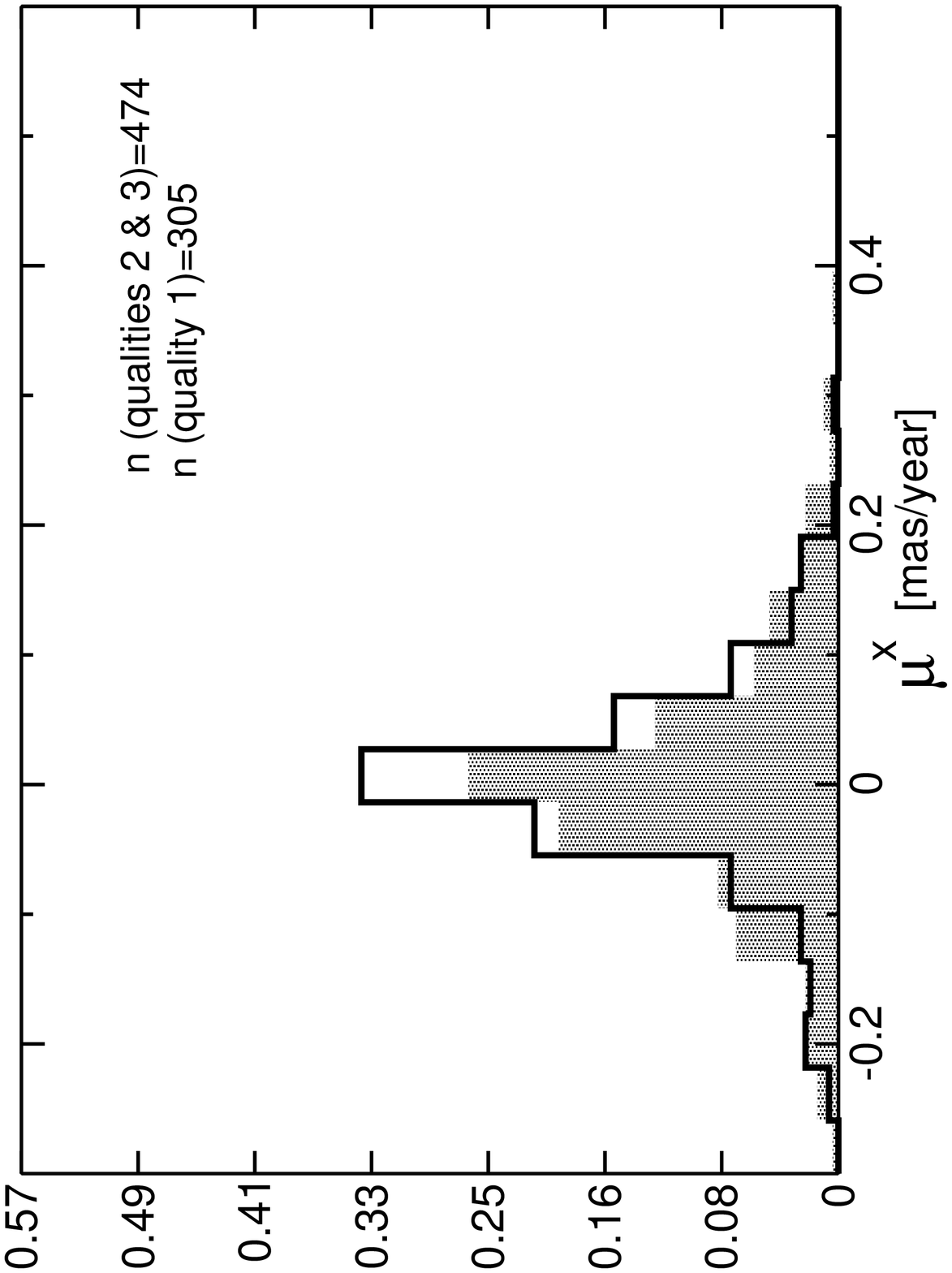,width=6.0cm}}}
\hspace*{0.6cm}\subfigure[]{\rotate[r]{\psfig{figure=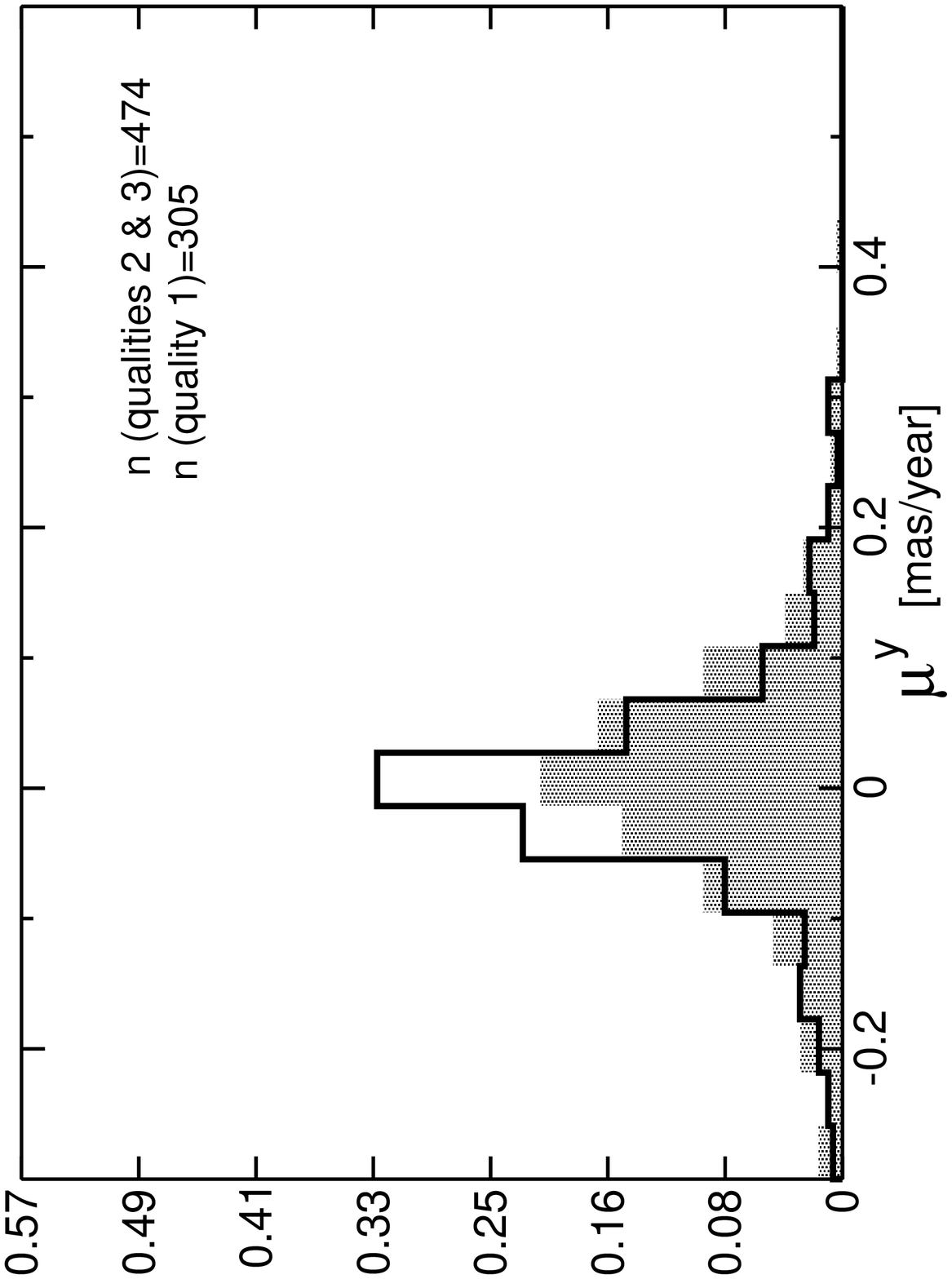,width=6.0cm}}}\\
\vspace*{0.6cm}
\subfigure[]{\rotate[r]{\psfig{figure=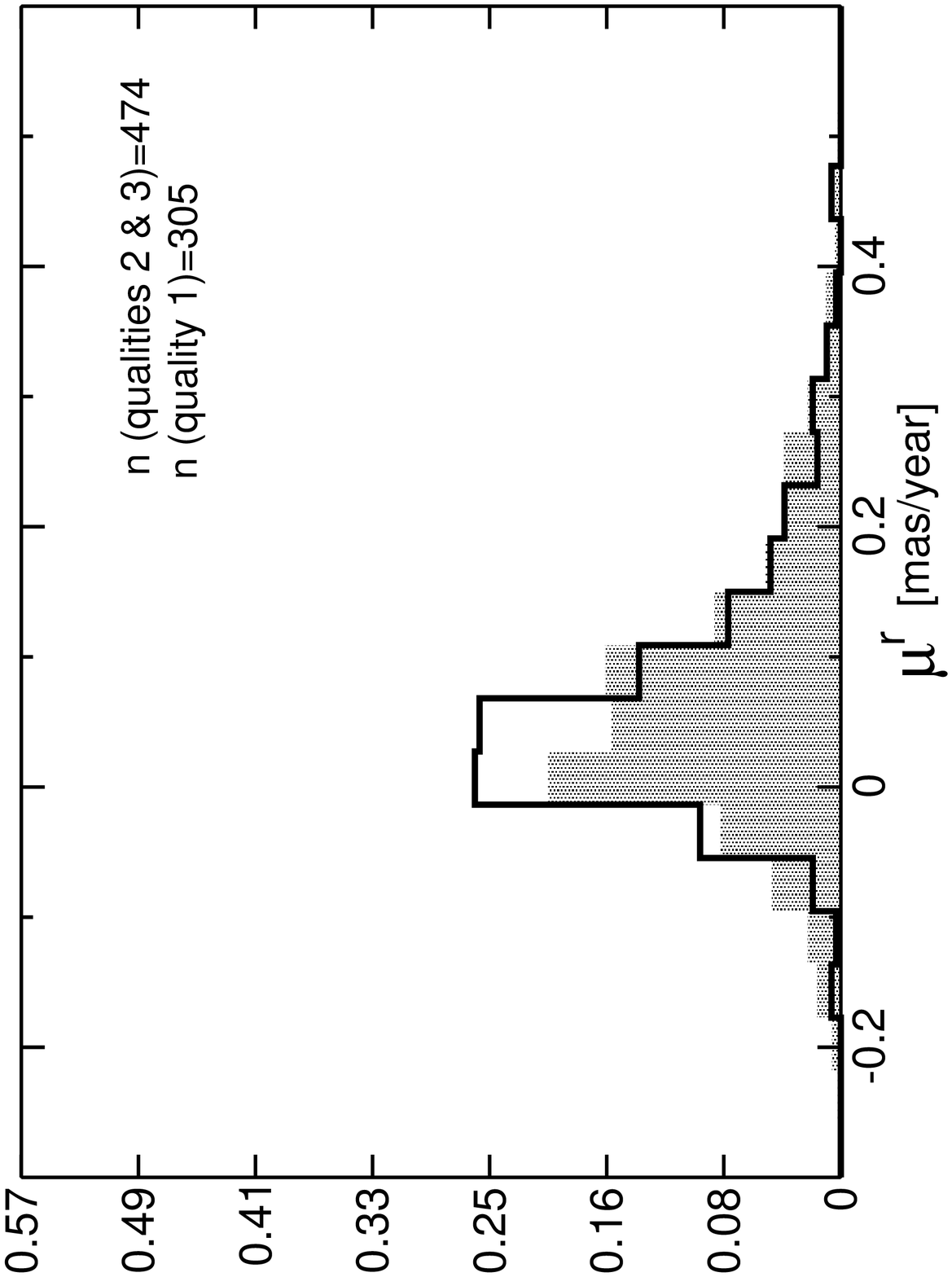,width=6.0cm}}}
\hspace*{0.6cm}\subfigure[]{\rotate[r]{\psfig{figure=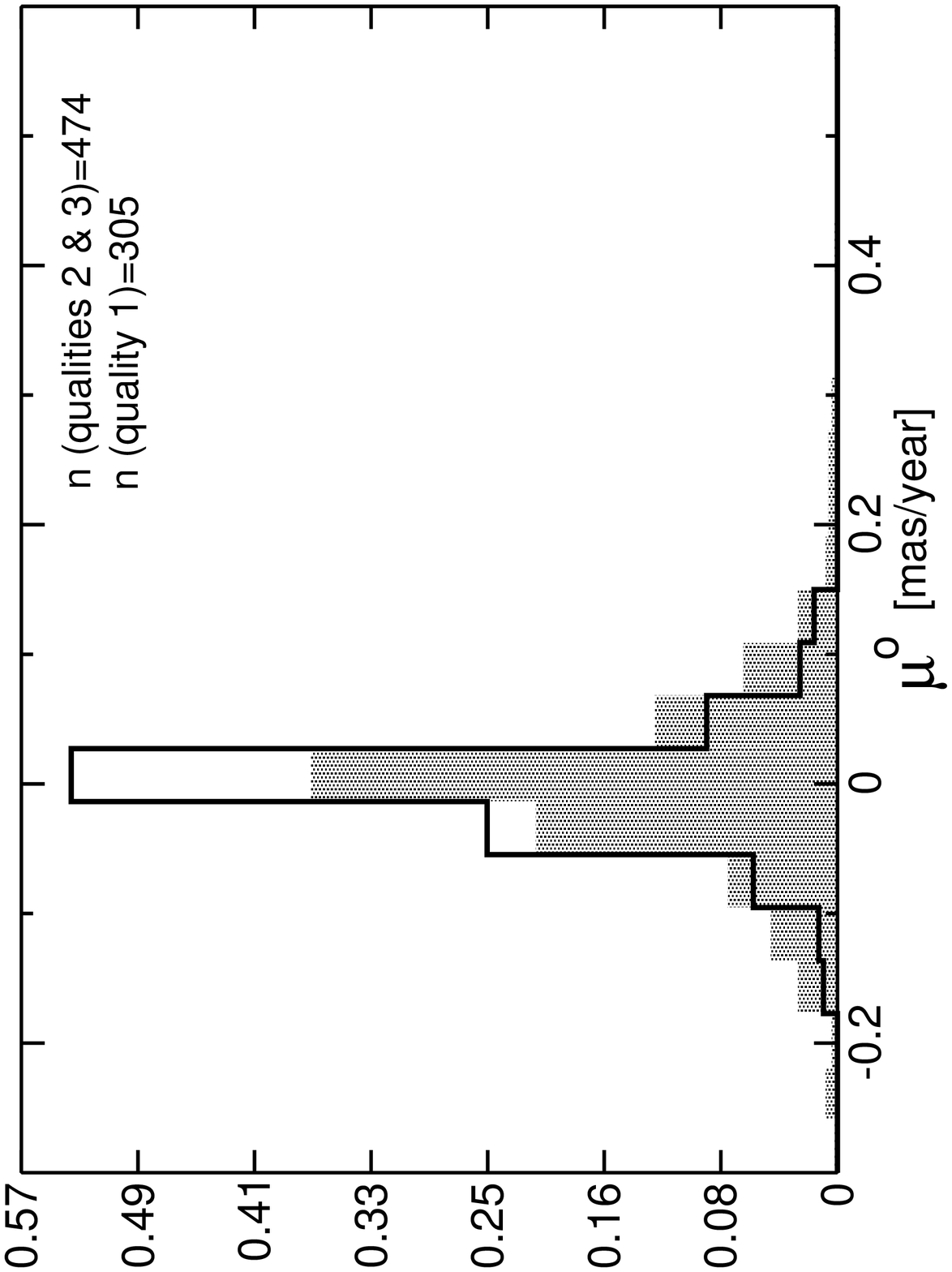,width=6.0cm}}}
\end{center}
\vspace*{-0.5cm}
\caption{Histograms showing the distribution of apparent proper motions in
orthogonal pairs of values: (a) $\mu^x$ with (b) $\mu^y$,
and (c) $\mu^{r}$ with (d) $\mu^{o}$. In each case,
Q1 components are shown with the solid thick black line, and
Q2 \& Q3 are in grey scale. The fractional distribution
was computed separately for each subset, therefore the area under
each curve sums to unity. All further histograms in this paper have
been prepared in this way.}
\label{fig:muxyhis}
\end{figure*}

We now show various diagrams which were designed to check the consistency of our
kinematic modeling and to detect any calculation-induced biases. These
tests mainly serve as diagnostics of possible measurement errors
but in addition already provide some insight into the astrophysics of
these sources.

In Fig.~\ref{fig:muxy} we display the Cartesian motion pairs of the
apparent proper motions: $(\mu^x,\mu^y)$ and $(\mu^r,
\mu^{o})$. The Q1 components are plotted separately from
the Q2 \& Q3 components. Fig.~\ref{fig:muxyhis} shows histograms of
the same data, and allows a better impression of the inner, crowded
part of the distributions. As expected, there is no preferred direction on the sky.  The $x,y$ motions form a circular
distribution around (0,0), with median apparent proper motion values in
the Q1 components of $\mu^x$=0.002$\pm$0.056 mas/yr and $\mu^y$=0.003$\pm$0.056 mas/yr. The corresponding values for the Q2 \& Q3 components are $\mu^x$=-0.006$\pm$0.083 mas/yr and $\mu^y$=0.003$\pm$0.087 mas/yr. 

\begin{figure}[htb]
\begin{center}
\includegraphics[clip,width=6.3cm,angle=-90]{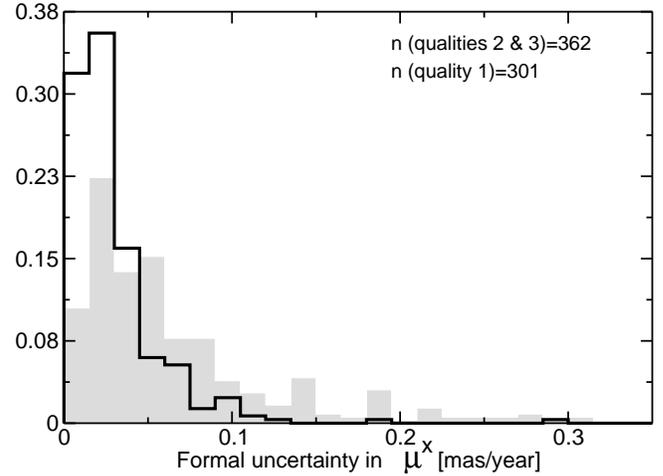}\\
\end{center}
\caption{Histogram showing the distribution of uncertainties in the components of apparent proper motion,
for Q1 only (solid thick black line) and Q2 \& Q3 (grey-scale). The number 
of the jet components in this figure compared to the previous figures 
changed because uncertainties are only calculated for components which 
have been observed in at least two epochs. Please see text for details.}
\label{fig:sigmu}
\end{figure}

     We can also look at the population of components per quadrant in the $\mu^x$--$\mu^y$ plane to see whether their distribution is consistent with placement in four ``unbiased'' bins. The expected standard deviation from a binomial distribution with a ``probability of success'' $p=1/4$ can be expressed as $\sigma_b = \sqrt{np\,(1-p)} = \sqrt{3n}/4$. Considering the placement of components per quadrant regardless of their proper-motion
uncertainties we obtain counts of (68,76,81,71), with a mean of 74 components and $\sigma_b = 7.4$.  The Q2 \& Q3 components show
somewhat more variation in the number of components per quadrant: (107,107,123,130), mean = 116.75, and
$\sigma_b = 9.4$.  The number of components in these counts is
somewhat lower than the total population in Table~\ref{table:overview} because
components having either $\mu^x=0$ (7 components) or $\mu^y=0$ (9 components) to the precision of
Table~\ref{table:proper} were excluded.
If we consider only components falling within a quadrant by at least $1\sigma$ in both $\mu^x$ and
$\mu^y$,
then the distribution of Q1 components is (29,36,35,31), with a mean
of 32.75 components and $\sigma_b$ of 5.0.   For the Q2 \& Q3 components, the distribution is  
(22,29,26,30), mean 26.75 components and $\sigma_b=4.5$.
 Fig.~\ref{fig:sigmu} shows the distribution of $\sigma_{\mu^x}$.
 The median of the uncertainties is $\sigma_{\mu^x}= 0.022$ mas/yr for the
 Q1 components, and $\sigma_{\mu^x}= 0.049$ mas/yr for the Q2 \& Q3 components.  These can be compared to the standard deviations of the $\mu^x$ distribution: 0.078 mas/yr for the Q1 components and 0.115 mas/yr
 for the Q2 \& Q3 components.

   By contrast, and as expected, the distributions are markedly different
   after conversion to the component specific ({\it radial, orthogonal}) frame: clearly,
   there is a predominance of outward radial motions in Fig.~\ref{fig:muxy}(c) and (d) (which we did not
   impose in any way as an {\it a priori} constraint).
   There were some
   components showing inward-directed motion, which will be discussed in
   more detail in \S~\ref{sub:inward}.

We can also examine the prevalence of clockwise or counter-clockwise motion of jet components with respect to the core. Fig.~\ref{fig:muxyhis} shows a roughly symmetrical distribution of the
orthogonal motions in positive and negative values, as one would expect if there is no intrinsic "handedness".  Following a similar
binomial-sampling exercise as above, we can quantify this symmetry (here,
$p = 1/2$ and thus $\sigma_b = \sqrt{n}/2$).  Considering the placement of
components in clockwise or counter-clockwise orthogonal motion without
regard to $\sigma_{\mu^o}$ results in counts of (140,158) with
$\sigma_b = 8.6$ for the Q1 components and (221,241) with $\sigma_b = 10.7$
for the Q2 \& Q3 components.  Thus each sub-group has about a 1-$\sigma$
offset from the expected symmetry.  However, if we consider only the
components that have values of $\mu^{o}$ at least
$\sigma_{\mu^{o}}$ different from 0, then the symmetry is improved:
(62,67) with $\sigma_b=5.7$ for the Q1 components, and (106,102) with
$\sigma_b=7.2$ for the Q2 \& Q3 components.  This also shows that over
half of the $\mu^{o}$ for all quality classifications aren't
significantly different from 0.

However, the Q1 components do have a tighter $\mu^{o}$ distribution
than do the Q2 \& Q3 components:  standard deviations of 0.04 mas/yr for
Q1 and 0.10 mas/yr for Q2 \& Q3.  Along with Fig.~\ref{fig:sigmu}, this
supports the appropriateness of the assignment into quality classes, and strongly suggests that we concentrate our statistical analysis on
the Q1 components. Nonetheless, the evidence is that the lower
quality components are sampling the same motion statistics.
We will revisit this issue in \S\ref{sub:qual1}. 

\section{The database of apparent transverse velocities ($\beta_{\rm app}$)}
\label{sec:beta}

Apparent transverse velocities ($\beta_{\rm app}$) have been calculated using the 1-year WMAP data (Spergel et al., 2003); the values adopted are $h=0.71^{+0.04}_{-0.03}$; $\Omega_m h^2=0.135^{+0.008}_{-0.009}$; $\Omega_{\rm tot}=1.02\pm0.2$. Differences in the apparent velocities
due to differences between the 1-year and 3-year WMAP parameters are
less than $0.1c$ for $\mu=0.1$ mas\,yr$^{-1}$ out to $z=4$; this is
small with respect to the formal uncertainties in the apparent velocities.

The values calculated for the apparent velocities are listed in 
Table~\ref{table:proper}. As noted in Table~\ref{table:overview}, components in
sources without reliable spectroscopic redshifts are omitted from
further analysis. This reduces the number of sources under consideration from 266 to 237, and the total number of jet components from 779 to
699. There are 272 Q1 components in 150 sources, of
which 186 are in 109 quasars, 16 are in 12 BL Lacs, and 70 are in 29 galaxies. 

\begin{figure}[htb]
\begin{center}
\includegraphics[clip,width=5.0cm]{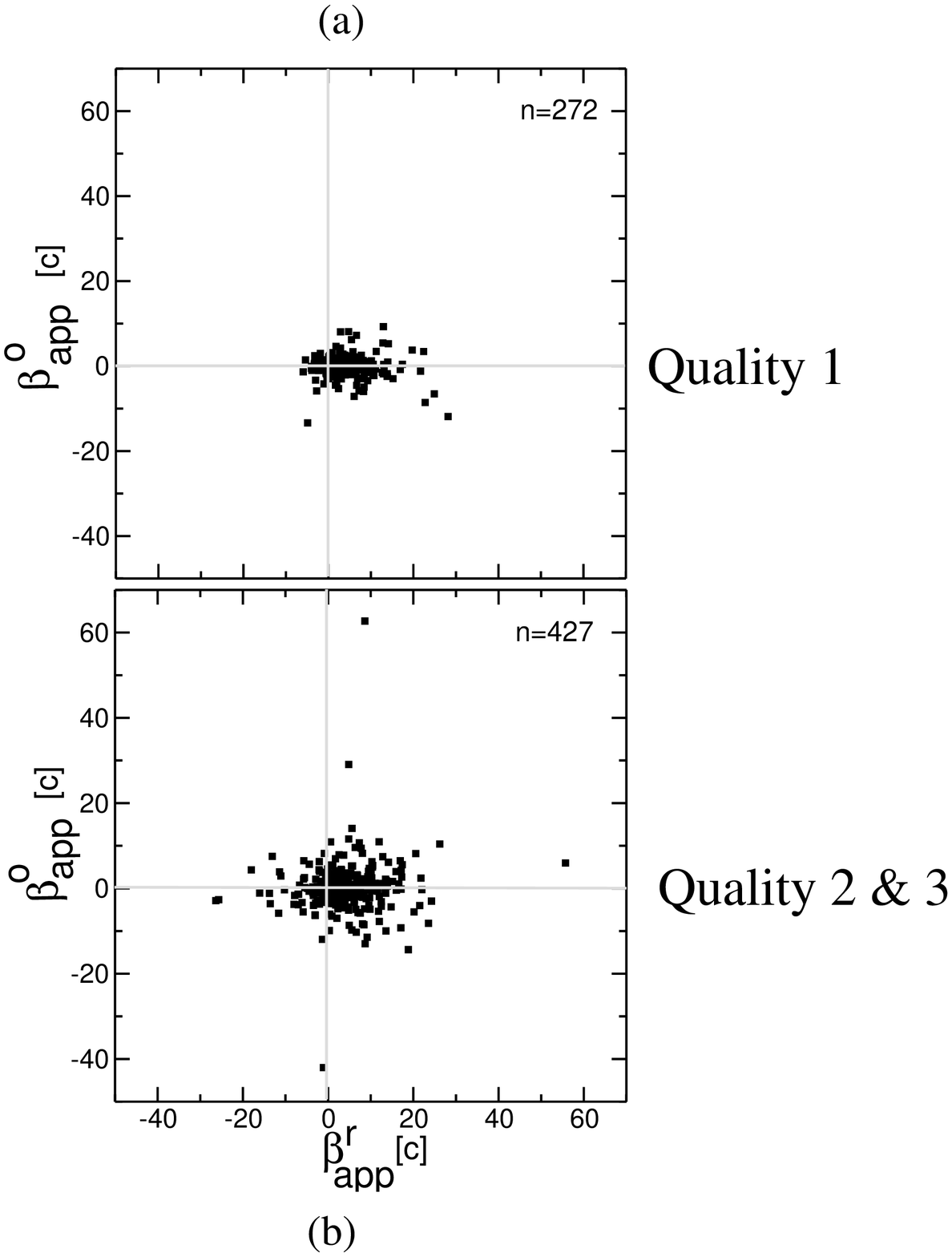}\\
\end{center}
\vspace*{-0.4cm}
\caption{The radial and orthogonal values $(\beta_{\rm app}^{r},\beta_{\rm app}^{o})$ of the apparent velocities for all components in all sources. Panel (a) shows the Q1 components, panel (b) the Q2 \& Q3 components.}
\label{fig:betaro}
\end{figure}

\begin{figure}[htb]
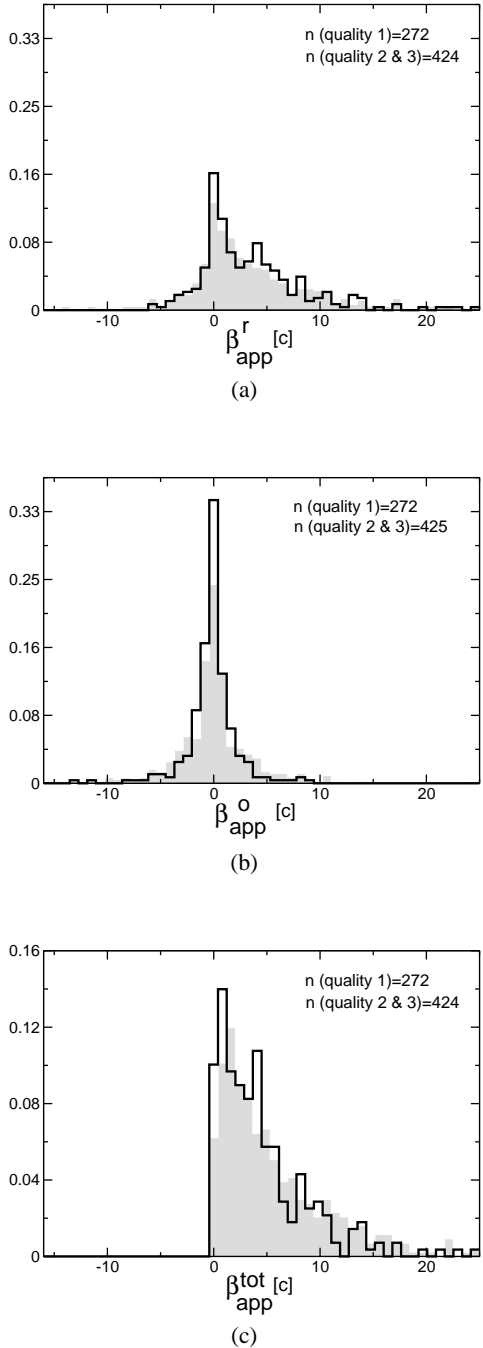

\begin{center}
\vspace*{0.6cm}
\subfigure[]{\rotate[r]{\psfig{figure=betar.histo.ps1,width=5.0cm}}}\\
\vspace*{0.5cm}
\subfigure[]{\rotate[r]{\psfig{figure=betao.histo.ps1,width=5.0cm}}}\\
\vspace*{0.5cm}
\subfigure[]{\rotate[r]{\psfig{figure=betatot.histo.ps1,width=5.0cm}}}\\
\end{center}
\vspace*{-0.4cm}
\caption{The distributions of (a) $\beta_{\rm app}^r$, 
(b) $\beta_{\rm app}^{o}$, and (c) $\beta_{\rm app}^{\rm tot}$ for 
all components. Some of the outliers visible in Fig.~\ref{fig:betaro} 
fall outside the plotted limits. Quality 1 components are shown as a thick 
solid black line, while the distribution of Q2 \& Q3 components is 
shown in grey. }
\label{fig:betarohisto}
\end{figure}

Fig.~\ref{fig:betaro} shows the radial and orthogonal apparent velocity
pairs $(\beta_{\rm app}^r,\beta_{\rm app}^{o})$. Histograms
allow the inner part of the distributions to be viewed more clearly,
and are shown in Fig.~\ref{fig:betarohisto}. These diagrams for
the apparent velocities are directly equivalent to Fig.~\ref{fig:muxy} and
Fig.~\ref{fig:muxyhis} for the apparent proper motions. Again, we see the
predominance of outward radial velocities, but now note some inward velocities, and generally much smaller, but not always insignificant orthogonal
velocity components, with equal prevalence of clockwise and
anti-clockwise directions.
Fig.~\ref{fig:betarohisto} also shows the distribution of the total
apparent velocities, $\beta_{\rm app}^{\rm tot}=\sqrt{(\beta_{\rm
app}^r)^2+(\beta_{\rm app}^{o})^2}$; these are the
velocities used in most previous superluminal motion studies. Note
that, being a quadratic sum, the total apparent velocity is always
positive and has a Ricean noise bias against very low values.

\subsection{Apparent velocity consistency analysis}
\label{sub:qual1}

\begin{figure*}[htb]
\begin{center}
\vspace*{0.6cm}
\hspace*{-0.9cm}
\subfigure[]{\rotate[r]{\psfig{figure=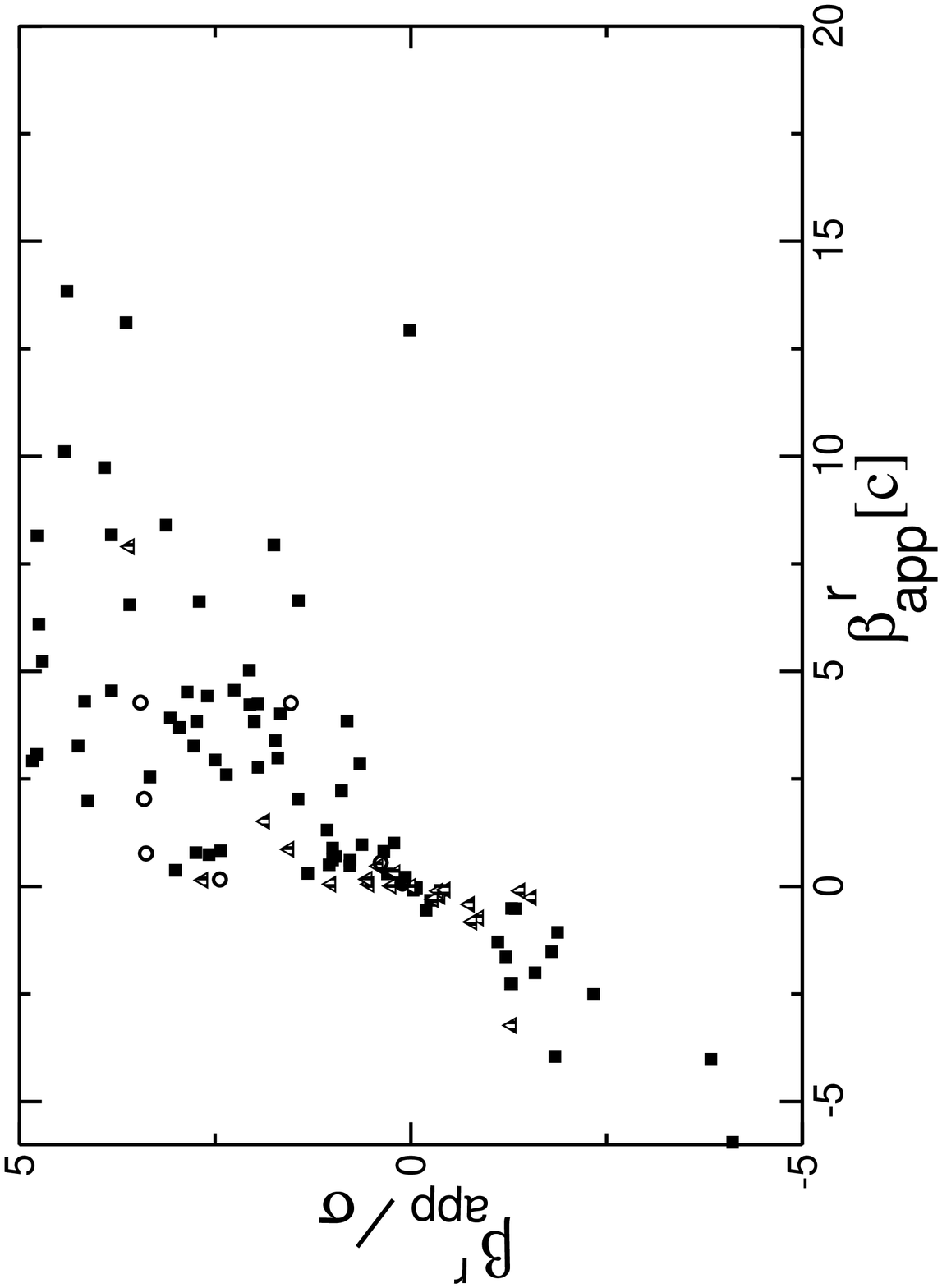,width=5.0cm}}}
\hspace*{0.9cm}
\subfigure[]{\rotate[r]{\psfig{figure=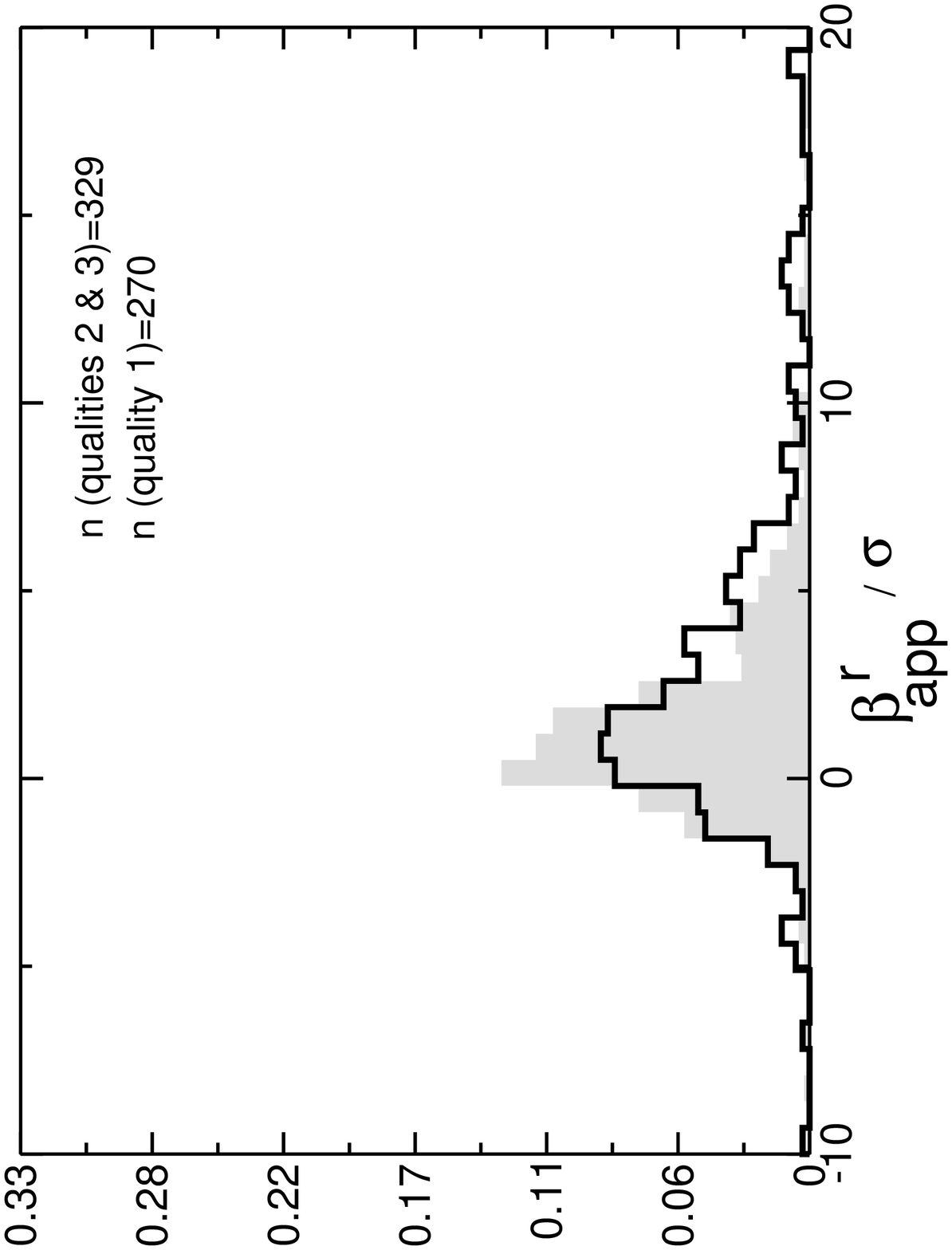,width=5.3cm}}}\\
\vspace*{0.7cm}
\hspace*{-0.9cm}
\subfigure[]{\rotate[r]{\psfig{figure=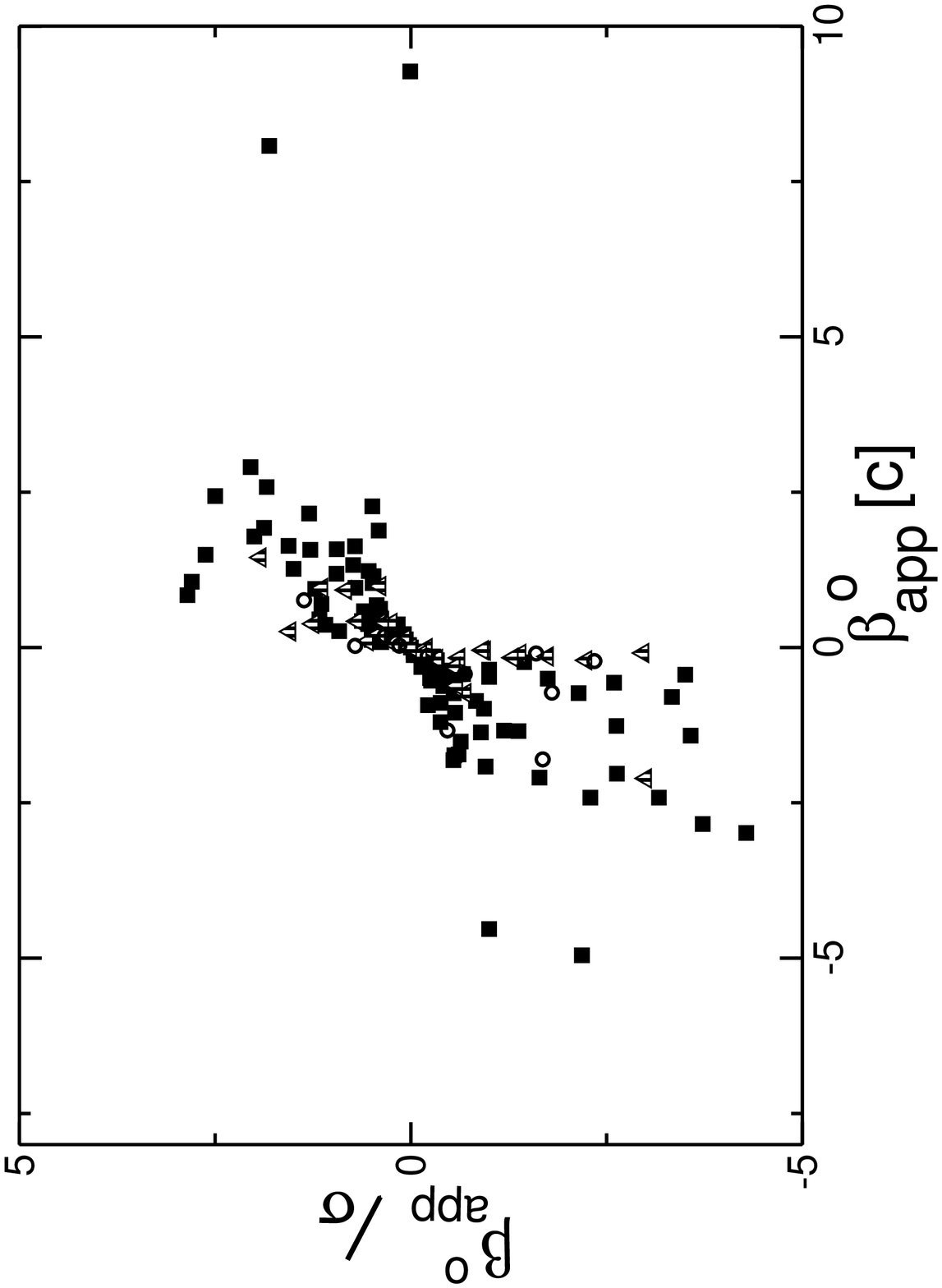,width=5.0cm}}}
\hspace*{0.9cm}
\subfigure[]{\rotate[r]{\psfig{figure=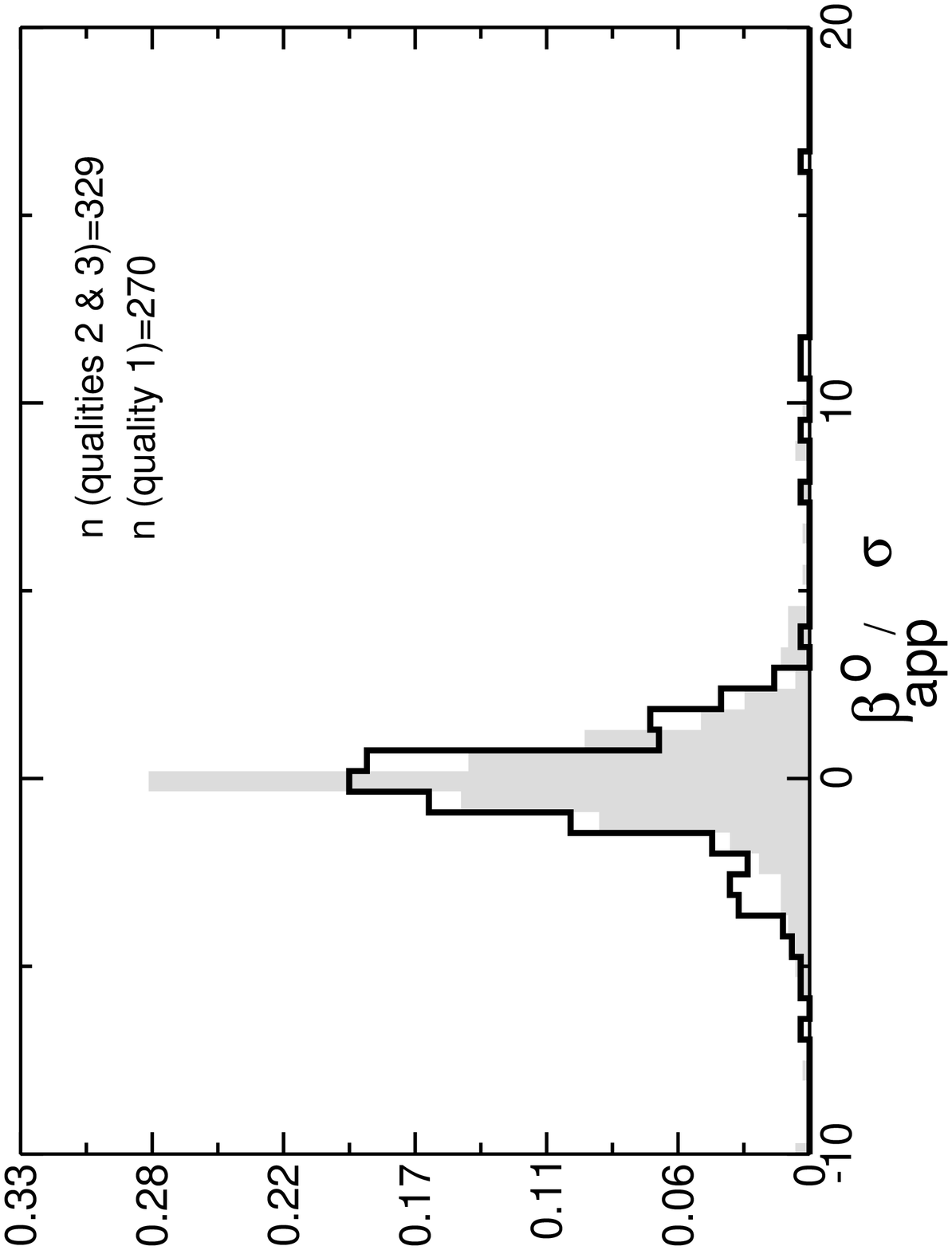,width=5.3cm}}}\\
\vspace*{0.7cm}
\hspace*{-0.9cm}
\subfigure[]{\rotate[r]{\psfig{figure=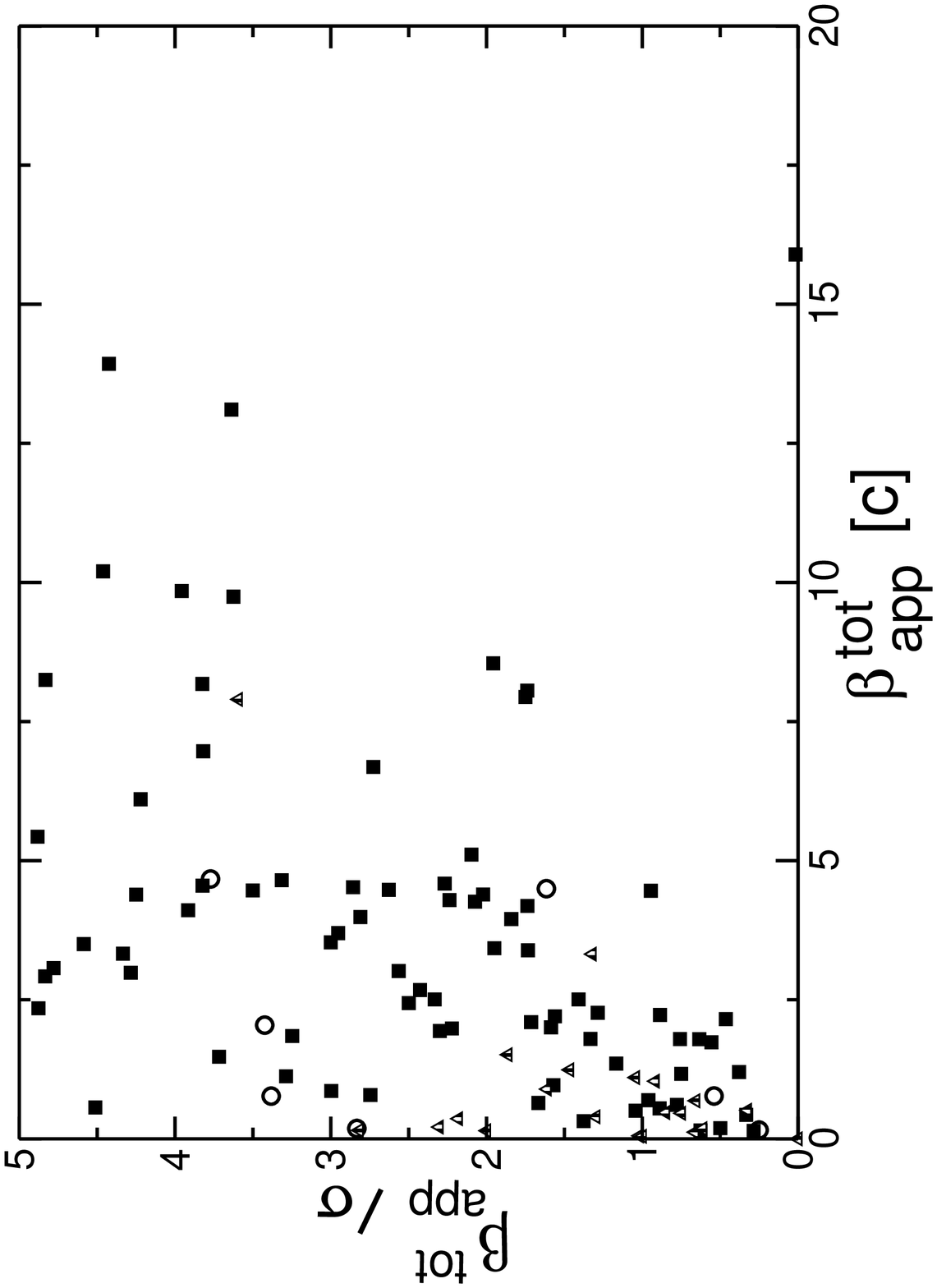,width=5.0cm}}}
\hspace*{0.9cm}
\subfigure[]{\rotate[r]{\psfig{figure=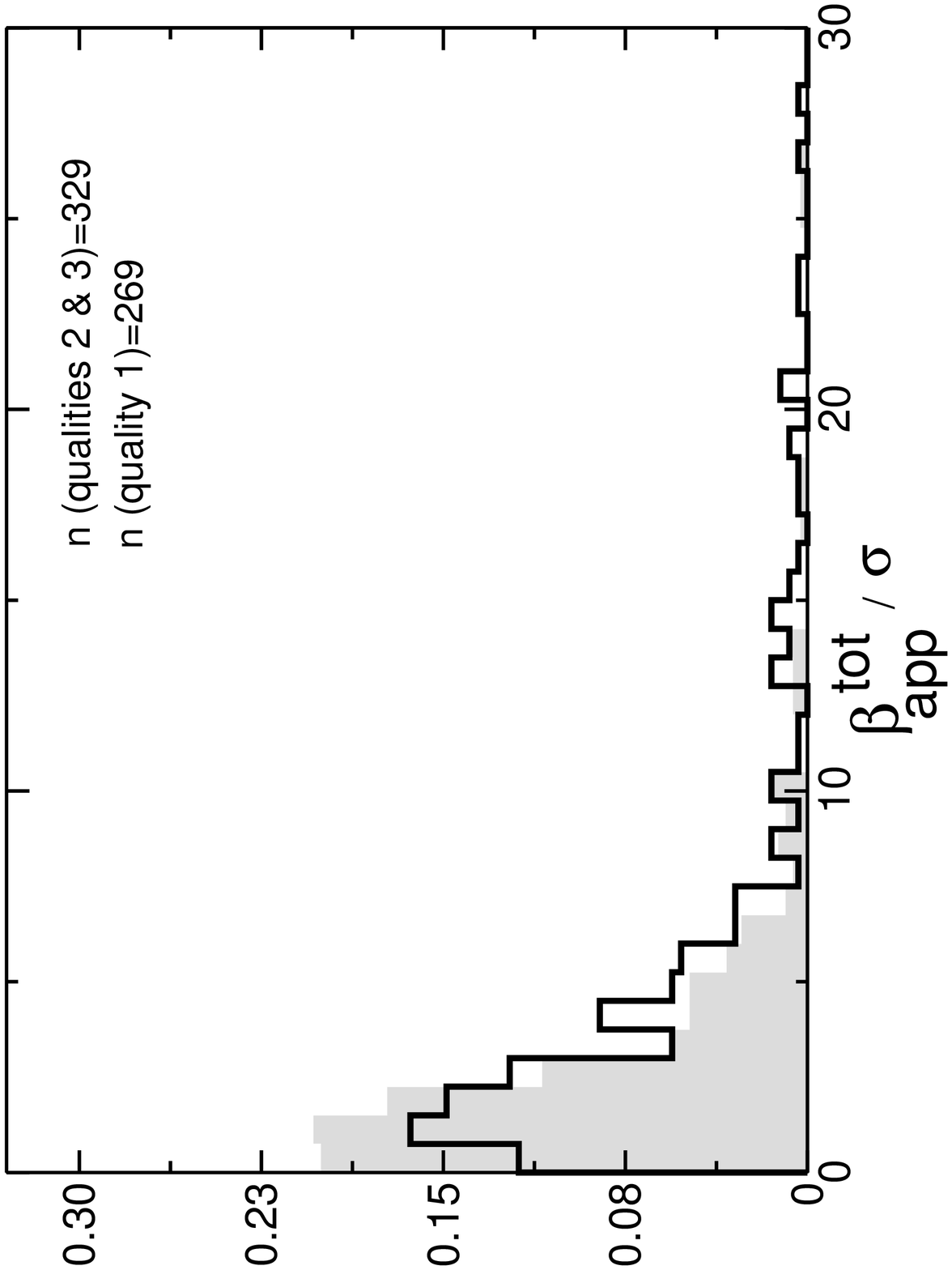,width=5.3cm}}}\\
\end{center}
\vspace*{-0.4cm}
\caption{Overview of the formal significance level of the apparent velocities: the ratio of each velocity $\beta_
{\rm app}$ over its formal uncertainty $\sigma_{\beta_{\rm app}}$.
On the left: scatter diagrams of the formal significance level (ratio of velocity over uncertainty) for all individual
Q1 components, as a function of their velocity. A substantial number of components with velocities significant
at more than $\pm$5$\sigma$ has been cut off.
On the right: histograms of the distribution of the formal significances for all Q1 components (black line), compared
to all Q2 \& Q3 components (in grey). Significance levels in excess of $\pm$5$\sigma$ are now included.
Panels a,b are for the radial velocity, c,d for the orthogonal velocity, and e,f for the quadratically summed total
velocity.}
\label{fig:sigmabeta}
\end{figure*}

Fig.~\ref{fig:sigmabeta} gives an impression of the significance level of the measured velocities. For better comparison we removed the ``spikes'' at 0 which result from components having no defined error. The median formal velocity uncertainty is $\sigma_{\beta_{\rm app}}=\,1.1 c$ for Q1 components, and twice as large, $\sigma_{\beta_{\rm app}}=\,2.0 c$ for Q2 \& Q3 components.
The apparent velocities in excess of a few $c$ have therefore
been measured with high relative significance: most of these components really
are significantly superluminal. On the other hand, a substantial number of
lower velocity values have only a modest relative significance and most of these motions are either subluminal or, at most, mildly superluminal. Some of the low
velocities have, however, been measured with fairly high relative significance, as can be seen from Fig.~\ref{fig:sigmabeta}. Obtaining more precise velocities for a
larger number of slow components would require observations over a much longer
time span and/or with much higher angular resolution.

Figs.~\ref{fig:betaro} and \ref{fig:sigmabeta} show that the distribution of apparent velocities for Q1 components
is tighter than that of the Q2 \& Q3 components,
as it was for the proper motions. But now, the
difference seems to be ``just'' due to outliers. The orthogonal apparent velocity
distributions are centered on zero both in the Q1 components
(median $\beta_{\rm app}^{o}=-0.1\, c$), and in the Q2 \& Q3 components (median $\beta_{\rm app}^{o}=-0.0\, c$), as expected. The median apparent radial velocity for the Q1 components, $\beta_{\rm app}^{\rm
r}=2.4\, c$, is actually slightly larger than the median for the
Q2 \& Q3 components, $\beta_{\rm app}^{r}=2.0\, c$. But this difference is not significant, in view of the formal uncertainties in the velocities, discussed above.
We have decided to carry out all further
analysis in this paper using only the Q1 components, where 
the measurement uncertainties are not only a factor two smaller, but also
more uniformly defined.\\


\section{Motion variations within sources}
\label{sec:within}

In many sources we have been able to track multiple components and we find that
motions within a source are often not consistent with a single uniform
velocity applicable to all components along a jet.  To complement the analysis in section 4 we therefore investigated 
the scatter of apparent velocity, among the components composing
individual jets, via the median of the distribution of standard deviations
in $\beta_{\rm comp}$ taken over all jets. For jets comprising at least three
Q1 components, these medians are 1.97$c$, 2.02$c$, and 0.98$c$ for 
$\beta^{\rm tot}$, $\beta^r$, and $\beta^o$, respectively.
For jets having at least three components of any quality class, these
medians become 2.42$c$, 2.90$c$, and 1.61$c$.  
These scatters can be compared to
the full $\beta$ distributions seen in Fig.~\ref{fig:betarohisto}.
The scatter of apparent velocity in individual jets (comprised of Q1 components only) is thus twice as great as the formal velocity uncertainty of 1.1 $c$ and thus the variations are real.

We now proceed to investigate various intra-jet motion differences.  We begin with a study of accelerations along the jet 
(\S~\ref{sub:accel}) and the related question of an apparent dearth of
high-velocity components at small projected distances from the core 
(\S~\ref{sub:vel-r}).  We then turn to bending along the jet
(\S~\ref{sub:bending}). We conclude in \S~\ref{sub:representative} by discussing ways to obtain a
single ``representative'' velocity per source, as required for properly
weighted population studies. 

\subsection{Accelerations along individual jets}
\label{sub:accel}

Evidence for increasing velocity in pc-scale jets has been found before: e.g.,
M87 (Biretta et al. 1995); 3C84 (Dhawan et al., 1998) and Cygnus A (Krichbaum
et al., 1998). It is possible that the apparent acceleration is not
intrinsic but rather results from a change in the jet direction or a change
in the jet pattern velocity unrelated to the jet bulk velocity. In NGC 315,
Cotton et al. (1999) found an increasing jet velocity in the inner 5 pc from
the core based on proper motion measurements as well as on the sidedness ratio.
In FR I radio sources there is observational evidence that jets decelerate on larger scales than is relevant here, in the sense that they are relativistic near the core and become non-relativistic within a few kpc (e.g. 3C 449, Ferreti et al., 1999).

Accelerations and decelerations along the lengths of the jet as well as
transverse velocity profiles are expected from theoretical considerations
(e.g., Blandford 2005) and jet simulations (e.g., T\"urler et al. 2000). The jet parts that we see
are most likely shock waves and this introduces additional kinematic
complexity, since the direction and the speed of the emitting plasma behind a shock
front must differ from the kinematic speed of the shock. 

With, in general,  only 3 epochs per source in our observations, we cannot properly investigate
acceleration of individual components ($d\beta/dt$).  Instead we characterize
the acceleration along a jet (or counter-jet)
in its entirety ($d\beta/dr$) by performing
a weighted least-squares fit for
a linear model to $\beta_{\rm app}(r)$.  
The linear coordinate corresponds to the angular $r$ along a 
jet/counter-jet axis via the angular-diameter distance $D_A = D_M/(1+z)$.
As such, this is a global 
property of the jet, rather than a local property of each component.
For computing these accelerations, we have used 
$\beta^r_{\rm app}$ ({\it e.g.,} 1106+380 in the plots in Appendix~\ref{app:veltab}). 
This is a very
simple model, and some of the jets with many components are clearly
not well described by a single acceleration ({\it e.g.,} 0633+734, 1755+578).
In order to be included in this analysis, a jet must pass three hurdles:
(1) the host source must have a redshift (in order to compute the acceleration
in units of [c/lyr]); (2) it must have at least three components
of the specified quality class (to ensure the least-squares fit for the
acceleration has at least one degree of freedom); (3) these three components
must have been detected in at least three
epochs (to ensure that the kinematic fit that produced their proper motions
[\S~\ref{sub:mucalc}] also
had at least one degree of freedom). The source's core was
not included as a component. Fits were carried out
using components of all quality-classes (number of jets per source class:
quasars=70, galaxies=25, BL Lacs=10; total=105) and within Q1 components only 
(Q=14, G=8, B=1; total=23). Galaxies seem
relatively over-represented in the Q1 sample compared to quasars, simply because a smaller fraction of quasars have three Q1 components in a
given jet. In the Q1 sample as a whole, all but two jets have either 3 or 4 components.

\begin{figure}[htb]
\begin{center}
\vspace*{0.3cm}
\subfigure[]{\rotate[l]{\psfig{figure=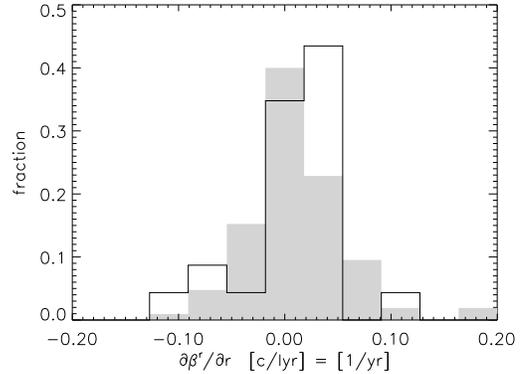,width=5.5cm}}}\\
\vspace*{0.5cm}
\subfigure[]{\rotate[l]{\psfig{figure=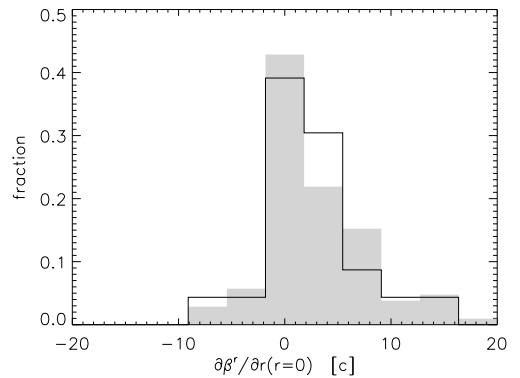,width=5.5cm}}}\\
\end{center}
\vspace*{-0.4cm}
\caption{Histograms of the parameters of the weighted least-squares
	 fit to $\beta_{\rm app}^r(r)$.
	 Panel (a) shows the acceleration term and panel (b) the 
	 velocity intercept at $r=0$.
         The Q1-only jets are represented by the thick line and those
         without quality-class constraint by the shaded area.}
\label{fig:braccel}
\end{figure}

\begin{figure}[htb]
\begin{center}
\vspace*{0.3cm}
\subfigure[]{\rotate[l]{\psfig{figure=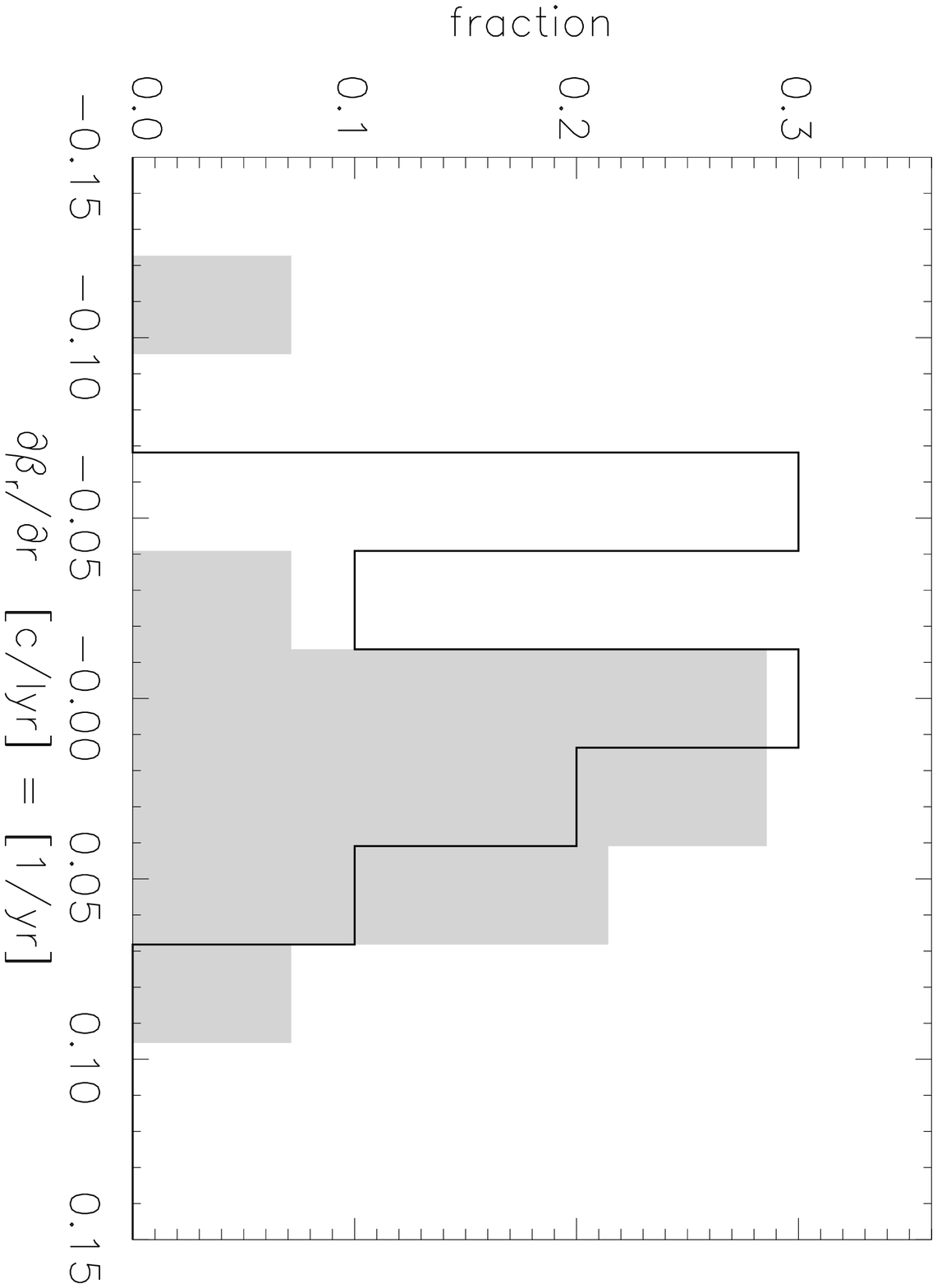,width=5.5cm}}}\\
\vspace*{0.5cm}
\subfigure[]{\rotate[l]{\psfig{figure=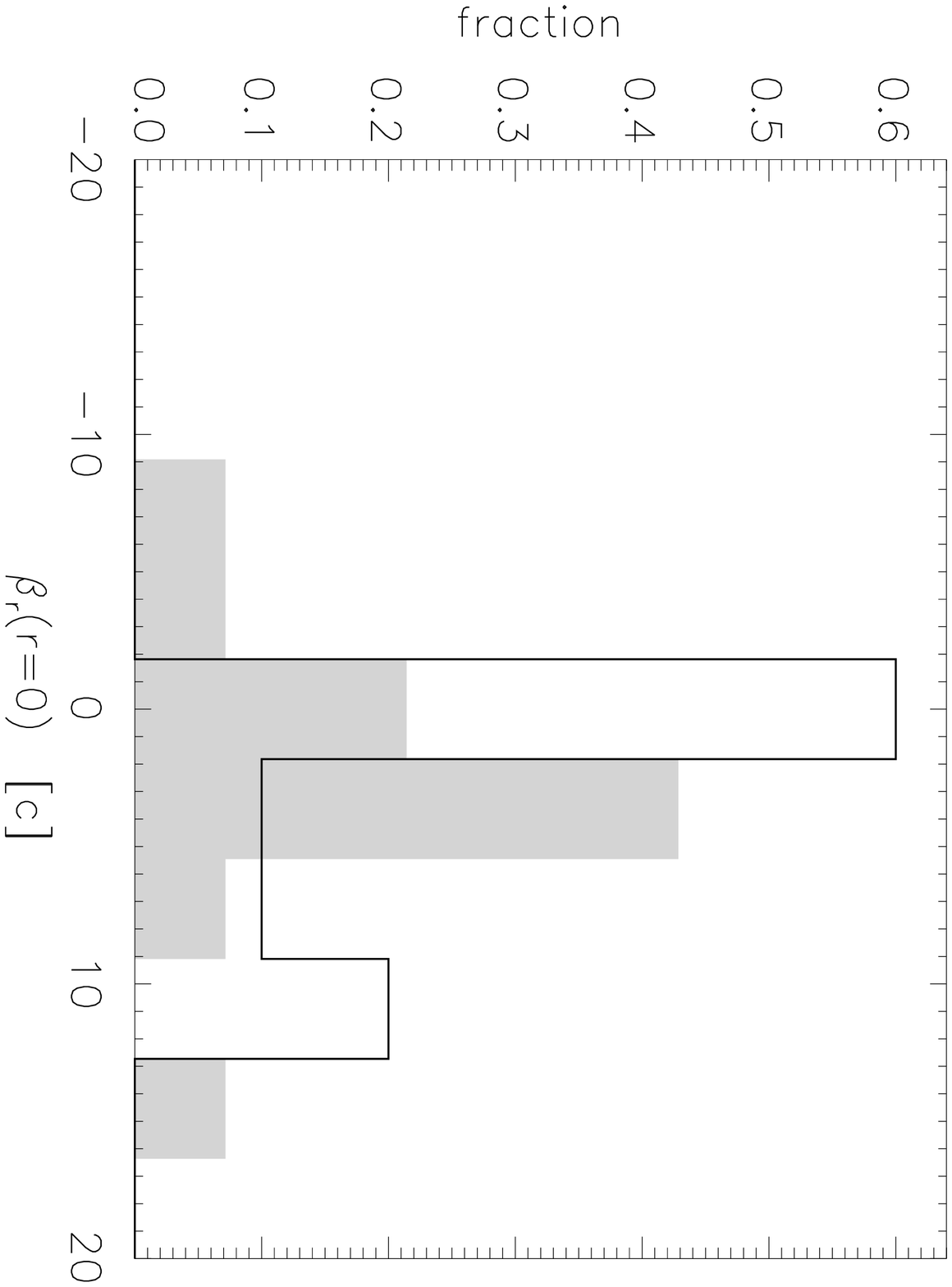,width=5.5cm}}}\\
\end{center}
\vspace*{-0.4cm}
\caption{Histograms of the parameters of the weighted least-squares
	 fit to $\beta_{\rm app}^r(r)$ for the Q1-only jets from
         quasars (shaded) and galaxies (thick line).
	 Panel (a) shows the acceleration term and panel (b) the 
	 velocity intercept at $r=0$.}
\label{fig:accSRC}
\end{figure}

Fig.~\ref{fig:braccel} plots histograms of both the acceleration and
velocity-intercept ($\beta^r$ at $r=0$) obtained from the linear fit 
to $\beta^r_{\rm app}(r)$.
There is a slight trend towards a positive
median acceleration and a clear signal of positive velocity-intercept, both
statements being more significant for the Q1-only sample.
The positive velocity-intercept condition merely states that the fitted line 
$\hat\beta^r_{\rm app}(r)$
extrapolated back to the core still has a positive value.  The implication is that, statistically, there is greater acceleration in the inner-most regions which the resolution of our observations is not able to sample properly.  
Fig.~\ref{fig:accSRC} shows histograms of the acceleration and 
velocity-intercept for quasars and galaxies from the Q1-only sample.
As a class, the quasars show more evidence for accelerations having a more positive $\partial\beta^r/\partial r$ and $\beta^r\,(r=0)$.

\subsection{A dearth of high-velocity components close to cores?}
\label{sub:vel-r}

\begin{figure}[htb]
\begin{center}
\subfigure[]{\includegraphics[clip,width=4.0cm,angle=-90]{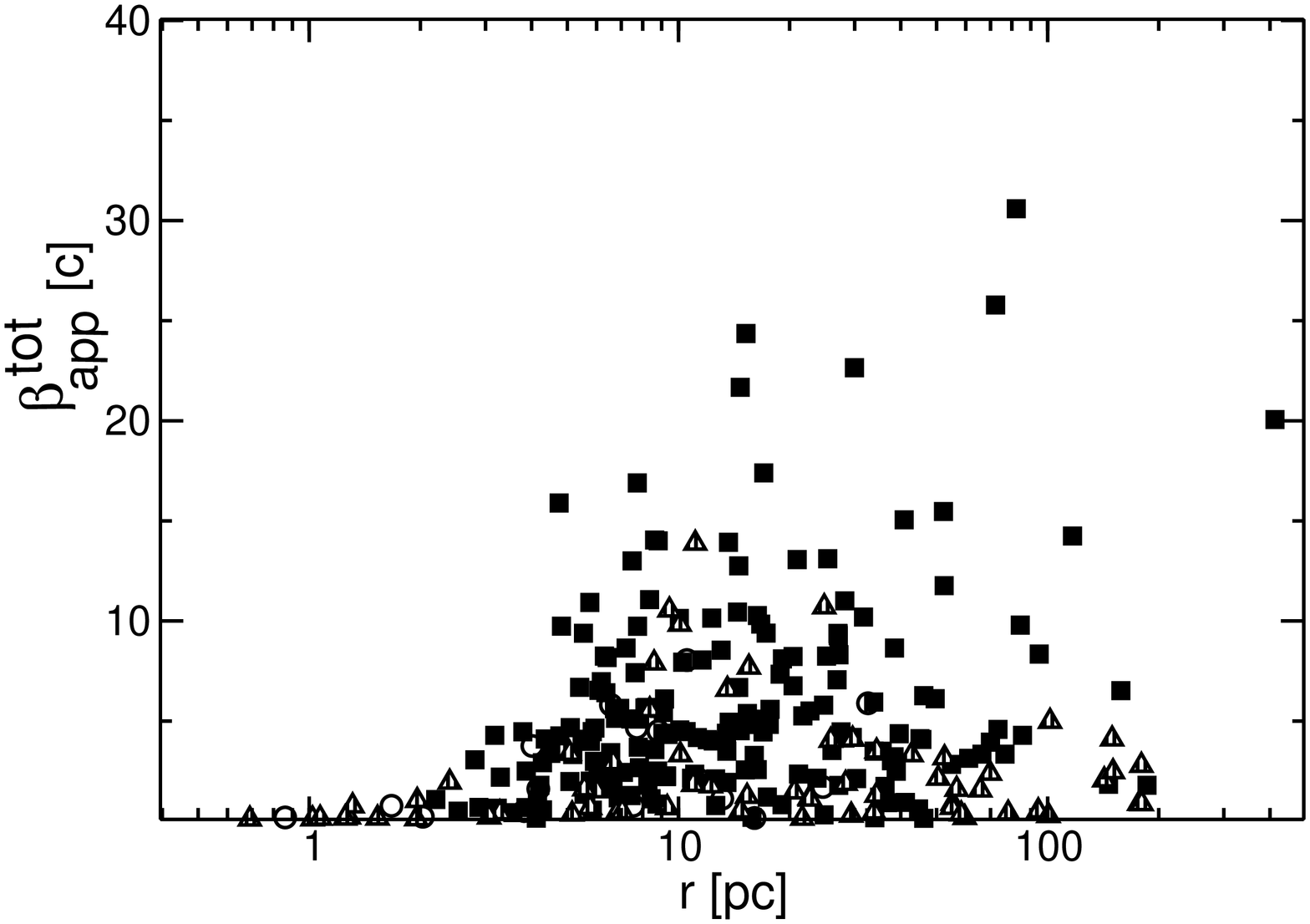}}\\
\vspace*{0.3cm}
\subfigure[]{\includegraphics[clip,width=4.0cm,angle=-90]{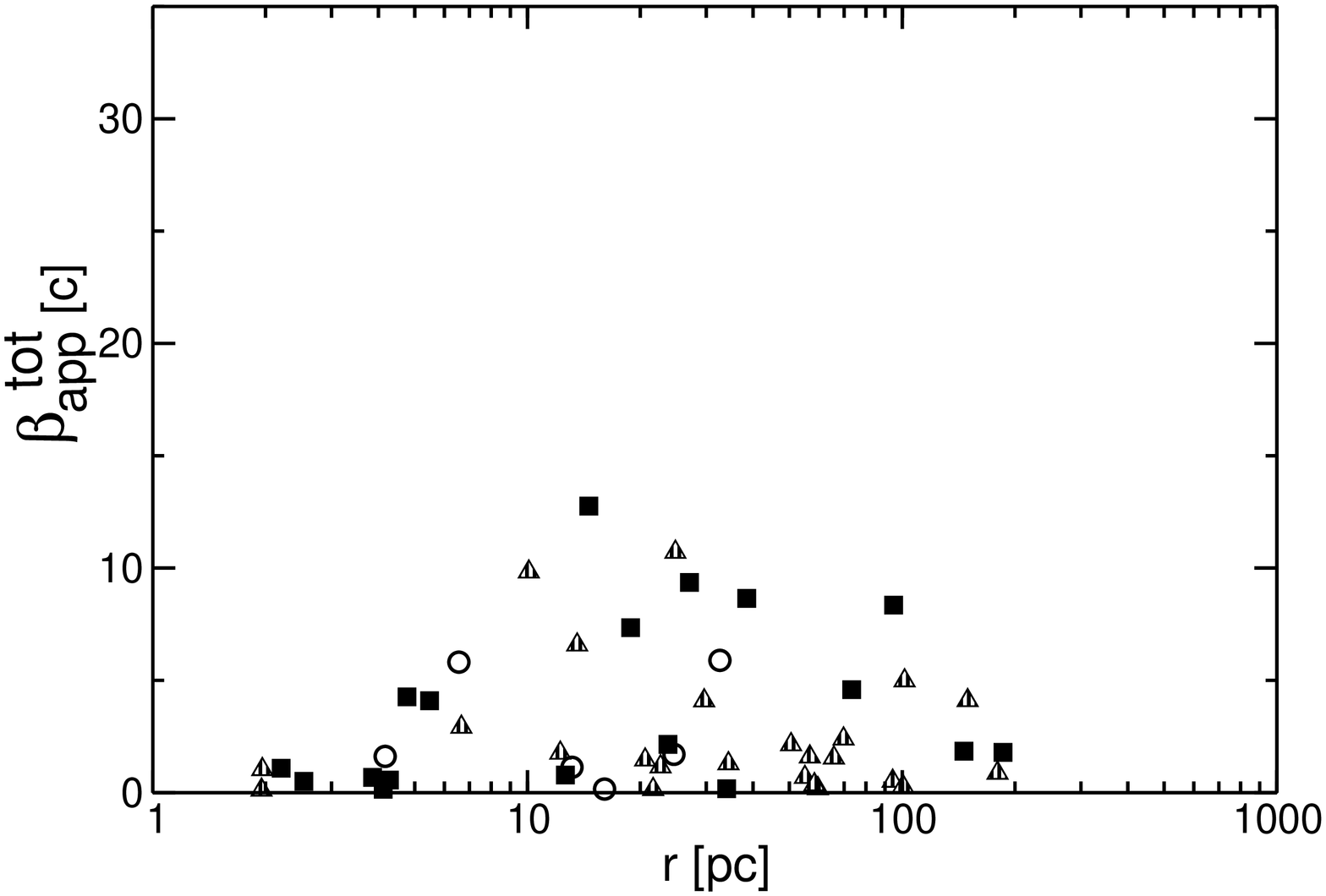}}\\
\vspace*{0.3cm}
\subfigure[]{\includegraphics[clip,width=4.0cm,angle=-90]{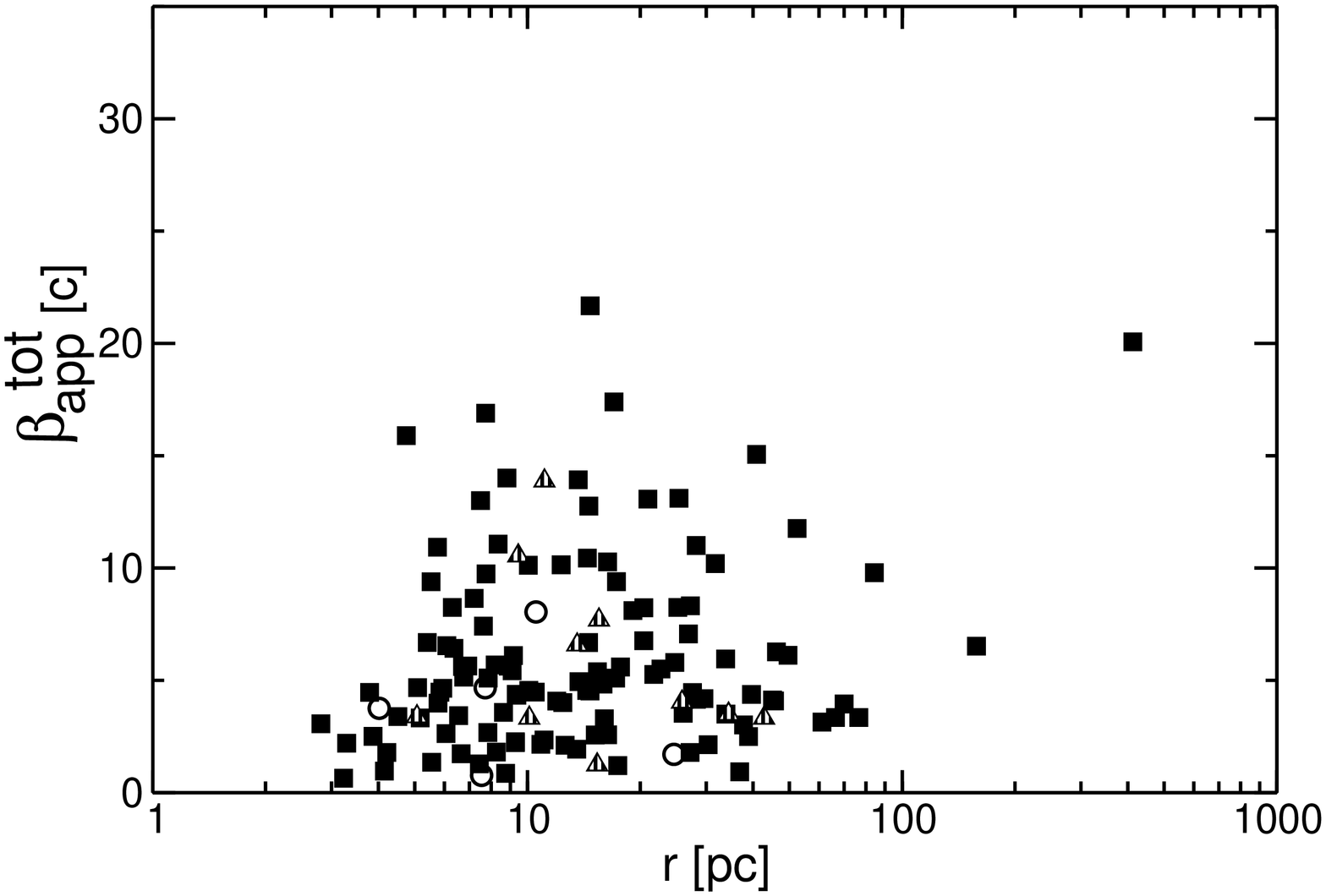}}\\
\vspace*{0.3cm}
\subfigure[]{\includegraphics[clip,width=4.0cm,angle=-90]{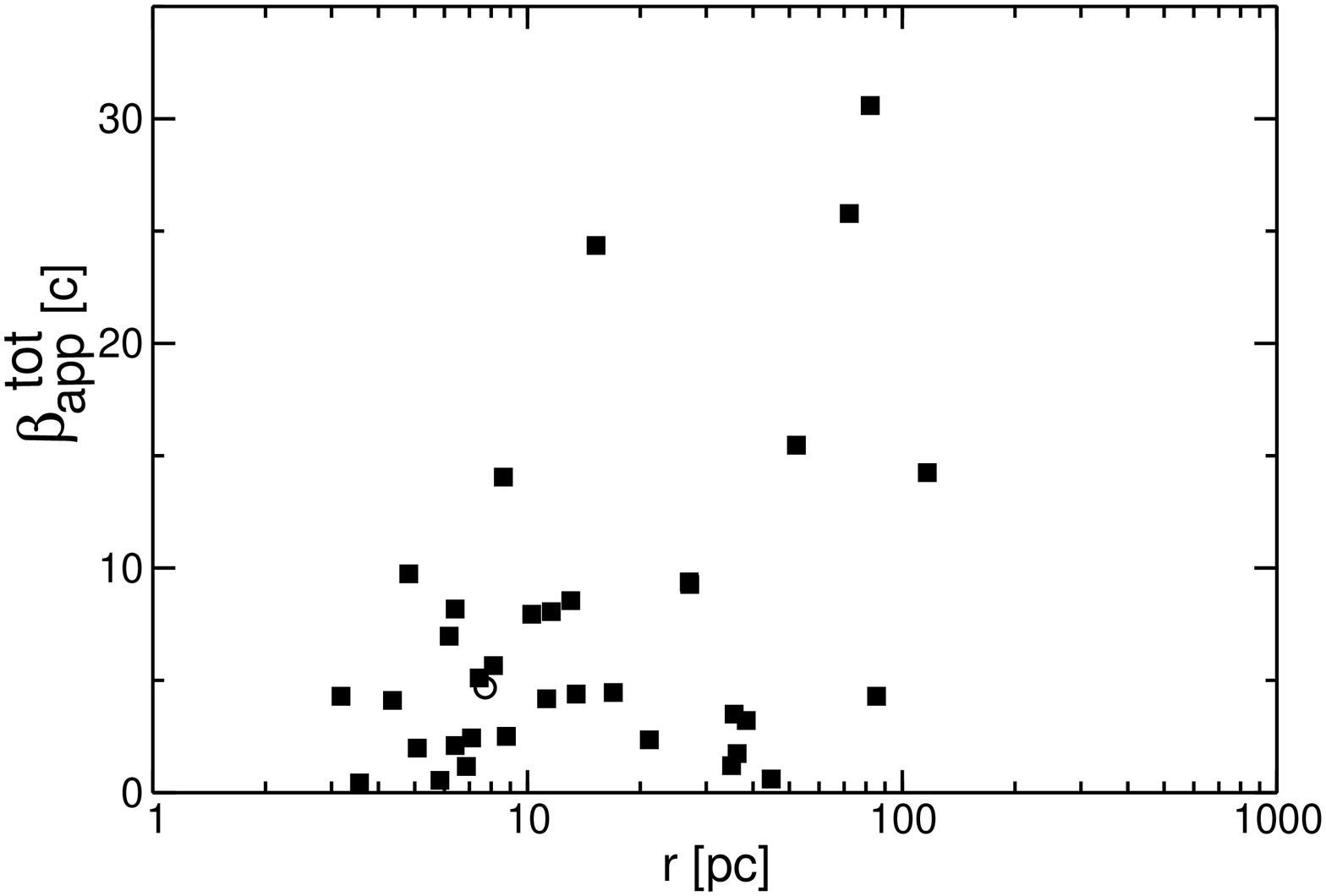}}\\
\end{center}
\caption{$\beta_{\rm app}^{\rm tot}$ as function of the projected
  radial distance from the core. Panel (a) shows all Q1 components. The other panels show Q1 components only for sources in successively higher apparent luminosity intervals:
(b) $10^{33}$--$10^{34}$ erg$\,$s$^{-1}\,$Hz$^{-1}$
 (c) $10^{35}$--$10^{36}$ erg$\,$s$^{-1}\,$Hz$^{-1}$, and (d) $10^{36}$--$10^{37}$ erg$\,$s$^{-1}\,$Hz$^{-1}$.}
\label{fig:beta-r}
\end{figure}

To investigate the tentative evidence for a prevalence of some acceleration
along resolvable parts of jets in a different way in Fig.~\ref{fig:beta-r}a  
 we plot the velocities of all Q1 components as a function of their projected
radial distance from the core. There is a lack of fast components
at distances below a few pc. In some sources high velocities have been measured close to the core at a wavelength of 7mm. We discuss this in some detail in \S~\ref{sub:beta-freq}. We believe, that the observed dearth here is largely due to at least two selection effects.

The first effect is related to our measurement method. The typical time
spanned by our series of observations is 3--5 years. 
Any component with a significantly superluminal
apparent velocity will have moved more than a few pc during the
monitoring interval, and its average radial position
will therefore not be within a few pc the core. Only
slower components will have stayed around near the core long enough for us to
have registered them there. 
It is important to stress, that this effect only
acts to move the faster components to larger radii in our dataset; it does {\it
not} bias the statistics of the velocity values themselves; their distribution,
(\S~\ref{sec:vel-dis}), is completely unaffected.

\begin{figure}[htb]
\begin{center}
\hspace*{-0.8cm}
\subfigure[]{\rotate[l]{\psfig{figure=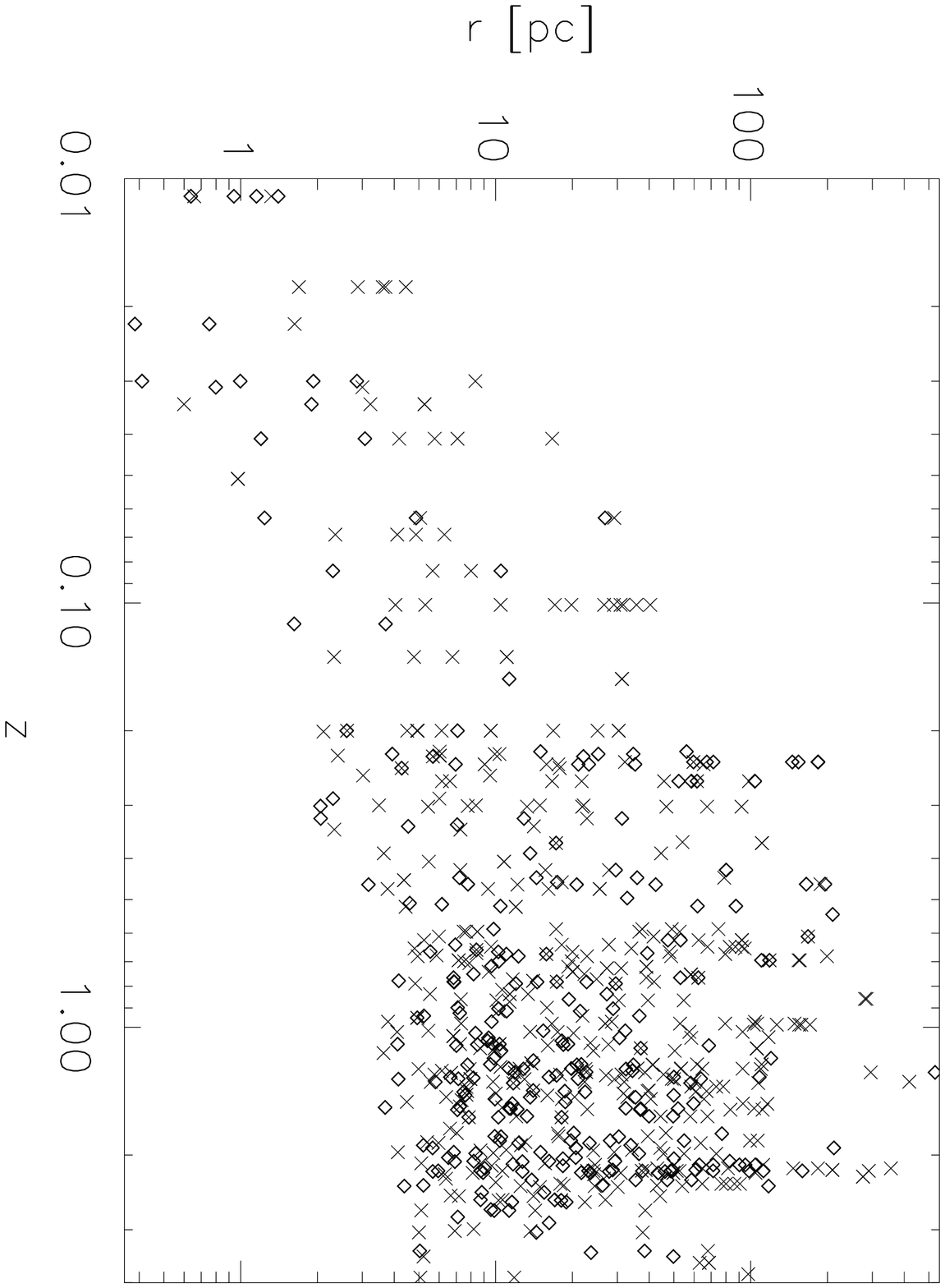,width=7.2cm}}}\\
\end{center}
\vspace*{-0.4cm}
\caption{Plot of $r(z)$ for the Q1 components (diamonds) and the
         Q2 \& Q3 components (crosses).}
\label{fig:rVz}
\end{figure}

The second effect is more physical in nature. In
\S~\ref{sec:main-cor} we show that the apparent velocities are positively
correlated with apparent luminosity and with redshift (two quantities
being themselves strongly correlated in flux-density limited samples). The
smaller radial distances are resolvable only at lower redshifts,
and therefore in the less luminous sources (usually galaxies,
see \S~\ref{sec:main-cor}). In these sources we find that larger
velocities do not occur at all, at any radius from the core. 
Fig.~\ref{fig:rVz} illustrates that the closest components can only be
detected in sources with the smaller redshifts. A fairly sharp curve
delimits the lower-bound of $r(z)$ in our sample of components. The dominance of
luminosity over radial distance in determining the distribution of velocities
is demonstrated by taking successively higher selections in luminosity for
Fig.~\ref{fig:beta-r}b,c,d. 

\subsection{Bending along individual jets}
\label{sub:bending}

The highest resolution VLBI observations show that in general curvature seems to play a more prominent role in the innermost regions of the AGN (e.g. Krichbaum et al. 1994; Britzen et al. 2000). More specifically, evidence for precession is found in an increasing number of sources (e.g., Britzen et al. 1999a, 2005a\&b; Walker et al. 2001; Lister 2001). Jet components in AGN exhibit either ballistic motions away from the core or curved paths suggestive of streaming motions along a funnel (e.g. Homan et al. 2001; Lister 2001). Some of these bent jets resemble helical structures in
projection, presumably originating in precession of the jet nozzle (e.g.,
Britzen et al. 2005b). The precession can have different physical causes for example   the precession of the accretion disk in binary black holes (e.g.,
Britzen et al. 2001a) or a misalignment between the rotation axes of the accretion disk and of a single Kerr black hole (Caproni et al. 2004). Components can be
ejected at different position angles (e.g., Britzen et al. 1999a; Qian et al.
2001; Bach et al. 2005), initially with ballistic trajectories, and then later
on following intrinsically curved paths (Denn, Mutel \& Marscher 2000; Stirling et al. 2003). 

As opposed to the ``global'' nature of the acceleration along the jet,
as discussed above, we can treat bending in a manner ``local'' to each
component.  Using only components that meet the quality criteria,
we compute the bending at component $C_i$ as the angle between the line segments
joining components $C_{(i-1)}$--$C_i$ and $C_i$--$C_{(i+1)}$.  Here, the
core is considered, so that a bending for the innermost component can
be computed.  However, no bending is associated with the outermost one.
The equation for the bending angle in this scheme is:\\

\begin{equation}
   \theta_{\rm bnd} = \cos^{-1} \left\{
            \frac{x_ax_b + y_ay_b}{\sqrt{x_a^2 + y_a^2}\sqrt{x_b^2 + y_b^2}}
                           \right\}
\end{equation}
where 
\begin{equation}
x_{\rm a} = X_{\rm i} - X_{\rm (i-1)},\\
x_{\rm b} = X_{\rm (i+1)} - X_{\rm i}
\end{equation}
(similarly for $y_a$ and $y_b$),
and an expression for the associated uncertainty comes from standard
error-propagation (with no correlations in the components' $X$ and $Y$
parameters, {\it cf} \S~\ref{sub:mucalc}):\\
\begin{equation}
   \sigma^2_{\theta_{\rm bnd}} = \frac{y_a^2\sigma_{x_a}^2 + x_a^2\sigma_{y_a}^2}{(x_a^2 + y_a^2)^2} + 
   \frac{y_b^2\sigma_{x_b}^2 + x_b^2\sigma_{y_b}^2}{(x_b^2 + y_b^2)^2}
\end{equation}
where the uncertainties for $x_a$, $x_b$, $y_a$, and $y_b$ come from 
quadrature sums of the uncertainties in the two components whose difference forms
these various intermediate variables.
The uncertainty of the core position is taken as 0.  
The calculated angle is
always in the range 0--180 degrees, regardless of whether the bending is
clockwise or counterclockwise for outward motions as exemplified in
2255+416 (Fig.~\ref{fig:maps1}) by the component sequences (C2-C3-C4) and
(C4-C5-C6), respectively.

\begin{figure}[htb]
\begin{center}
\vspace*{0.5cm}
\subfigure[]{\rotate[r]{\psfig{figure=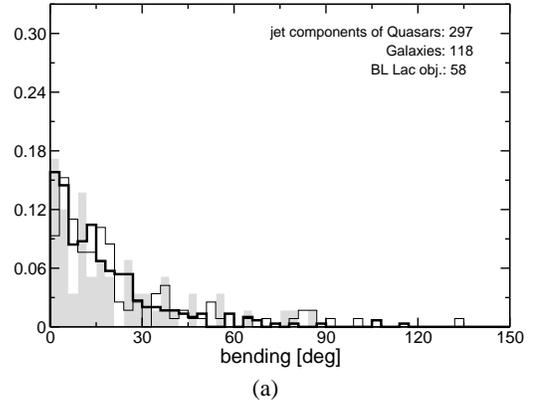,width=5.0cm}}}\\
\vspace*{0.5cm}
\subfigure[]{\rotate[r]{\psfig{figure=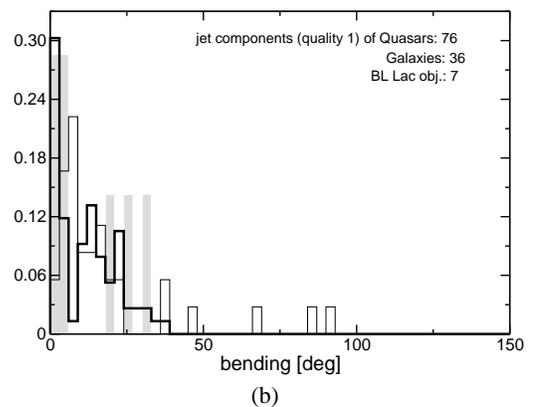,width=5.0cm}}}\\
\end{center}
\caption{(a) Distributions of measured local angles of bending (see
  text for details). Quasars: thick solid line, galaxies: thin solid
  line, BL Lac Objects: grey-scale. Panel (a) shows results when using
  all components, panel (b) when using only Q1 components.}
\label{fig:bend}
\end{figure}

\begin{figure}[htb]
\begin{center}
\vspace*{0.2cm}
\subfigure[]{\rotate[l]{\psfig{figure=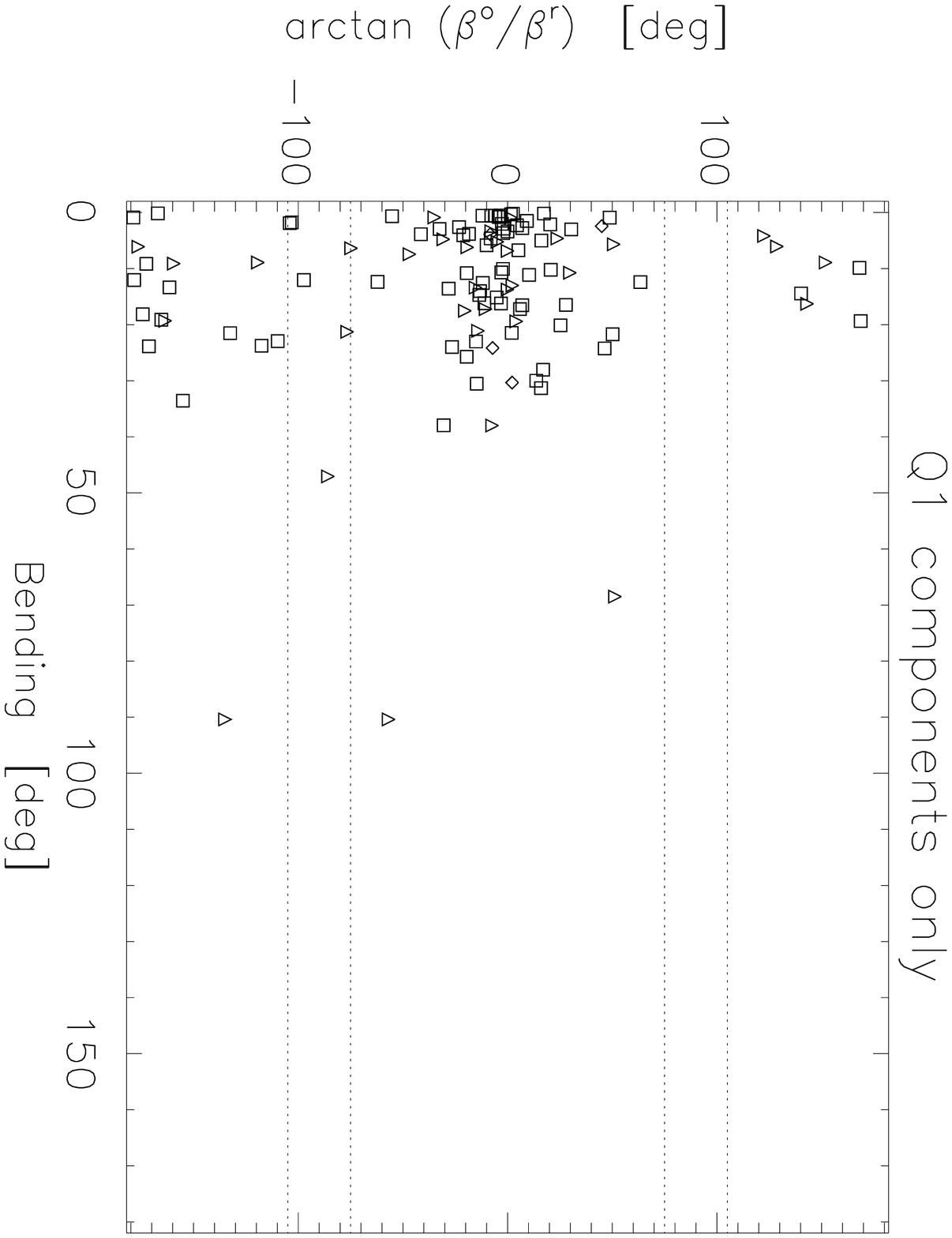,width=5.0cm}}}\\
\vspace*{0.1cm}
\subfigure[]{\rotate[l]{\psfig{figure=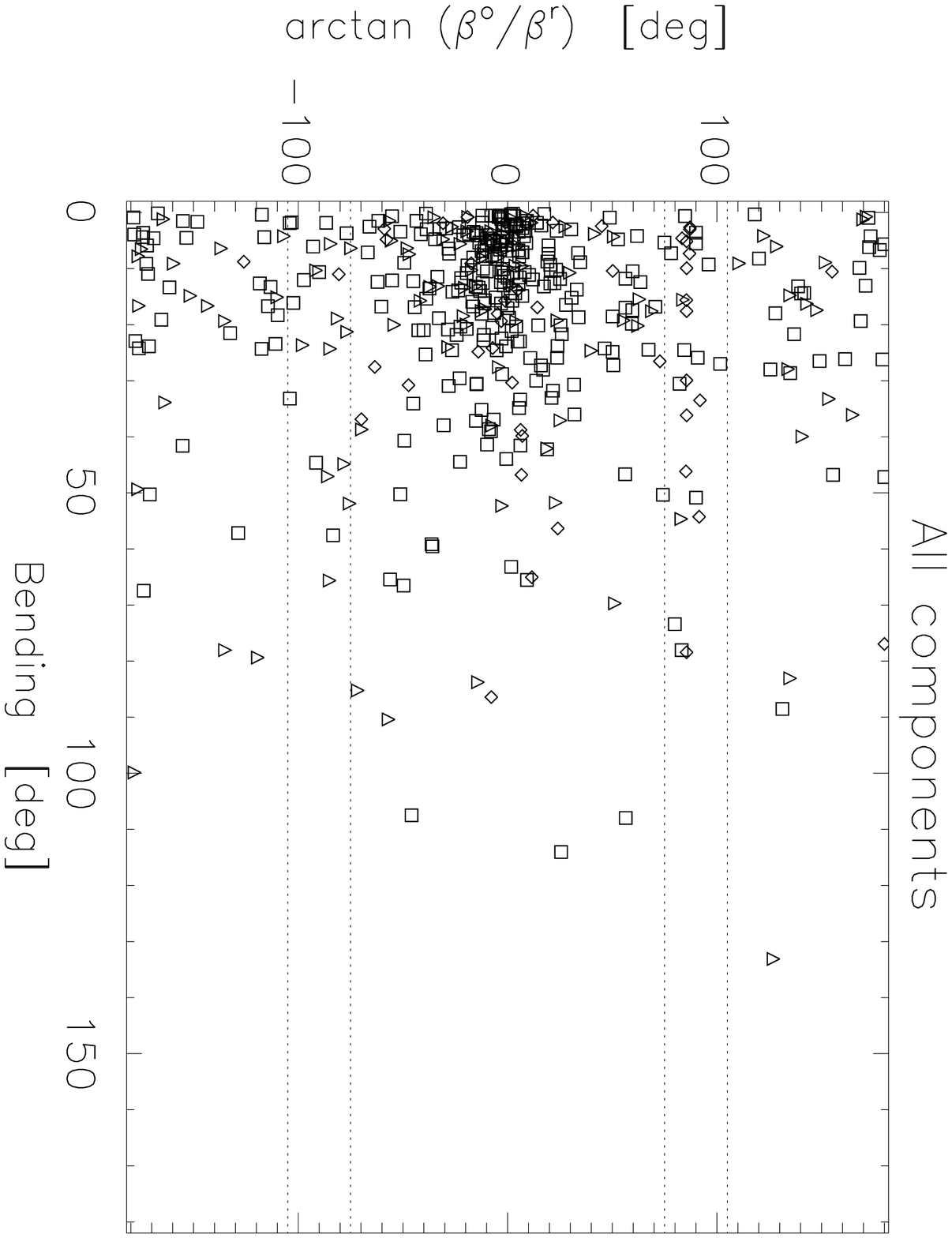,width=5.0cm}}}
\vspace*{0.2cm}
\end{center}
\caption{$\tan^{-1}(\beta^o/\beta^r)$ as a function of the bending angle $\theta_{\rm bnd}$
for the Q1 components (a) and all components (b). Quasars are denoted by squares,
galaxies by triangles, and BL Lacs by diamonds.}
\label{fig:bends}
\end{figure}

  This bending is characterized above as a property of the individual
components, but it of course also depends on each component's neighbors
that meet the quality criterion.  The Q1-only sub-set can be expected to have
fewer components, spaced farther apart along the jet thus producing 
smaller bending angles as defined above.  This seems borne out
in the two panels of Fig.~\ref{fig:bend}, especially for the quasars.

The size and quality of our data set allows us to take the bending analysis one step further. Fig.~\ref{fig:bends} shows the relation between
$\tan^{-1}(\beta^o/\beta^r)$ and the bending angle $\theta_{\rm bnd}$ for 
Q1-only components and for all components.  The former quantity relates to the
angle a component's velocity makes with its local radial direction: $0^\circ$
implies purely outward radial motion, $\pm180^\circ$ purely inward radial
motion, and $\pm90^\circ$ purely circumferential motion around the core.
Apparent circumferential motion is rare in that there are very few Q1 components ($\sim$4\%
of the components from galaxies and $<$2\% of those from quasars) within $15^\circ$ of $\pm90^\circ$. When
all quality classes are considered, more components show 
apparent circumferential motion, 5\% from quasars, $\sim$7\% from galaxies, and
preferentially more from BL Lacs ($\sim$21\%).  However, there is no
clear correlation between large bending angles $\theta_{\rm bnd}$ and
the angle a component's velocity makes with the local radial direction.

\subsection{Can a representative velocity for each source be determined ?}
\label{sub:representative}

As discussed in the introduction to this section,
there are many sources in which 
we 
have been able to track multiple components, and in any single source these 
components often have significantly different apparent velocities. For population studies, a single velocity per source must be derived to give each source equal weight, for example when comparing the velocity statistics of galaxies, where there are often many components per source, to quasars, which often have fewer components in their jets.

\begin{figure*}[htb]
 \begin{center}
   \vspace*{0.5cm}
      \hspace*{-0.5cm}\subfigure[]{\rotate[r]{\psfig{figure=betar.br.histo.ps1,width=4.3cm}}}
          \hspace*{0.3cm}\subfigure[]{\rotate[r]{\psfig{figure=betao.br.histo.ps1,width=4.3cm}}}
     \hspace*{0.3cm}\subfigure[]{\rotate[r]{\psfig{figure=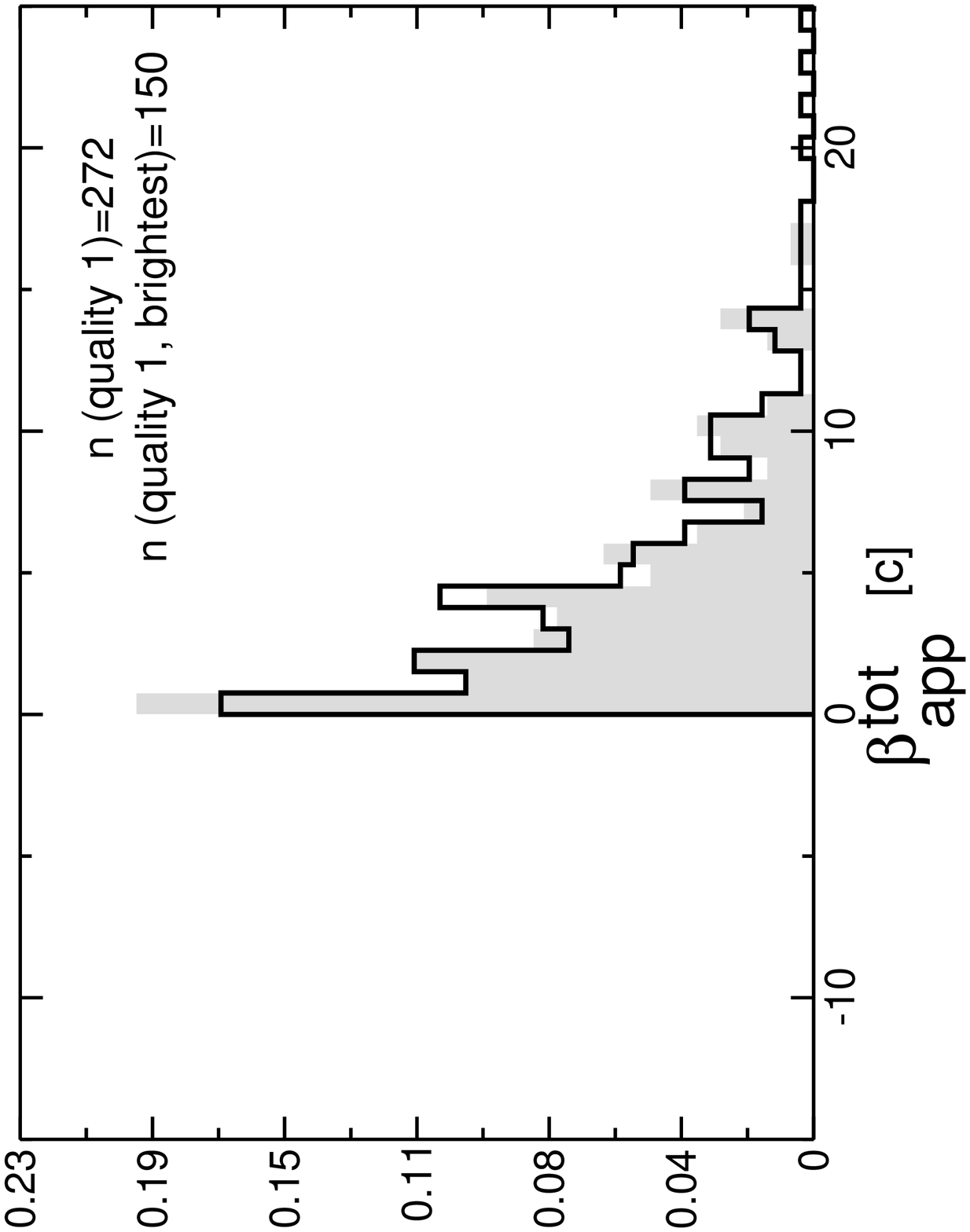,width=4.3cm}}}\\
          \vspace*{0.5cm}
         \hspace*{-0.5cm}\subfigure[]{\rotate[r]{\psfig{figure=betar.cl.histo.ps1,width=4.3cm}}} 
    \hspace*{0.3cm}\subfigure[]{\rotate[r]{\psfig{figure=betao.cl.histo.ps1,width=4.3cm}}}
             \hspace*{0.3cm}\subfigure[]{\rotate[r]{\psfig{figure=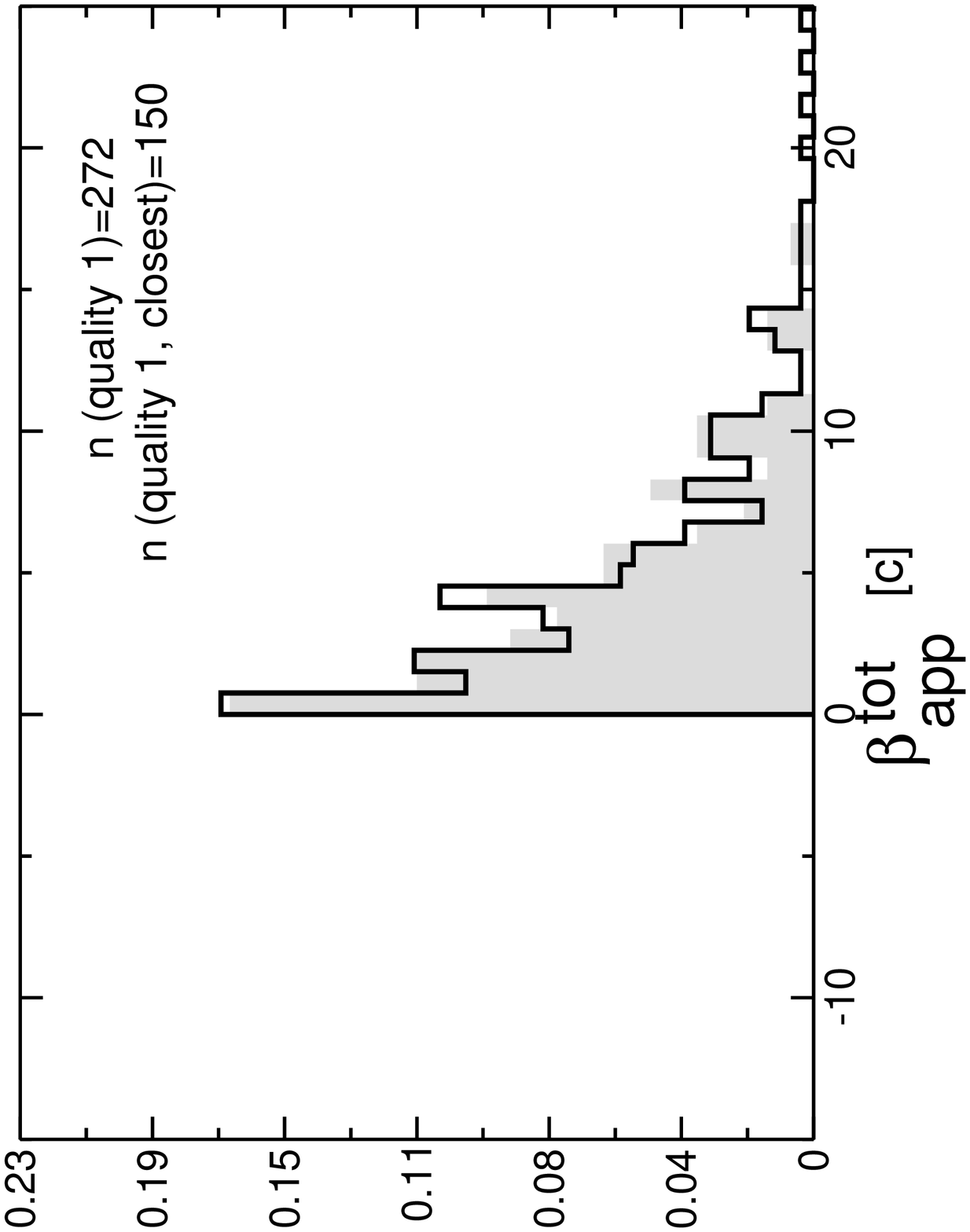,width=4.3cm}}}\\
               \end{center}
              \caption{The $\beta_{\rm app}^r$, $\beta_{\rm app}^{o}$, and $\beta_{\rm app}^{\rm tot}$ distributions obtained with different selections of a single component velocity per source, shown in grey: (a--c) the brightest Q1 component in each of 150 sources, and (d--f) the Q1 component closest to the core in each of 150 sources (these sub-samples overlap in all but 23 sources). For comparison, the distribution for all 272 Q1 components in the full sample is shown with a solid black line in all panels.}
			               \label{fig:ro}
		                  \end{figure*}

The selection
from our sample of either the brightest or the closest Q1 component per 
source yields a distribution of velocities which cannot be statistically 
distinguished from that of our full sample of Q1 components. 
It is important to point out, however, that the distribution of brightest Q1 component velocities terminates at $\sim18\, c$. Histograms of the relevant distributions are shown in Fig.~\ref{fig:ro}. Note that the selected sub-samples of the 150 brightest or 150 closest Q1 components are part of our full sample of all 272 Q1 components, and the histograms are far from independent; the brightest and closest sub-samples overlap for all but 23 sources between the brightest and the closest sub-sample.
Despite the similarities in Fig.~\ref{fig:ro}, we favor the use of the brightest rather than the closest component. With limited resolution the position of components close to the (usually
time-variable) core may be affected by deconvolution problems.
We have not given Q1 labels to components where we suspected ``core-blending''
in our sample. There are also selection effects which increase the fraction of slow
components at the smallest radial distances ($<$5 pc, see \S~\ref{sub:vel-r}). 
While the full sample of closest Q1 components spans a large range of radial distances, mostly above a few pc, and so should not be much affected, the use of the brightest components might nevertheless be preferable partly because of this effect. Of course, if jets bend, or vary in velocity
with time or along their length, then selecting the brightest
components would tend to bias towards features which are the most
strongly Doppler boosted into our line-of-sight. These do not
necessarily have either the fastest or any other characteristic
apparent velocity (for reference, see the discussion in Vermeulen \& Cohen 1994). 

We have thus shown that taking the velocity of the 
brightest components - this means one velocity per source - as representative may be adequate for population
studies, knowing  that their distribution matches that of all Q1 components. 

\section{The distribution of apparent radial velocities in the sample}
\label{sec:vel-dis}

\begin{figure}[htb]
\begin{center}
\vspace*{0.3cm}
\includegraphics[clip,width=5.5cm,angle=-90]{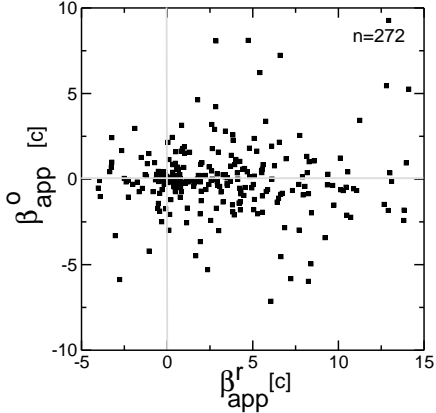}
\end{center}
\vspace*{-0.4cm}
\caption{The radial and orthogonal values $(\beta_{\rm
app}^{r},\beta_{\rm app}^{o})$ of the apparent velocities for the Q1
components, on an expanded scale with respect to
Fig.~\ref{fig:betaro}.
} \label{fig:betaro-inner}
\end{figure}

The radial velocities in Fig.~\ref{fig:betaro-inner} (an expanded version of
Fig.~\ref{fig:betaro}) and Fig.~\ref{fig:betarohisto} show 
a fairly broad distribution of velocities,
tailing off towards $\sim15\,\rm{c}$, with small sprinkling of radial
velocities extending up $\sim30\, c$. The distribution also extends, seemingly
continuously, to somewhat negative (i.e., apparently inward) radial velocities,
but these have tapered off at about $-5\,\rm{c}$. 
We find a pronounced narrow
peak of subluminal or, at most, mildly superluminal
sources/components. 

To obtain such a broad distribution of apparent velocities, predictions from
relativistic jet models (e.g., Vermeulen \& Cohen 1994; Lister \& Marscher
1997; Cohen et al.\ 2007) requires a broad distribution of the
Lorentz $\gamma$ factors of the moving features. 

\subsection{Negative radial velocities}
\label{sub:inward}

Of the 272 Q1 components in our sample, 56 in 42 different sources (26 of 186 quasars and 16 of 70 galaxies) have a (slightly) negative radial velocity. But only 15 of the negative Q1 component velocities (in 8 quasars and 7 galaxies)
 have a formal significance better than 2$\sigma$; the median velocity of those is $-2.6\,\rm{c}$ (see also Fig.~\ref{fig:sigmabeta}). We believe it is likely that some of the observed negative radial velocities really represent features moving towards the core, for example due to reconfinement shocks (e.g., Britzen et al.\ 2005a). Apparent negative velocities in jets have also been discussed by  Wehrle et al., (2001), Kellermann et al. (2004), and Piner et al. (2007). A theoretical model of apparently backward-moving knots has been presented by Istomin \& Pariev (1996) and Gomez et al. (1997). In many cases, however, the negative velocities either are simply the result of our measurement procedure, or have a geometrical rather than a directly physical explanation in the source. We first identify three possibilities relating to our adopted measurement procedure: 

1) CSOs (Compact Symmetric Object): there are a number of well-known CSOs and similar two-sided sources, often galaxies, where, rather than to the quite inconspicuous or absent core at 5 GHz, we have referenced to one of the hot spots, often at one extreme end of the structure. Components in the jet moving towards that hot spot will then have the sign of their velocity reversed; the hot spot is likely to be almost stationary or at least moving much more slowly than components in the jet. Examples of such sources include the quasar 2021+614 and the galaxies 1946+708 and 2352+495. As expected, the absolute values of the velocities in the two-sided sources are small.

2) Incorrect cores: even in some one-sided sources we may have, adopted a prominent compact jet component as the reference, particularly if it seemed to be (nearly) stationary. Many such stationary features exist in jets even amongst moving components, for example as a result of shocks or bends in jets; see also \S~\ref{sub:sublum}. With the choice of such a reference feature, any component between it and the true core will have the sign of its velocity reversed, compared to the regular convention where positive radial velocity implies motion away from the core. 

3) Blended cores: we also suspect that in some sources there was a gradual centroid displacement of the reference feature during the time spanned by our observations, as a result of blending between the true core and a component which was newly emerging and/or becoming relatively brighter. This can also lead to apparently negative velocities for components which in reality are stationary or slowly moving out.

Geometric explanations are also possible. Relativistic jets are preferentially seen at small angles to the line-of-sight as a result of Doppler favoritism. Thus, even slight intrinsic bends can create sections of the jet in which components appear to move back towards the core in projection. 

\subsection{Subluminal and stationary components}
\label{sub:sublum}

The uncertainties of the motions measured in our sample are such that we cannot discriminate well between ``stationary'' components and components moving at a subluminal or at most mildly superluminal velocity. It can be anticipated that samples selected on beamed emission (as the CJF was designed to be) should show relatively few, if any, subluminal motions (e.g., Vermeulen \& Cohen 1994). On the contrary, however, we find that there is a sharp peak at low velocities. The prevalence of subluminal or even stationary components is enhanced even over the broad velocity distribution which, by its broadness, is itself already difficult to accommodate within simple beaming models. In the sample of Q1 components, a subluminal radial velocity has been measured for 42 of 272 components in 36 sources: 26 components in 25 quasars, 14 components in 9 galaxies, and 3 components in 2 BL Lac objects. 
For Q1 \& Q2 \& Q3 26 subluminal components in 20 sources have a relative significance of $2\sigma$ or better. 

Slow components occur relatively often, but by no means exclusively, in
galaxies. However, this cannot be taken as direct evidence for standard
unification models, in which the relativistic jets in galaxies are supposed to
be viewed at larger angles to the line-of-sight than in quasars. For example the CJF sample contains a number of well-known CSOs which are always identified with galaxies; the prototype is 2352+495 (e.g., Wilkinson et al. 1994; Taylor et al. 1996b; Readhead et al. 1996). It has been firmly established that CSOs are not dominated by beamed emission, and that components in their jets are typically subluminal; speeds of 0.2--0.4$\,$c are common (e.g., Polatidis \& Conway 2003).

We find that slow jet components often coexist with superluminal
components, and can occur in sources at any observed luminosity. 
Such (near-)stationary features
probably have a geometrical explanation in some cases, and a truly physical
cause in others. Some slow components are likely to represent sites of enhanced
Doppler beaming, where a curved jet points most closely along the line of
sight, as in the source 4C39.25 (Alberdi et al.\ 2000). Others are probably
associated with a stationary shock in the jet, for example a standing
recollimation shock caused by pressure imbalances at the boundary between the
jet fluid and the external medium, where the energy density enhancement
produced downstream can give rise to stationary radio knots (e.g.,
Mart$\acute{\rm i}$ \& M\"uller 2003). If a jet bends abruptly, whether because
of internal instabilities, or because the jet is influenced by gradients in the
external medium, perhaps even to the extent of being deflected, it is also
likely to develop some stationary features associated with shocks; see for
example the work by G$\acute{\rm o}$mez et al.\ (2001), G$\acute{\rm o}$mez
(2005) and Agudo et al.\ (2001).
A proper characterization of ``stationary'' components requires a denser
sampling of the observations in time than is available for CJF, as demonstrated
by the oscillations in the position of a ``stationary'' bright jet component in
1803+784, uncovered by Britzen et al.\ (2005a). 

\section{Correlations with luminosity and other source parameters}
\label{sec:main-cor}

In this section we first demonstrate that in the flux-density limited CJF sample there is a correlation between the distribution of apparent velocities and the observed radio source luminosities (\S~\ref{sub:corr-L}). In \S~\ref{sub:class} we show the impact of this correlation on the overall velocity distributions of CJF quasars, galaxies, and BL Lac objects, and briefly discuss this, together with properties such as spectral index and core dominance, in the light of orientation-unification models. The correlation with observed luminosity must also be taken into account when comparing samples with different selection criteria (see \S~\ref{sec:comp}).

\subsection{A correlation with the observed 5 GHz core luminosity}
\label{sub:corr-L}

\begin{figure*}[htb]
\begin{center}
\subfigure[]{\rotate[r]{\psfig{figure=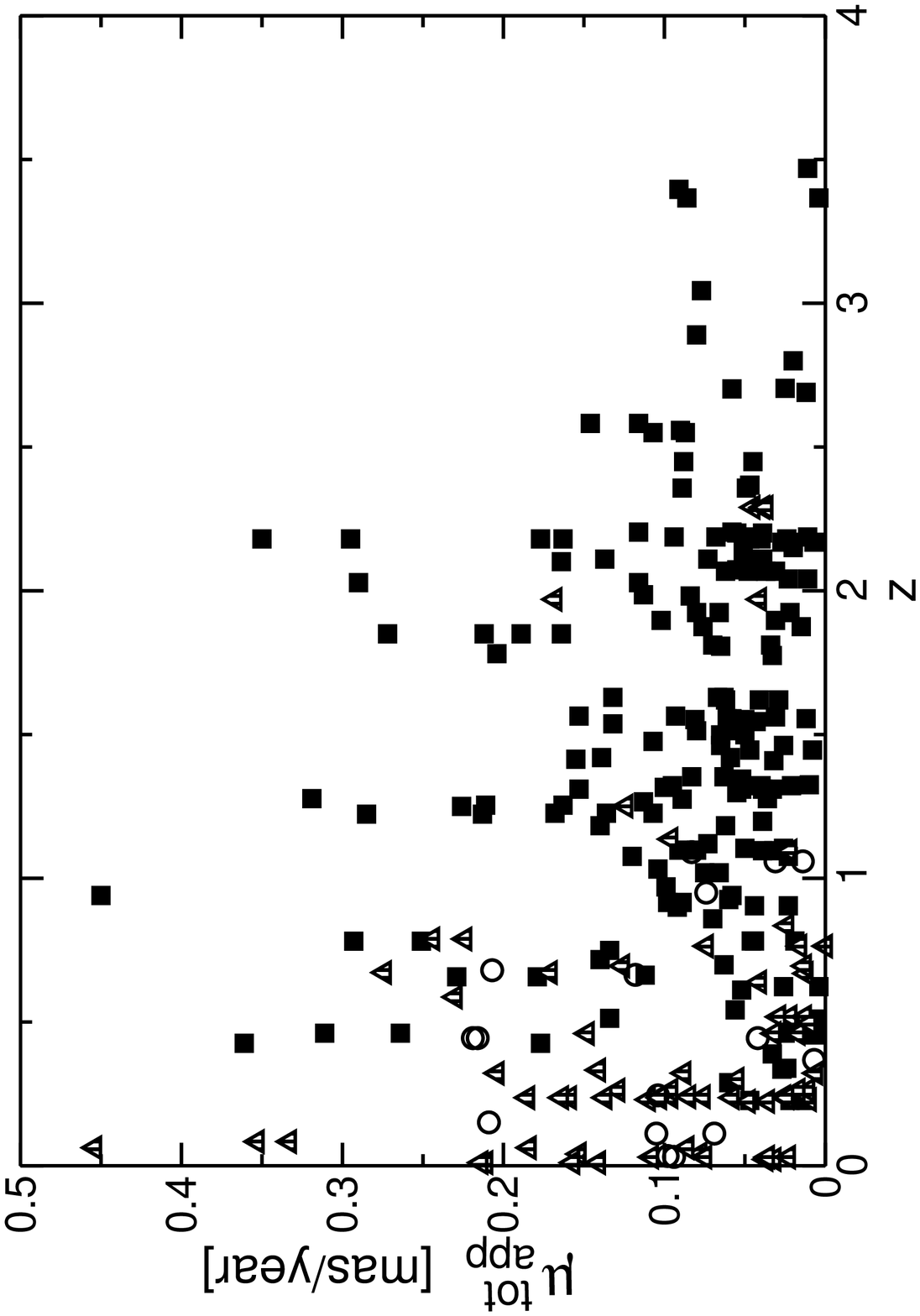,width=6.2cm}}}
\hspace*{0.3cm}
\subfigure[]{\rotate[r]{\psfig{figure=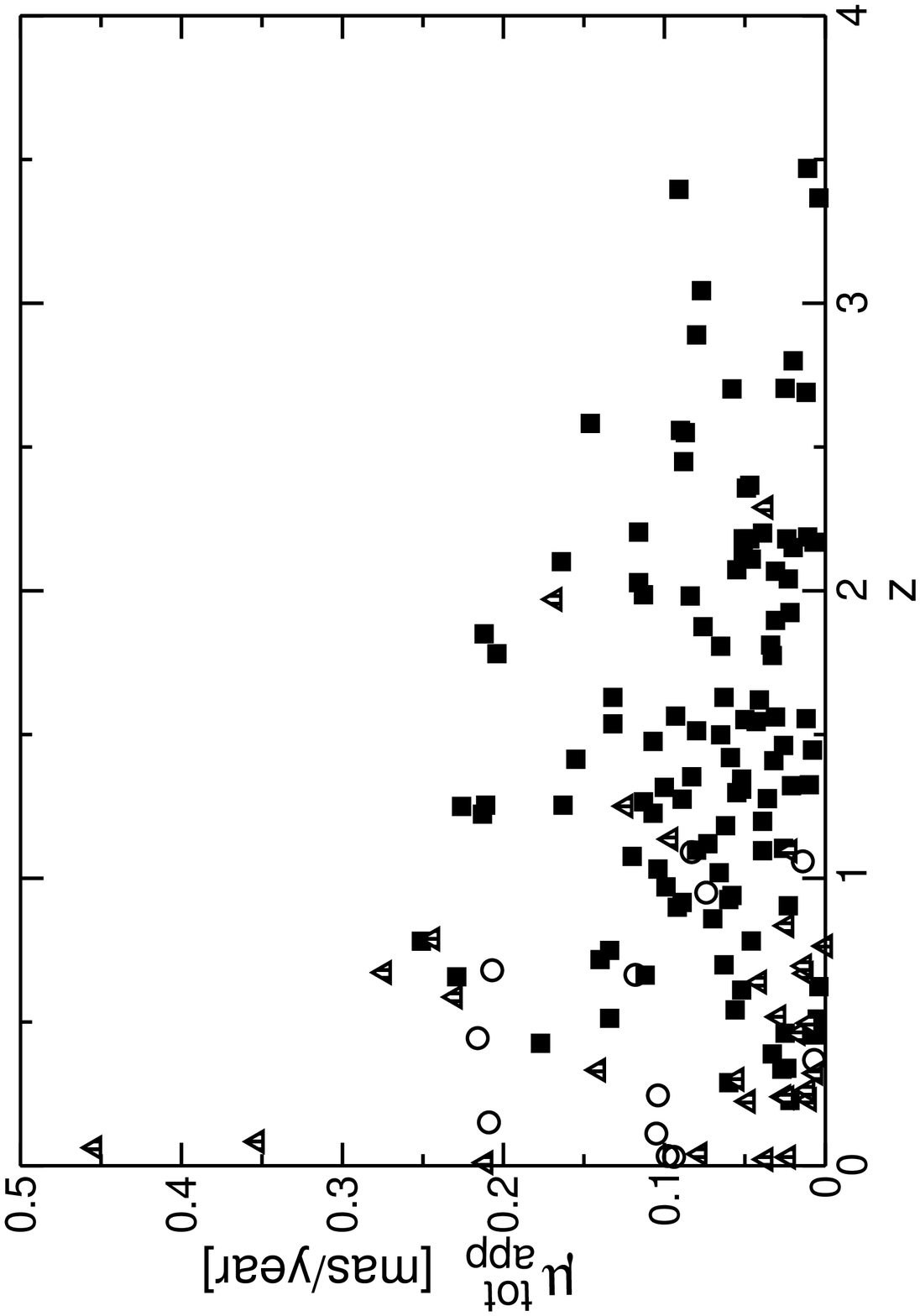,width=6.2cm}}}\\
\subfigure[]{\rotate[r]{\psfig{figure=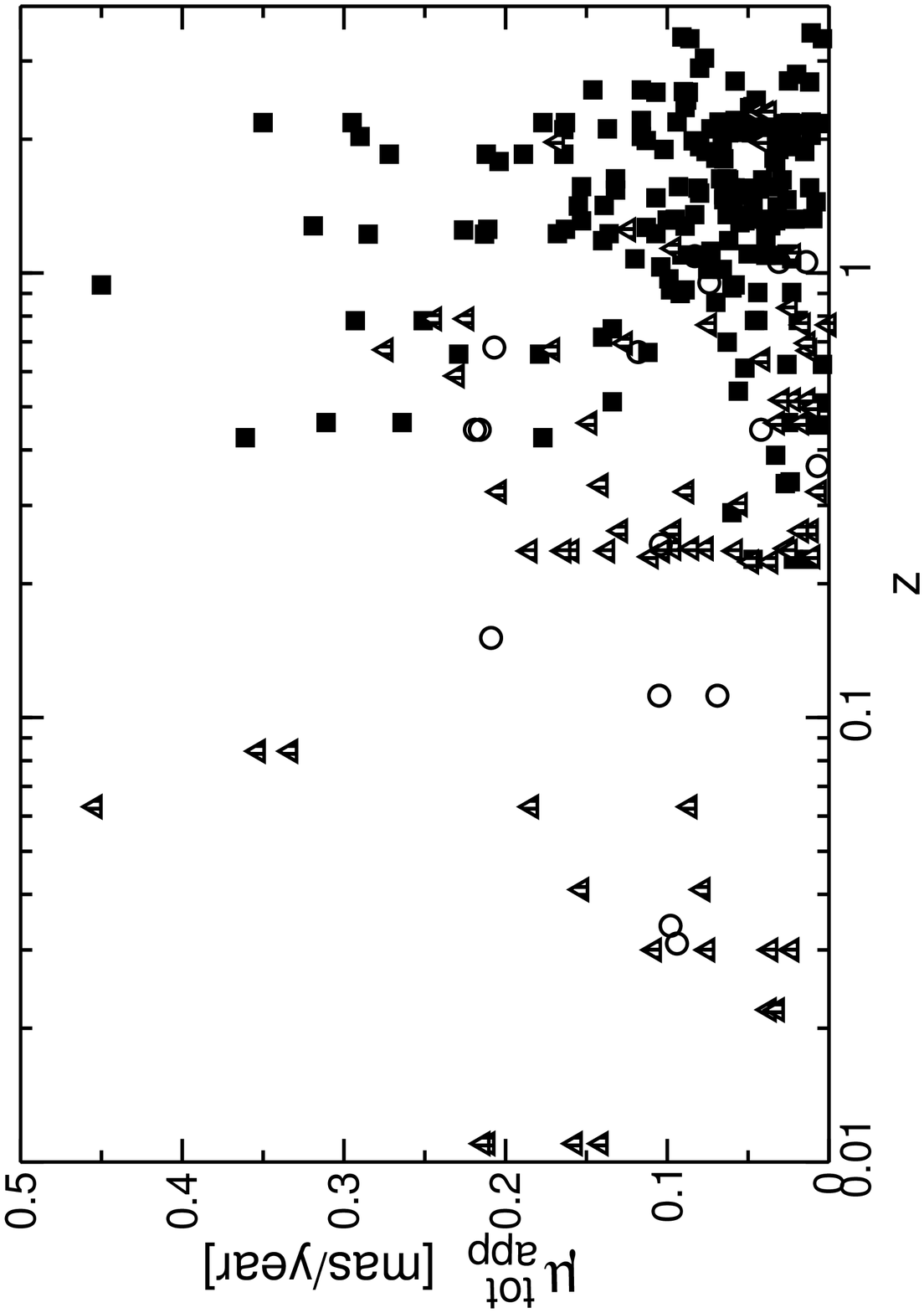,width=6.2cm}}}
\hspace*{0.3cm}
\subfigure[]{\rotate[r]{\psfig{figure=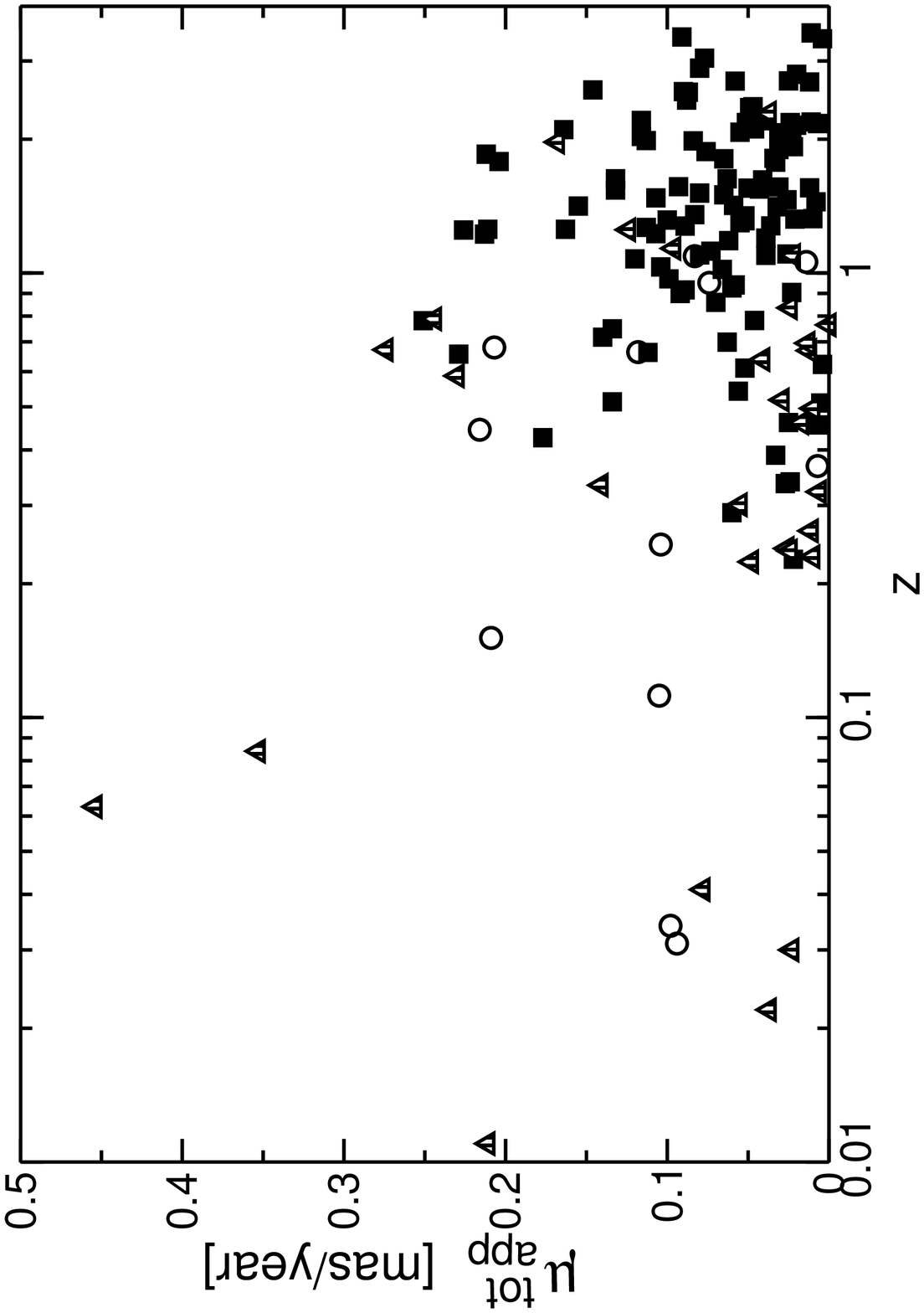,width=6.2cm}}}\\
\subfigure[]{\rotate[r]{\psfig{figure=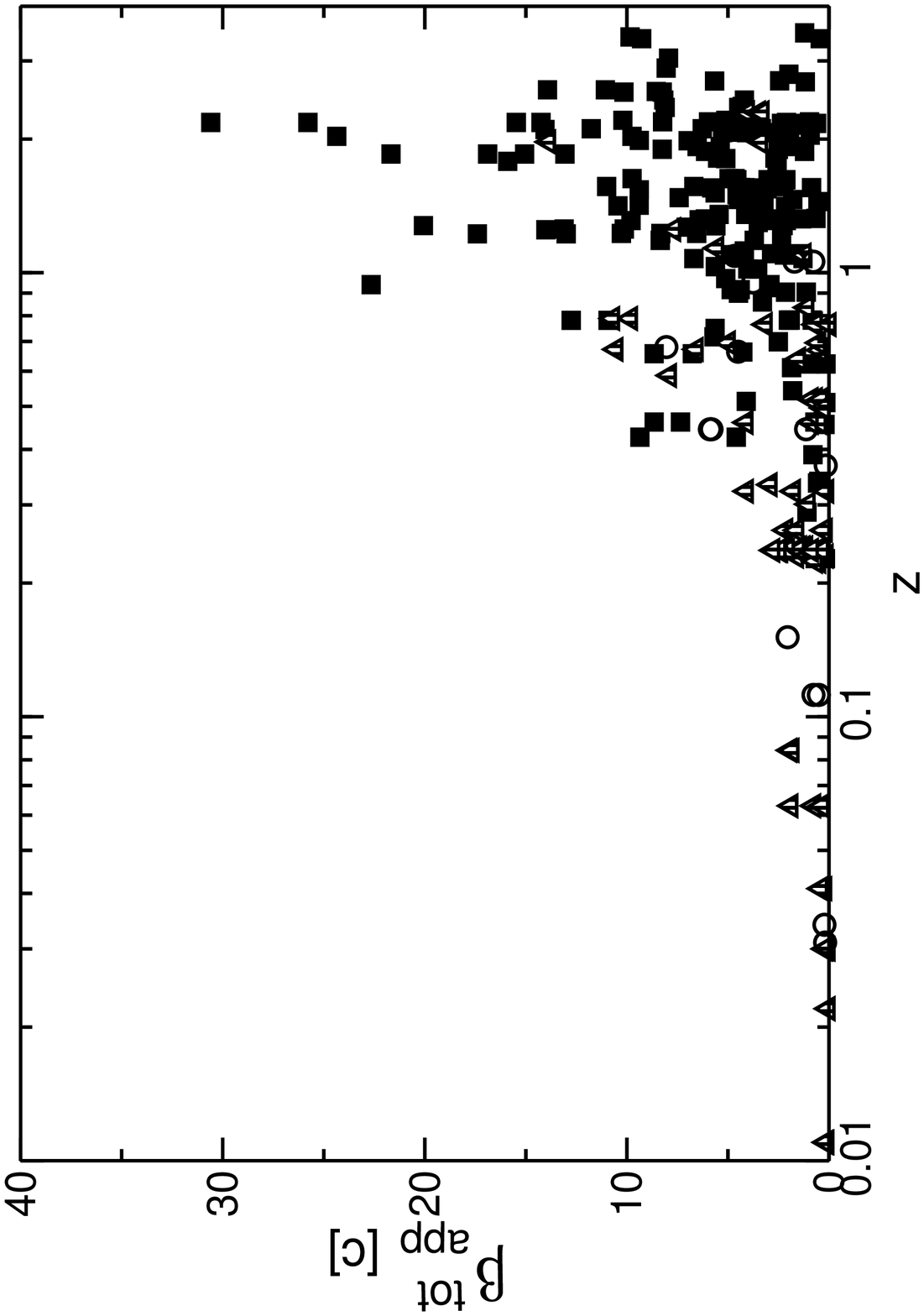,width=6.2cm}}}
\hspace*{0.3cm}
  \subfigure[]{\rotate[r]{\psfig{figure=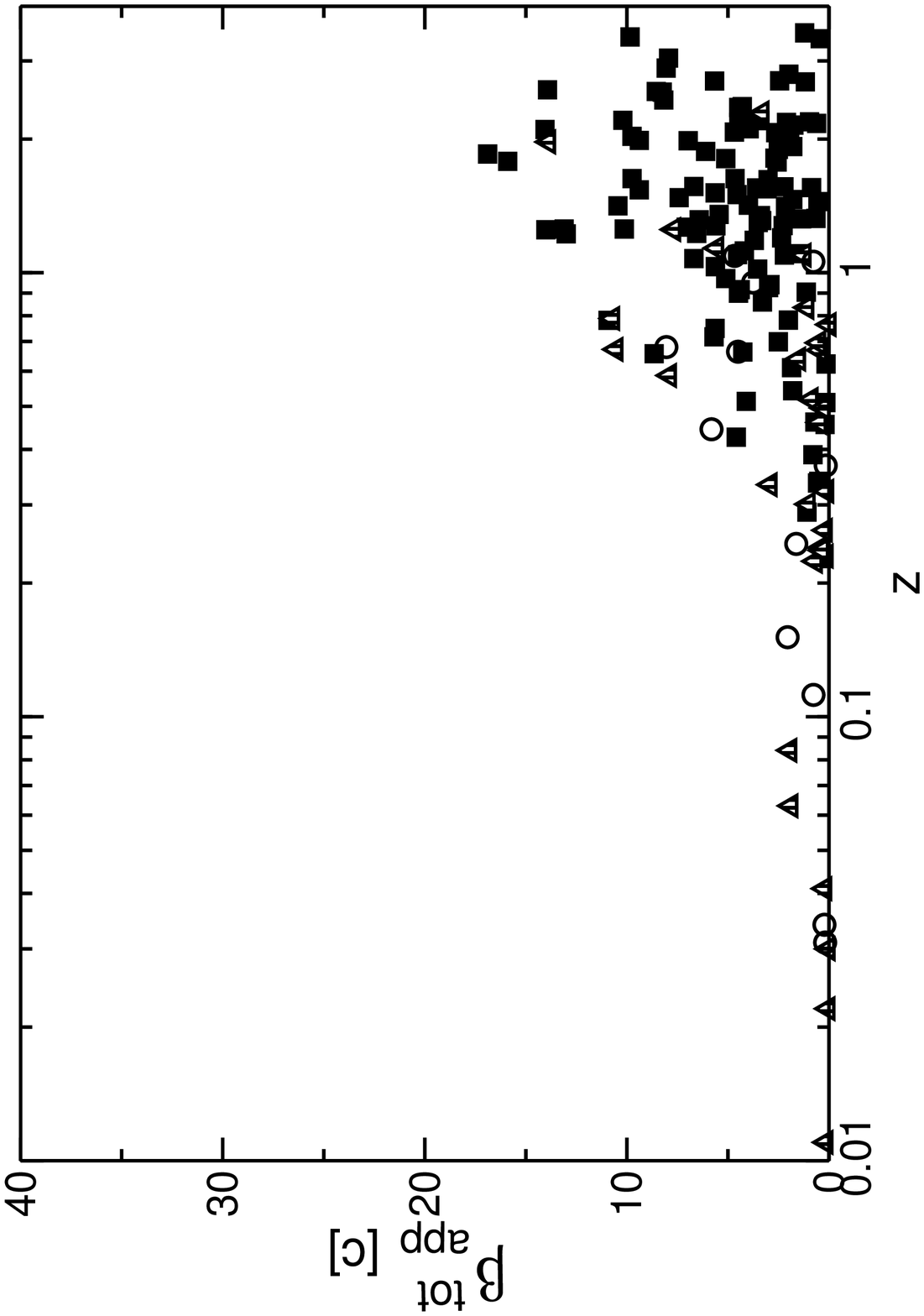,width=6.2cm}}}
    \vspace*{-0.1cm}
      \end{center}
        \caption{Motions as a function of redshift. On the left are all Q1 components, on the right the brightest Q1 component per source: (a,b) the well-known ``$\mu$--$z$'' diagram of apparent proper motions ($\mu^{\rm tot}$) against redshift; (c,d) the same relationship with a logarithmic redshift axis; (e,f) the proper motions converted to speed ($\beta_{\rm app}^{\rm tot}$) as a function of redshift (logarithmic axis). The quasars are indicated by filled black squares, the BL Lac objects by open circles, and the galaxies by striped triangles.  }
	   \label{fig:beta-z}
	     \end{figure*}

With the large, homogeneously analyzed flux-density limited CJF sample
available, the $\mu$--$z$ and $\beta_{\rm app}^{\rm tot}$--$z$ diagrams, shown in
Fig.~\ref{fig:beta-z} reveal rather different features than their predecessors (e.g., Cohen et al. 1977, 1988),
which were based on smaller, heterogeneous samples. The left-hand panels of
Fig.~\ref{fig:beta-z} contain all Q1 components. Each source can have different
numbers of components, so in order to give all sources (i.e., all
independent redshifts) equal weight, we also include on the right-hand side of
Fig.~\ref{fig:beta-z} and subsequent figures diagrams which include only the
brightest Q1 component per source.

\begin{figure}
\begin{center}
\subfigure[]{}\includegraphics[clip,width=5.0cm,angle=-90]{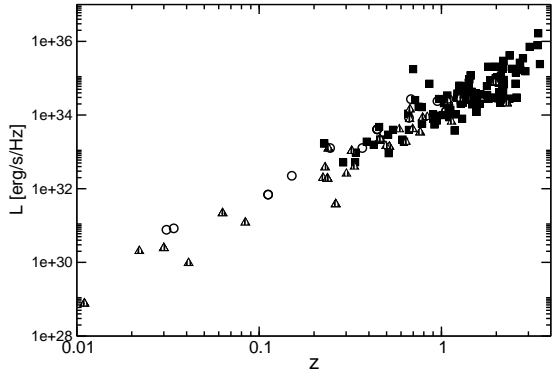}
\end{center}
 \caption{The apparent luminosity, calculated based on the 5 GHz core
 flux density, as function of the source redshift for all sources with a Q1 jet component. The quasars are indicated by filled black squares, the BL Lac objects by open circles, and the galaxies by striped triangles.} 
\label{fig:lum-z}
\end{figure}

\begin{figure*}
 \begin{center}
\subfigure[]{\includegraphics[clip,width=5.0cm,angle=-90]{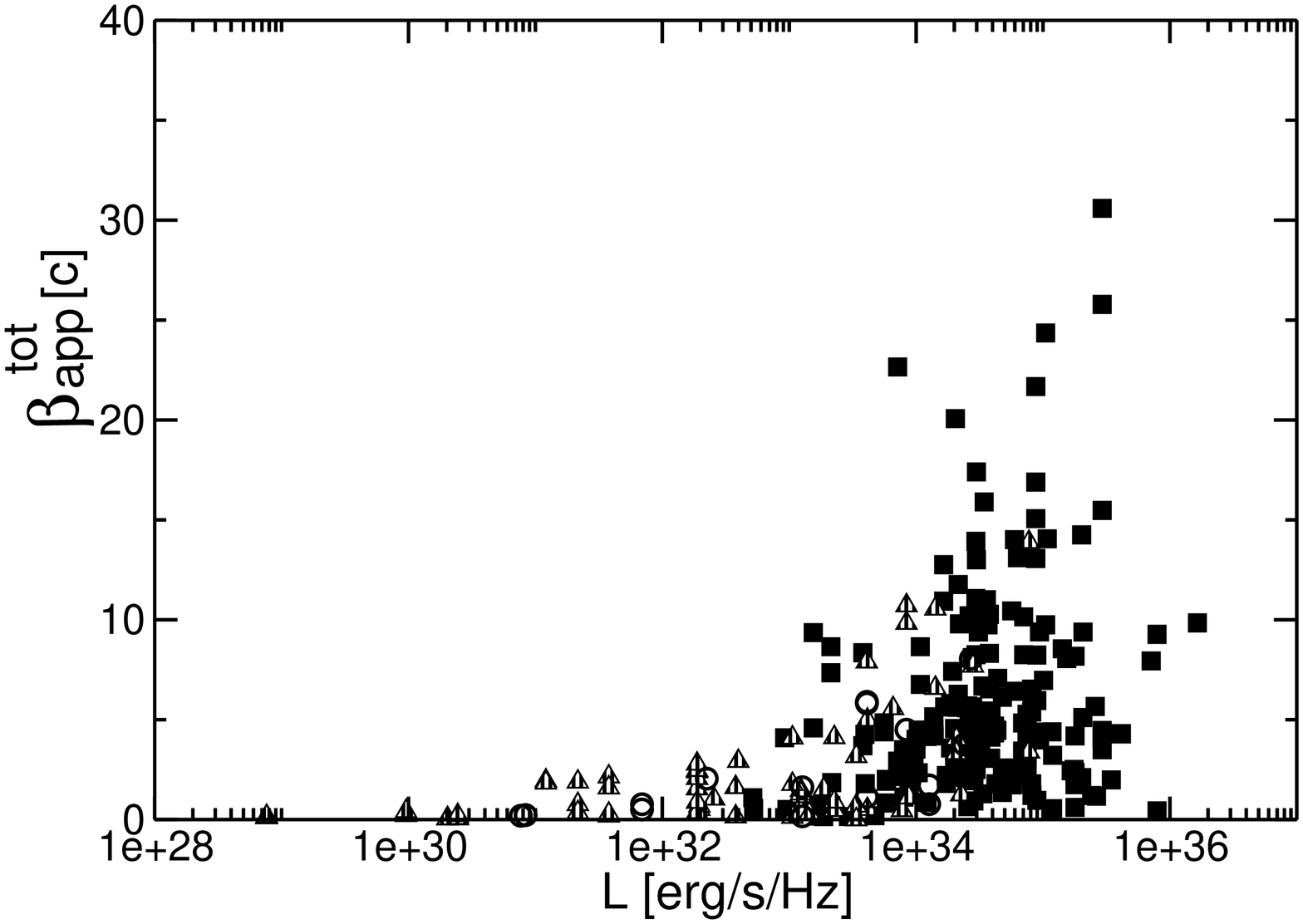}}
\hspace*{0.7cm}
\subfigure[]{\includegraphics[clip,width=5.0cm,angle=-90]{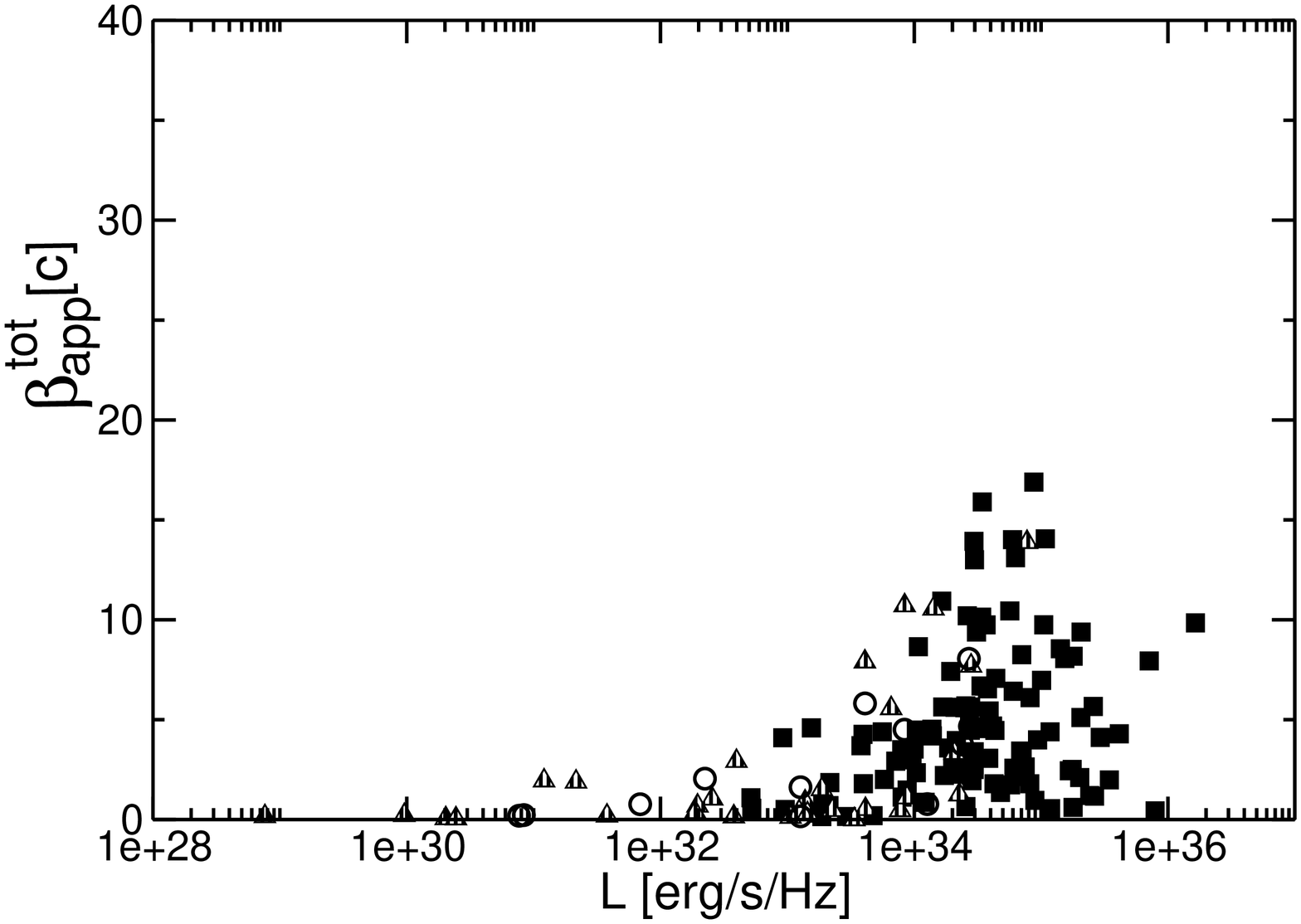}}
 \end{center}
 \caption{The total apparent velocities as a function of the
 observed source core
 luminosity at 5 GHz: (a) for all Q1 components; (b) for the brightest Q1 component in each source. The quasars are indicated by filled black squares, the BL
 Lac objects by open circles, and the galaxies by striped
 triangles.}
\label{fig:lum-beta}
\end{figure*}

The CJF sample shows a well-defined upper envelope in $\beta_{\rm app}^{\rm tot}$--$z$,
particularly sharply visible for the brightest Q1 components. However, that
upper limit, far from being constant, is a strong function of redshift.  
The CJF sample has a strong correlation between redshift
and source luminosity, inevitable for a flux-density limited survey. This is shown in Fig.~\ref{fig:lum-z}. We have derived
the apparent radio luminosity ($L=F_{m}4\pi D_L^2$)
from the observed 5 GHz radio core
flux density $F_{m}$, measured through model-fitting in the last available observing
epoch; procedures and results are given in Paper~I.  
The luminosity distance is computed as
$D_L = D_M(1+z)$, using the $D_M$ from in \S~\ref{sec:beta}.
Fig.~\ref{fig:lum-beta}
shows the correlation of apparent velocities with the observed 5 GHz
luminosities, for all CJF Q1 components on the left, and for the brightest Q1
components on the right.
Globally, the same distribution is seen in $\beta_{\rm app}^{\rm tot}$--$L$ as for the diagrams with redshift. Th
e upper envelope is again particularly well-defined for the brightest Q1 components. It runs in the sense that at
 low luminosities, there are no fast motions at all, while the largest velocity encountered increases with lumino
 sity. However, even though clearly subject to small number statistics, we note that the highest velocities may not occur at the very highest luminosities; a turnover of the envelope is part of ``aspect curves'' as summarized below. Despite what seems to be a rather clear envelope, the motion values do not crowd up strongly against it. Conversely, they seem to be a bit sparser close to the limit, while the rest of the ``allowed'' velocity range at any luminosity seems covered fairly uniformly almost down to zero motion.

 The shape of the $\beta_{\rm app}^{\rm tot}$--$L$ has meanwhile been confirmed remarkably consistently by Cohen et al.\ (2007) in the 2cm VLBA survey, which has an overlap of only 24 sources with the CJF. Cohen et al.\ (2007) give an extensive discussion of beaming models which can reproduce the global shape of the upper envelope remarkably well with so-called ``aspect curves'', that trace how the observed luminosity and speed would vary as a given source is turned from having its jet pointed directly at the observer (maximal luminosity, no speed), through small angles (slightly reduced luminosity, maximal speed), to large angles (strongly reduced luminosity, low speed). However, Cohen et al.\ (2007) have concluded that, in reality, to explain the detailed distribution of velocities as a function of observed luminosity, many different aspect curves are required.

\begin{figure*}
 \begin{center}
\subfigure[]{\includegraphics[clip,width=5.0cm,angle=-90]{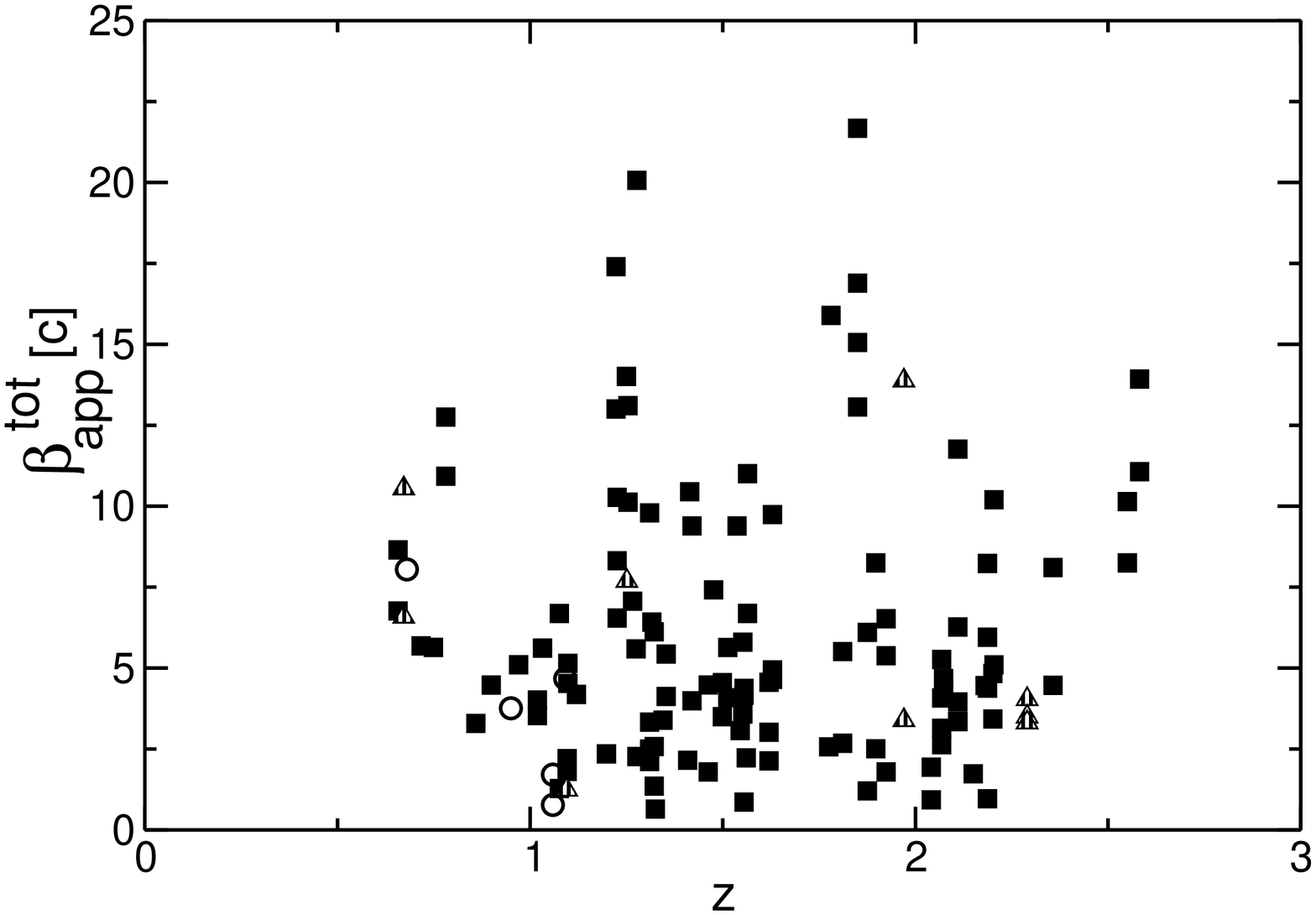}}
\hspace*{0.7cm}
\subfigure[]{\includegraphics[clip,width=5.0cm,angle=-90]{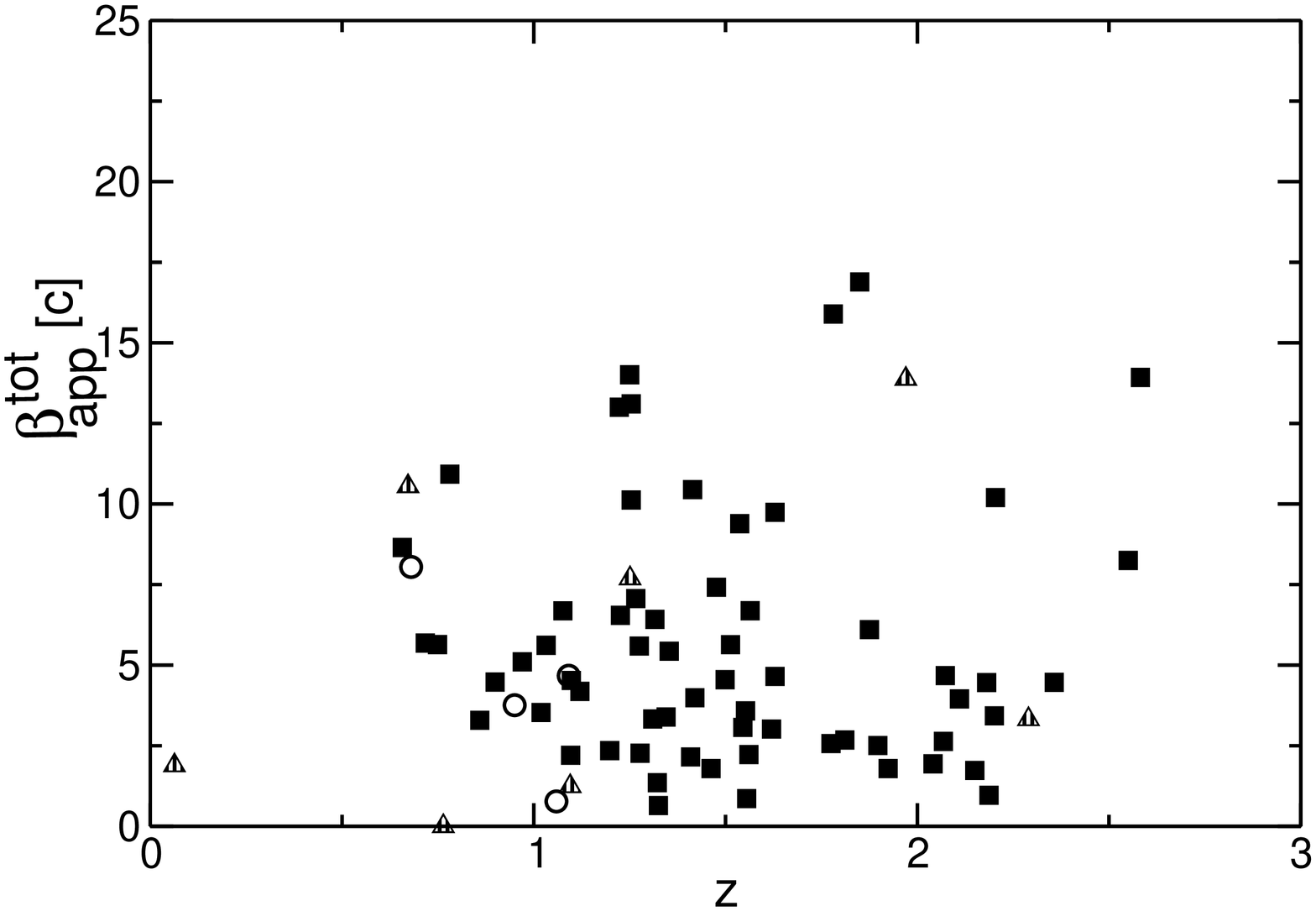}}
  \vspace*{-0.3cm}
   \end{center}
 \caption{The total apparent velocities as a function of redshift, restricted to sources for which the observed core luminosity is in the range $10^{35}$--$10^{36}$ erg$\,$s$^{-1}\,$Hz$^{-1}$. (a) shows all Q1 components, (b) only the brightest Q1 component per source.
The quasars are indicated by filled black squares, the BL
 Lac objects by open circles, and the galaxies by striped
 triangles.}
\label{fig:beta-z-lum}
\end{figure*}

Since the observed "high frequency" (i.e. $ \ge 5$ GHz)  luminosity is a quantity which is physically closely tied up with the relativistic jets that also display the moving components, we are confident that luminosity, rather than redshift, is the variable which truly correlates with velocity. A secondary argument in favor of this assertion is that much of the gradient in velocity as a function of redshift takes place over cosmologically fairly late times, certainly compared to ``the great quasar era'' (1.5$<$z$<$3). Breaking the luminosity-redshift degeneracy directly requires a restricted range of luminosities with a sufficient number of sources spread over a reasonably large range in redshift, or an orthogonal cut restricted in redshift. Fig.~\ref{fig:beta-z-lum} shows the best such selection afforded by the CJF sample: these are $\beta_{\rm app}^{\rm tot}$--$z$ diagrams in which only sources in the restricted range of luminosities $10^{35}$--$10^{36}$ erg$\,$s$^{-1}\,$Hz$^{-1}$,
are included. There is no longer a definite sloping upper envelope to the velocities as a function of redshift which supports the idea that, indeed, the primary correlation is with observed luminosity.

      \clearpage
       \pagebreak

\subsection{Other correlations}
\label{sub:class}

\begin{figure}[htb]
\begin{center}
\vspace*{0.3cm}
\subfigure[]{\includegraphics[clip,width=5.3cm,angle=-90]{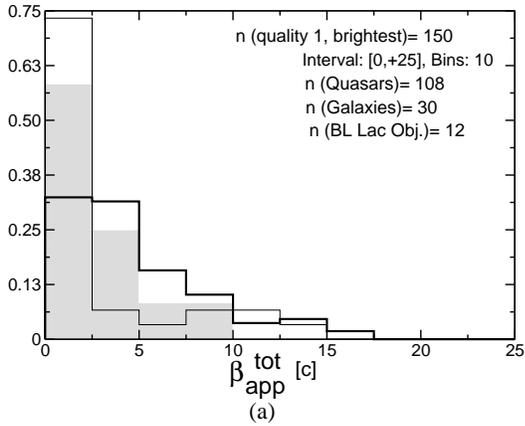}}\\
\end{center}
\caption{$\beta_{\rm app}^{\rm tot}$ for the different source classes
based on the brightest sub-sample.
The quasars are indicated
by a thick solid black line, the BL Lac objects in grey, and the
galaxies with a thin black line.}
\label{fig:beta-class}
\end{figure}

Apparent superluminal velocity studies have often indicated differences in the statistics between objects classified as quasars, galaxies, and BL Lacs (e.g., Gabuzda et al. 1994; Vermeulen 1995; Jorstad et al. 2001; Piner et al. 2007). We show histograms of the distributions in these different classes for  the CJF survey in Fig.~\ref{fig:beta-class}. The CJF galaxies clearly have slower apparent jet component velocities than the quasars, on average. We have verified that this conclusion holds when the known CSOs are left out.

However, we believe this trend may not be intrinsic to the classification of the host object. Fig.~\ref{fig:lum-beta} shows an increase in $\beta_{\rm app}^{\rm tot}$ with observed luminosity for all classes of sources; the same is true as a function of redshift (Fig.~\ref{fig:beta-z}). Over the range where the quasars and galaxies overlap in redshift and luminosity, their motion distributions cannot be distinguished. Indeed, superluminal velocities up to $\ge10\, c$ exist in some galaxies. Classification of the optical host type (we have taken these from the literature in many cases) is often an uncertain undertaking: Much further work is needed to clarify how the ratio between stellar and non-stellar optical continuum, radio continuum, and jet component velocities are inter-related.

There are too few BL Lac objects with a known redshift in the CJF sample to be
able to study their velocity distribution in comparison to quasars or galaxies.
In the past, different studies have reached opposite conclusions on whether BL
Lac objects typically have slow or fast jet components (e.g., Gabuzda et al. 1994; Wehrle et al. 1992; Ghisellini et al. 1993; Vermeulen 1995; Jorstad et al. 2001). We suspect this
can arise when different definitions of BL Lac objects are used.

We have also investigated whether the velocities in CJF sources depend on their 1.4--5 GHz radio spectral index, or on the ratio at 5 GHz between the VLBI-core flux density and the single-dish (i.e. total) flux density ("Core-dominance") .  In the context of quasar-galaxy orientation unification models, these parameters are sometimes taken as proxies for 
the amount of beaming, or for the jet angle to the line-of-sight. But for the CJF sample, neither of these properties is correlated with the component velocities.

\section{Comparisons between motion surveys}
\label{sec:comp}

Compared to the CJF survey, previous compilations of apparent velocities contained far fewer sources, and usually were rather more heterogeneous, both in the sample composition, and in the observing setup. In contrast to our CJF results, these older lists typically contained a large fraction of highly superluminal velocities, many in excess of $10c$, and ranging up to $40c$ and more (after conversion to current cosmology). There may also have been a preference to study highly variable objects (e.g., Wehrle et al. 1992), which may tend to select for high Doppler factors (e.g. Homan et al.\ 2006). The earlier observations typically involved sources which are several times brighter than most CJF sources. Those sources at significant redshifts are often amongst the most luminous ones around. The correlation discussed in \S~\ref{sub:corr-L} implies a significantly enhanced chance to have a fast velocity at the high end of the luminosity distribution.

\subsection{Comparison with the VLBA 2cm Survey and the RRFID Survey}
\label{sub:2cm}

Piner et al. (2007) have recently published the kinematics of a significant sample of 77 AGN taken from 8  GHz VLBA images in the RRFID (Radio Reference Frame Image Database). These data are based on the first 5 years of the database (1994-1998). The overlap between the CJF and this stage of the RRFID survey is 17 sources. A larger sample of velocities has been published for 110 sources from the VLBA 2cm survey  (Kellermann et al.\ 1998, 2004). While this is still only about half as many sources as in CJF, the observing and data analysis were carried out homogeneously, and a large number of observing epochs were obtained for each source. The overlap between the  samples is 24 sources with measured motions. 

The VLBA 2cm survey has brighter sources (limit 1.5 Jy on the extrapolated 2cm flux density), no spectral index limit, and covers a much wider area of sky compared with the CJF sample. A cursory inspection shows that the velocities in these common sources often differ significantly between the two surveys. But the observations often do not pertain to the same components because the measurements often did not span the same time period and because the two surveys differ in spatial resolution by about a factor of three. Disentangling the situation in the individual sources is beyond the scope of the present work, and requires concurrent multi-frequency observations in the future. We can, however, compare the velocity statistics of the two full  samples.

We compare the histograms of the apparent velocity distributions i.e. Fig.~\ref{fig:betarohisto} in this paper, and Fig.~5 in Kellermann et al.\ (2004). In direct analogy to the distribution of velocities discussed in for the CJF distribution, we find that the VLBA 2cm survey distribution shows evidence for two populations: one with slow or at most mildly relativistic components, the other with a broad range of velocities extending to highly superluminal values. Indeed, Kellermann et al.\ (2004) have already proposed that the ``stationary'' components may well represent a different phenomenon. The measured apparent speed distribution for the 94 best-measured components (Fig. 4) by Piner et al. (2007) also shows a peak at low apparent speeds together with a tail extending out to about 30$c$.

Returning to the VLBA 2cm survey, the distribution of the superluminal velocities appears to be flat out to almost $10c$, whereas the distribution in CJF already turns down beyond 4--5$c$. Likewise, while velocities in the range 15--25$c$ are very rare in CJF, they are more common in the VLBA 2cm survey. Thus, the mean velocity in the VLBA 2cm survey is substantially higher than in the CJF survey; discounting the probably separate population of ``stationary'' components, the ratio is roughly a factor of 1.5--2.

The correlation discussed in \S~\ref{sub:corr-L} between the apparent velocities and the observed luminosities, is also prominent in the VLBA 2cm survey; compare Fig.~\ref{fig:lum-beta} to Fig.~6 in Kellermann et al.\ (2004).
There is a larger proportion of highly superluminal velocities in the VLBA 2 cm survey than in the CJF sample. However, the luminosities in the VLBA 2cm survey are slightly larger, as one might expect given that the sources are brighter. 

\subsection{Increasing apparent velocity with increasing observing frequency}
\label{sub:beta-freq}

In the context of increasing apparent velocity with observing frequency a survey including results at 7mm has been published by Jorstad et al.\ (2001). Both the mean and the largest velocities are higher than in the VLBA 2cm survey and we have already noted that the VLBA 2cm survey apparent  velocities seem to be faster than the CJF-results. However, while Jorstad et al.\ (2001) have published an impressive body of work, their sample size is comparatively modest. And with its focus on sources detected in gamma-rays, it is likely to be less representative of a complete flux-density limited survey. It does provide evidence that sources prone to being detected in gamma-rays often show apparent superluminal motion (e.g., Bower et al. 1997; Barthel et al. 1995. There are plausible beaming models which can explain the clear trend for higher velocities to be seen at higher frequencies. For example, if individual jets contain matter moving with a range of velocities, perhaps in a fast spine and slower sheath configuration (e.g., ``two-fluid model'': Pelletier \& Roland 1998; Laing \& Bridle 2002), then optical depth effects could reveal faster material at higher frequencies. Alternatively the link between velocity and resolution could be taken as evidence that higher velocities occur closer to the central engine indicating that some deceleration already takes place on parsec scales. This, however, conflicts with the very tentative evidence for a greater prevalence of acceleration rather than deceleration along the jets seen in our CJF data. Overall, then, while the velocity-frequency trend is clear much further work is needed 
to understand the details of the effect and how it impinges on the theory of relativistic jets.  

\section{Conclusions}
\label{sec:conclusions}
In this paper we have described the statistical analysis of jet component motion in 237 CJF-sources. We have described in detail the necessary tests that have been performed to ensure that our treatment is free of any calculation-induced bias. We have investigated several different ways of handling various sorts of complications in the estimation of the kinematic parameters for individual jet components, for example splitting and merging components mergers. 

The program we developed allows us, from the computation of 2-D kinematic
models, to estimate both radial motions {\it and} orthogonal motions. The
overwhelming predominance of outward radial velocities in AGN is confirmed, as well as the
dominance of significant apparent superluminal motions. Our most important
results are briefly summarized as follows: \\

\noindent
a) Observational results from CJF alone\\
\begin{itemize}
\item Motion within sources:\\
In general, we find a spectrum of motion values in a given source: motions are not consistent with a single uniform velocity applicable to all components along a jet. The median of the distribution of standard deviation taken for jets comprising at least three Q1 components are 1.97$c$ ($\beta^{\rm tot}$), 2.02$c$ ($\beta^{\rm r}$), and 0.98$c$ ($\beta^{\rm o}$).\\

\item Accelerations along individual jets:\\
Although three epochs are generally not sufficient to properly investigate acceleration of individual components along the jet, we can characterize the acceleration along a jet (or counter-jet) in its entirety. For the Q1-only sample we find a slight trend towards a positive outward  acceleration. We also adduce some evidence for greater acceleration in the inner-most regions but these observations do not have the resolution to test this properly. The quasars reveal a more positive $\partial\beta^r/\partial r$ and $\beta^r\,(r=0)$.\\
\item A lack of fast components at distances below a few pc:\\
This effect might arise simply from the limited time sampling of the motions; this could easily be tested in a sub-sample observed with smaller time intervals. Alternatively the effect could be physical in nature relating to the fact that the smaller physical distances are resolvable only at the lower redshifts and thus in the less luminous sources, where we do not find large velocities.\\
\item Bending along individual jets:\\
Mild degrees of bending are common but we find very few Q1 components undergoing apparently large bends i.e. within  $15^\circ$ of $\pm90^\circ$. Only $\sim$4\%
of the components from galaxies and $<$2\% of those from quasars show such bends.\\
\item Representative velocities:\\
Despite the subtleties of the motions within a given jet, a single velocity per source is required for population studies. We find that the sample of the brightest Q1 components (one per source) yields a distribution of velocities which cannot be statistically distinguished from that of our full sample of Q1 components; this has therefore been used as the velocity sample for our various correlation studies.\\ 
\item Radial velocity distribution:\\
The distribution of radial velocities shows a broad distribution of velocities (apparent velocities up to 30 $c$) with a pronounced narrow peak of slow apparent velocities superimposed. These subluminal and stationary components occur relatively often: 42 of 272 components of the Q1 components reveal such "slow" radial velocities which may need to be explained in a different manner to the superluminal motions. For the latter the broad distribution of apparent velocities requires a broad initial distribution of the Lorentz factors of the moving features. \\
\item Negative radial velocities:\\
Negative or "backwards" superluminal motions are seen and in 15 cases these are significant. In some of these cases the phenomenon could be ascribed to geometrical effects in curved jets seen close to the line of sight or to actual backward motions of shocks. However we suspect that in the majority of cases the apparent backwards motions arise from problems with our choice of reference feature.\\
\end{itemize}
b) Correlations with other source parameters\\
\begin{itemize}
\item Correlation with the observed 5 GHz luminosity:\\
We have been able to break the degeneracy between redshift and luminosity and find a strong correlation between the 5 GHz luminosity and apparent velocity. A clear outcome of this analysis is that at the high end of the luminosity distribution the chance to have a fast velocity is significantly enhanced. The correlation with luminosity may contribute to the lack of fast components at the smallest core separations. \\
\item Apparent velocity with regard to the source class:\\
The CJF galaxies clearly show slower apparent jet component velocities than the quasars, on average. However this difference may well be due to the velocity-luminosity correlation since  over the range where quasars and galaxies overlap in redshift and luminosity, their motion distributions cannot be distinguished.  The number of BL Lac objects with a known redshift in the CJF sample is too small to allow a meaningful comparison with quasars or galaxies to be made.\\
\item No correlation with (1.4--5 GHz) radio spectral index or the core-dominance parameter:\\
Neither of these parameters shows any correlation with the component velocities.\\
\item Dependence of the average apparent velocities on the observational frequency of the motion survey:\\
A comparison of the results of the CJF with previous compilations of apparent velocities is hampered by the fact that they contained many fewer sources. The dependence of velocity on the luminosity additionally complicates a comparison. Nevertheless the mean velocity in the VLBA 2cm survey is substantially higher than in the CJF survey -- the ratio could be roughly a factor of 1.5--2. This trend of increasing apparent velocity with increasing observing frequency is supported by even higher velocities found by Jorstad et al. (2001) in a VLBI survey observed at frequencies up to 43 GHz. Simultaneous multi-frequency studies of well-chosen sub-samples are needed to clarify this phenomenon. \\
\end{itemize}

\begin{acknowledgements}
It is a pleasure to thank A. Witzel and T.P. Krichbaum for very inspiring discussions and many helpful comments. 
We thank the referee K. Kellermann for his advice and comments that significantly improved this manuscript.
M. Karouzos carefully proof-read the manuscript.
This work was supported by the European Commission, TMR Programme,
Research Network Contract ERBFMRXCT96-0034 ``CERES''.
S. Britzen acknowledges support by the Claussen-Simon-Stiftung.
This project has been supported by the DLR, project 50QD0101.
This research has made use of the NASA/IPAC Extragalactic Database (NED) which 
is operated by the Jet Propulsion Laboratory, California Institute of Technology, 
under contract with the National Aeronautics and Space Administration. 
The National Radio Astronomy Observatory is operated by Associated Universities, Inc., under cooperative agreement with the National Science Foundation. The European VLBI Network is a joint facility of European, Chinese, South African and other radio astronomy institutes funded by their national research councils.
This work is based on observations with the 100-m telescope of the MPIfR (Max-Planck-Institut für Radioastronomie) at Effelsberg.
\end{acknowledgements}

\begin{appendix}
\section{New Redshift Information}\label{app:z}

For 19 sources, our redshift determinations constitute an update to
previously published values (see Table~\ref{table:proper}, column 3 for
values). Details for these 19 sources are as follows:
\begin{flushleft}
{\sl 0018+729}\\
\end{flushleft}
An unpublished Palomar 200" Hale Telescope observation by Vermeulen \&
Readhead on 1994 August 15 does not confirm the presence of the
[\ion{O}{ii}]$\lambda$3727 line on the basis of which Snellen et al. (1996) reported a redshift $z$=0.821.
\begin{flushleft}
{\sl 0340+362}\\
\end{flushleft}
The probable emission line redshift published in Vermeulen \& Taylor
(1995) was confirmed in an unpublished Palomar 200" Hale Telescope
observation by Vermeulen \& Readhead on 1996 January 16, which revealed
the \ion{C}{iii}]$\lambda$1909 emission line in addition to some of the emission
lines already tentatively identified in Vermeulen \& Taylor (1995); the
final redshift measurement is $z$=1.485$\pm$0.002.
\begin{flushleft}
{\sl 0346+800}\\
\end{flushleft}
This source still has no secure emission line redshift of which we are
aware. However, an unpublished Palomar 200" Hale Telescope observation
by Vermeulen \& Readhead on 1996 January 14 shows a probable narrow
emission line near 7550~\AA\ and a possible faint broad emission feature
near 5500~\AA. If interpreted as [\ion{O}{ii}]$\lambda$3727 and \ion{Mg}{ii}$\lambda$2798,
respectively, these would tentatively suggest a redshift near $z$=1.03.
\begin{flushleft}
{\sl 0602+673}\\
\end{flushleft}
The probable redshift, $z$=1.95, already listed in Kellermann et~al. (1998), is based on an observation at the Keck I Telescope by Taylor, which shows irregularly shaped emission features consistent with \ion{C}{iv}$\lambda$1549 and \ion{C}{iii}]$\lambda$1909.
\begin{flushleft}
{\sl 0604+728}\\
\end{flushleft}
The emission line redshift, $z$=0.986$\pm$0.001, is based on an
unpublished Palomar 200" Hale Telescope observation by Vermeulen \&
Readhead on 1996 January 14. It shows narrow
[\ion{O}{ii}]$\lambda$3727 and [\ion{Ne}{v}]$\lambda$3426 emission lines, and more
tentative H$\delta\lambda$4102 and [\ion{Ne}{iii}]$\lambda$3869 emission
lines. \ion{Mg}{ii}$\lambda$2798 is also seen in emission, but has an associated
absorption line component as well; this associated absorption system is
also visible in \ion{Fe}{ii}$\lambda$2586, 2600 and $\lambda$2374,2382
lines.
\begin{flushleft}
{\sl 0738+491}\\
\end{flushleft}
The probable emission line redshift, $z$=2.32, is based on an
unpublished Palomar 200" Hale Telescope observation by Vermeulen \&
Readhead on 1996 January 14. It shows a sharp emission line near 4040~\AA,
which is probably Ly$\alpha\lambda$1216, and a broad, irregular
emission feature near 5140~\AA\ that probably corresponds to
\ion{C}{iv}$\lambda$1549.
\begin{flushleft}
{\sl 0800+618}\\
\end{flushleft}
The emission line redshift, $z$=3.044$\pm$0.002, is based on an
unpublished Palomar 200" Hale Telescope observation by Vermeulen \&
Readhead on 1996 January 14, which shows
Ly$\alpha\lambda$1216, \ion{Si}{iv}/\ion{O}{iv}$\lambda$1400, and \ion{C}{iv}$\lambda$1549 in
emission. There is also evidence for emission lines corresponding to
Ly$\beta\lambda$1026 and \ion{C}{iii}]$\lambda$1909. Furthermore, there is
separate, deep, possible damped, absorption line system at $z_{\rm
abs}$=2.963, in which at least the following lines are visible:
\ion{C}{iv}$\lambda$1549, \ion{Si}{ii}$\lambda$1527, \ion{C}{ii}$\lambda$1334, 
\ion{O}{i}$\lambda$1302,
\ion{Si}{ii}$\lambda$1216, Ly$\alpha\lambda$1216, and Ly$\beta\lambda$1026.
\begin{flushleft}
{\sl 0942+468}\\
\end{flushleft}
The emission line redshift, $z$=0.639$\pm$0.002, is secure. It is based
on narrow [\ion{O}{ii}]$\lambda$3727 and
[\ion{O}{iii}]$\lambda$4959, 5007 emission lines in an unpublished Palomar
200" Hale Telescope observation by Vermeulen \& Readhead on 1996
January 16. Engels et al. (1998) have published
another value, $z$=0.993, but this was based on assuming that a single
feature centered at 5577~\AA\ (where there is a very prominent night sky
line) would be \ion{Mg}{ii}$\lambda$2798.
\begin{flushleft}
{\sl 1125+596}\\
\end{flushleft}
The emission line redshift, $z$=1.799$\pm$0.003, is based on unpublished
Palomar 200" Hale Telescope observations by Vermeulen \& Readhead on
1996 January 14 and 16, which show an unambiguous combination of
emission lines: \ion{Si}{iv}/\ion{O}{iv}$\lambda$1400, \ion{C}{iv}$\lambda$1549, and
\ion{C}{iii}]$\lambda$1909. There is also evidence for an emission line
corresponding to \ion{Mg}{ii}$\lambda$2798. The resonance lines have an irregular
profile due to several unresolved associated absorption systems which
cut into the broad emission lines.
\begin{flushleft}
{\sl 1300+580}\\
\end{flushleft}
Unpublished observations by Vermeulen, Taylor, \& Readhead at the
Palomar 200" Hale Telescope on 1994 May 15 and 1996 January 16, and at the
Keck I Telescope on 1996 June 18, together show the
emission line redshift to be $z$=1.088$\pm$0.003, based on
\ion{Mg}{ii}$\lambda$2798 and [\ion{O}{ii}]$\lambda$3727 emission lines.
\begin{flushleft}
{\sl 1438+385}\\
\end{flushleft}
Observations at the Keck I Telescope by Taylor show a spectrum 
with broad \ion{He}{ii} at 4538~\AA\ ($z$=1.766),
broad \ion{C}{iii}] at 5311~\AA\ ($z$=1.783), weak \ion{Al}{iii} at 5133~\AA\ (z=1.762),
and weak \ion{Mg}{ii} at 7783~\AA\ (z=1.780). We estimate the redshift to be 1.775$\pm$0.01 based on the stronger lines. Earlier, Vermeulen et al. (1996) had published a probable emission line redshift of $z$=1.773.
\begin{flushleft}
{\sl 1459+480}\\
\end{flushleft}
The emission line redshift, $z$=1.059$\pm$0.002, is based on broad
\ion{Mg}{ii}$\lambda$2798 and narrow [\ion{O}{ii}]$\lambda$3727 emission lines in an
unpublished Keck~I Telescope observation on 1996 June 18 by Vermeulen
\& Taylor.
\begin{flushleft}
{\sl 1716+686}\\
\end{flushleft}
The probable redshift, $z$=0.339$\pm$0.001, is based on an observation at the Keck
I Telescope on by Taylor and strong [\ion{O}{ii}], H$\beta$, [\ion{O}{iii}], and H$\alpha$.
This redshift is inconsistent with the $z$=0.777 found
by K\"uhr (1980) and also with the $z$=0.798 published by
Stickel \& K\"uhr (1994). 
\begin{flushleft}
{\sl 1734+508}\\
\end{flushleft}
The emission line redshift, $z$=0.835$\pm$0.001, is based on very
prominent narrow [\ion{Ne}{v}]$\lambda$3426, [\ion{O}{ii}]$\lambda$3727, 
[\ion{Ne}{iii}]3869, and
[\ion{O}{iii}]$\lambda$4959, 5007 emission lines in an unpublished Keck I
Telescope observation on 1996 June 18 by Vermeulen \& Taylor.
\begin{flushleft}
{\sl 1809+568}\\
\end{flushleft}
The emission line redshift, $z$=2.041$\pm$0.002, is based on prominent
\ion{C}{iv}$\lambda$1549 and \ion{C}{iii}]$\lambda$1909 emission lines in an
unpublished Keck I Telescope observation on 1996 June 18 by Vermeulen
\& Taylor. 
\begin{flushleft}
{\sl 2054+611}\\
\end{flushleft}
The emission line redshift, $z$=0.864$\pm$0.002, is based on prominent
\ion{Mg}{ii}$\lambda$2798, H$\gamma\lambda$4340, and H$\beta\lambda$4861 emission
lines in an unpublished Keck I Telescope observation on 1996 June 19 by
Vermeulen \& Taylor. [\ion{O}{iii}]$\lambda$4959 is probably present at the edge
of the observed spectrum as well.
\begin{flushleft}
{\sl 2229+695}\\
\end{flushleft}
The emission line redshift, $z$=1.413$\pm$0.002, is based on broad
\ion{C}{iii}]$\lambda$1909 and \ion{Mg}{ii}$\lambda$2798 emission lines in an unpublished
Keck I Telescope observation on 1996 June 19 by Vermeulen \& Taylor.
\begin{flushleft}
{\sl 2238+410}\\
\end{flushleft}
The emission line redshift, $z$=0.726$\pm$0.001, is based on narrow
[\ion{O}{ii}]$\lambda$3727 and [\ion{O}{iii}]$\lambda$4959, 5007 emission lines in an
unpublished Keck I Telescope observation on 1996 June 18 by Vermeulen
\& Taylor. Ca H+K absorption at the same redshift is also visible.
\begin{flushleft}
{\sl 2319+444}\\
\end{flushleft}
The redshift, $z$=1.251$\pm$0.003, is based on a broad \ion{Mg}{ii}$\lambda$2798
emission line and a clear 4000~\AA\ break in the continuum, seen in an
unpublished Keck I Telescope observation on 1996 June 19 by Vermeulen
\& Taylor.\\

\section{Kinematic modeling procedures}
\label{app:fitmeth}

The starting point in all cases was the set of circular Gaussian
parameters estimated from the complex visibilities per epoch as
discussed in Paper~I.  These include the positions $(x,y)$ of the
components found by {\it difmap} with respect to the reference point, their uncertainties $(\sigma_x,
\sigma_y)$, and the elements of the correlation matrix associated with
these parameters.  
In the ideal case, we could estimate the kinematic parameters for all
of a source's components in a single fit, incorporating the {\it
difmap}-output correlation matrices as {\it a priori} covariance
matrices to reflect the fact that the visibilities and their ({\it
u-v}) distribution may constrain different parameters more or less
effectively in different epochs.  This approach should preserve most of
the ``information" contained (or not) in the original visibilities.

However, the existence of a ``splitter" component, where two distinct
{\it difmap} components are identified with a single jet component at
some epoch(s), can frustrate this approach.  The residuals of the
individual ``sub-components" with respect to the estimated position
for the associated single component dominate the chi-square statistic
for the kinematic model, which in turn becomes insensitive to
adjustments in the parameters of all other components.  Additionally,
the use of an {\it a priori} covariance matrix that includes non-zero
cross-component elements ensures some of the specious adjustment to
``splitter" components communicates to the other better-behaved
components.  Thus, the first variant approach to the kinematic
modeling was to fit independent models for each component within a
source rather than for each source as a whole.  This meant not taking
into account any cross-component correlation-matrix elements in forming
the {\it a priori} covariance matrices.  Further, the effective
position of a ``splitter" at an epoch was controlled by the relative
sizes of the constituent sub-components' uncertainties.  In some cases,
the resulting time series of component position was not well-described
by monotonic motion. We also tried other weighting schemes for the $(\sigma_x,
\sigma_y)$.
Each of the above variants became an independent choice in the tactics
for estimating the kinematic models for the sources:
\begin{itemize}
\item{ } include all components in a source in a single fit or fit 
independently per component
\item{ } use the {\it difmap} correlation matrix as an {\it a priori}
covariance matrix or not
\item{ } incorporate $(\sigma_x, \sigma_y)$ as estimated from {\it difmap}, treat them as
uniform across all epochs, or set them proportional to reciprocal flux-density
\end{itemize}
\noindent The complete set of kinematic models could be computed for
all sources given a combination of the above three choices.  

In order to gain confidence in the incorporation of ``splitters" into
the statistics of component kinematics, we had to investigate whether
using any of the alternate tactics 
led to any biases.  
First, we looked at only sources composed of Q1
components.  These could be fit using the ideal method as described above.
We then recomputed the kinematics fits for these Q1-only sources using
different tactics more accommodating to ``splitters".  Specifically, we would
thus be comparing separate fits using the first and the last option in
the each of the above three bullets.
Let us call
these the ``Joint" and ``Independent" fits.  To be able to include the 
lower-quality components reliably, we would need to demonstrate at least that
the ``Independent" fit introduces no bias compared to the more rigorous ``Joint"
fit for these Q1-only sources.

\begin{figure}[htb]
\begin{center}
\subfigure[]{\psfig{figure=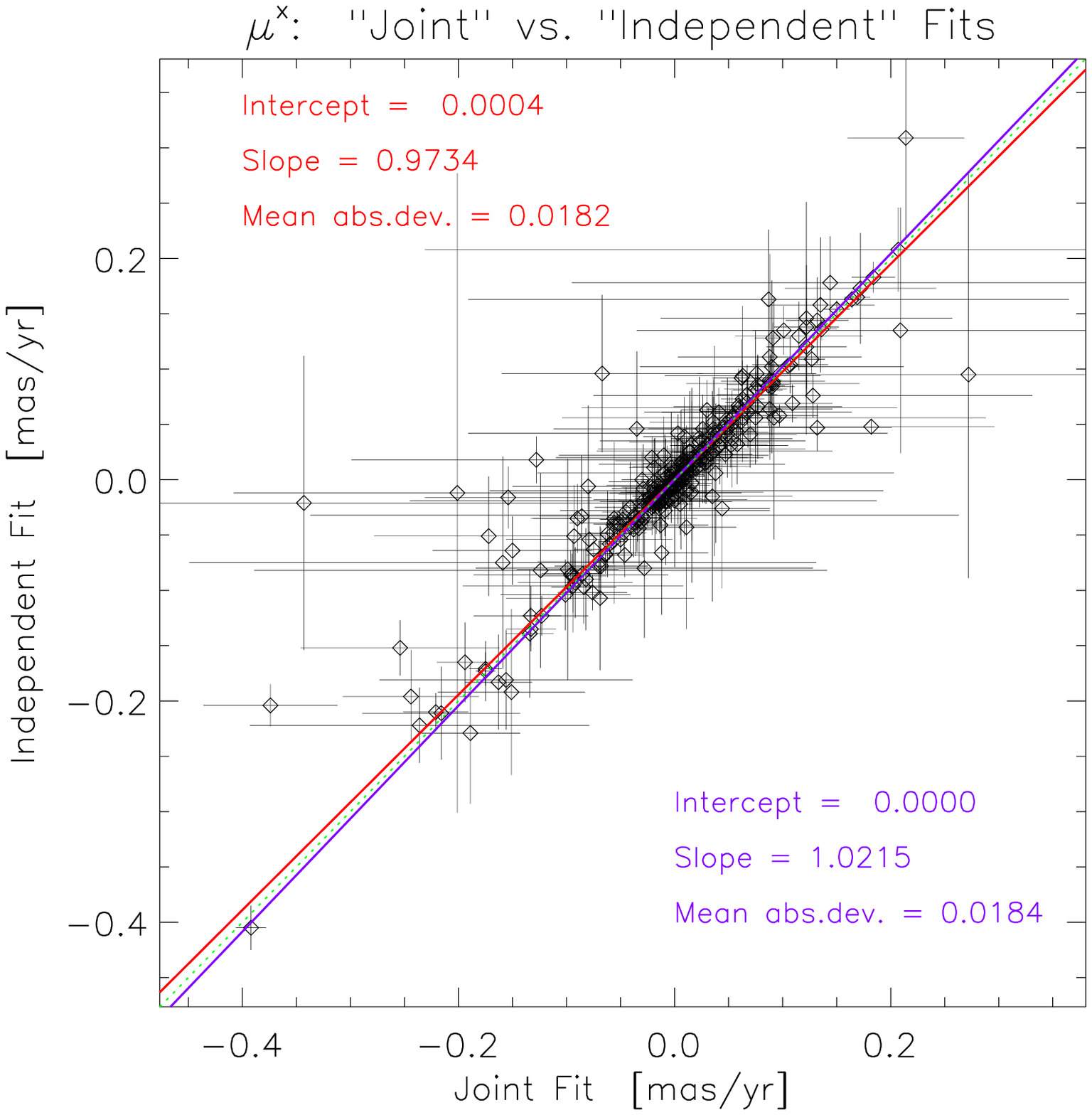,width=8cm}}
\vspace*{0.5cm}
\subfigure[]{\psfig{figure=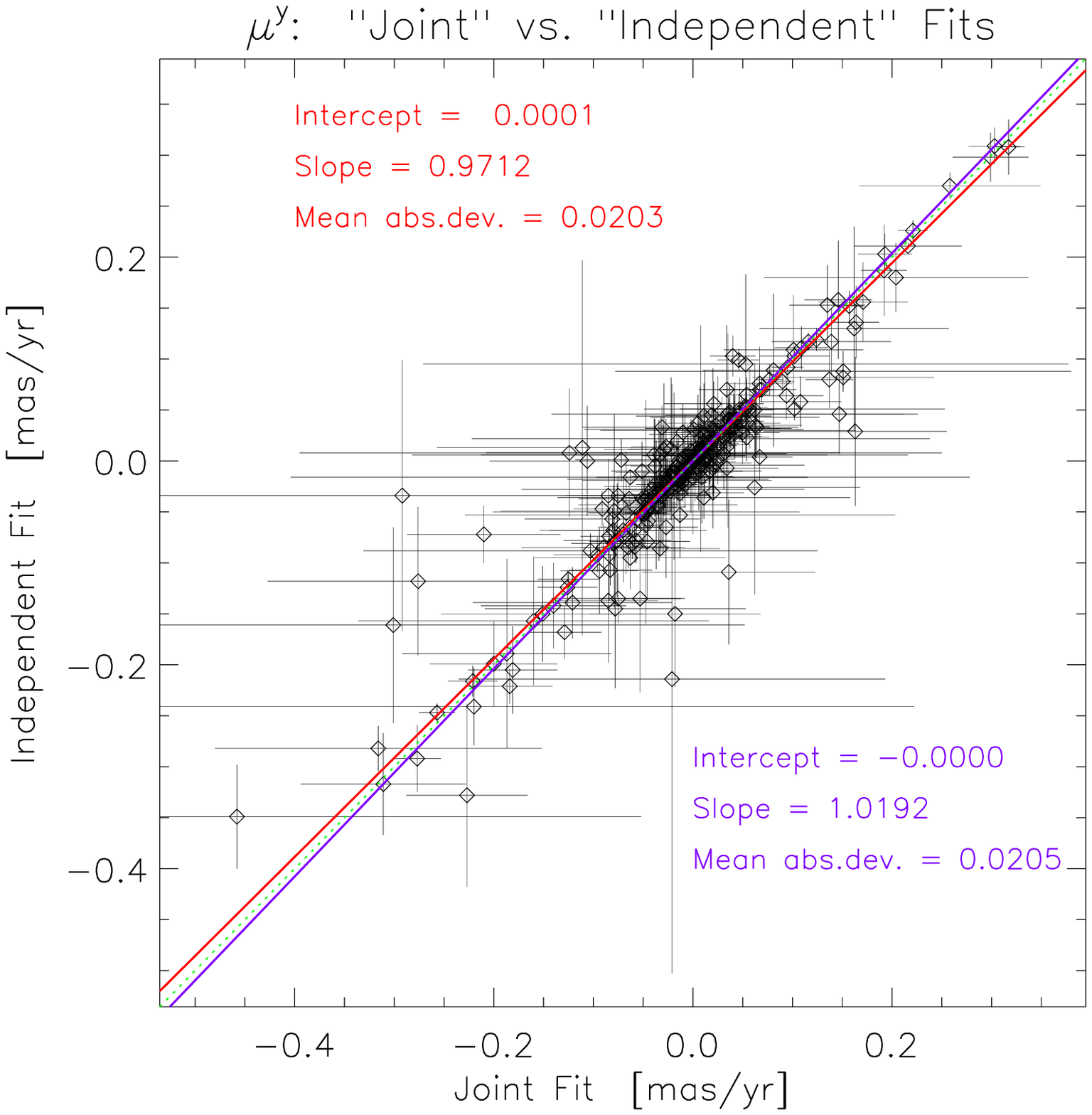,width=8cm}}
\end{center}
\vspace*{-0.8cm}
\caption{Comparison of proper motion values computed using different
tactics in the kinematic modeling.  Each point represents a single
component, with uncertainties along the abscissa
from the ``Joint" fit and along the ordinate from the
``Independent" fit.  Separate 1-D absolute-deviation fits are shown,
one using the ``Independent" fit as the dependent variable (red) and the
other using the ``Joint" fit as the dependent variable (blue).  The dotted
green line marks the (ideal) diagonal.  Panel (a)
shows $\mu^x$ and panel (b) shows $\mu^y$.}
\label{fig:flag}
\end{figure}

\begin{figure}
\begin{center}
\psfig{figure=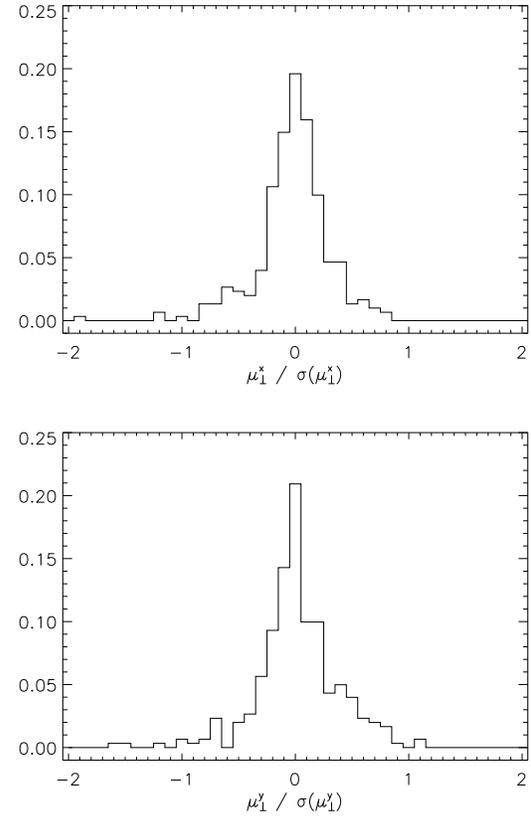,width=8cm}
\end{center}
\vspace*{-0.8cm}
\caption{Histograms of $\mu_\perp$ normalized by $\sigma_{\mu_\perp}$.
Here, $\mu_\perp$ is the perpendicular distance from each point in the 
panels of figure~\ref{fig:flag} to the diagonal.
The top panel shows $\mu_\perp^x$ and the bottom shows $\mu_\perp^y$.} 
\label{fig:perphist}
\end{figure}

Fig.~\ref{fig:flag} shows the correlation between the proper-motion estimates
for these two sorts of fits -- panel (a) for $\mu^x$ and panel (b) for 
$\mu^y$.  Each point corresponds to a single component, whose uncertainty in the abscissa direction come from the ``Joint" 
fit and in the ordinate direction from the ``Independent" fit.
We made two separate one-dimensional absolute-deviation straight-line fits
to these points, one using the ``Independent" fit as the dependent 
variable (red) and the other using the ``Joint" fit as the dependent variable
(blue).  Fit statistics are annotated in the appropriate color.
A dotted green line marks the diagonal, upon which all points
would ideally lie.
We then computed the perpendicular distance on these plots from each point
to the diagonal ($| \mu_{\rm joint} - \mu_{\rm indep} | / \sqrt{2}$) and
the uncertainty in this $\mu_\perp$ via conventional error propagation
assuming no correlation between the ``Joint" and ``Independent" uncertainties.
This perpendicular distance is significantly more strongly peaked near zero
than is either of the uncertainties in the two kinds of fits, and there are
no obvious trends in the two-dimensional plot of $\mu_\perp$ vs.\ either
uncertainty (here with $\mu_\perp$ keeping a sense of above or below the
diagonal). Fig.~\ref{fig:perphist} shows histograms of $\mu_\perp$ normalized 
by $\sigma_{\mu_\perp}$ -- distributions
that are consistent with 0-mean processes, significantly
more compact and peaked (kurtosis = 21.2 for $\mu_\perp^x$, 11.8 for 
$\mu_\perp^y$) than Gaussian.

Thus, we should be able to use the ``Independent" fit tactics in order
to be able to incorporate components from sources having lower-quality
ones without biasing the statistical results from the higher-quality
sources.  We scale the resulting uncertainties from these fits to a
reduced-chi-square of unity, which should result in 
reasonably conservative values.  A disadvantage of this approach
is that some information contained in the {\it difmap} fits from
paper~I for sources containing only Q1 components will not be carried
forward through the kinematic modeling.  Furthermore, the
``Independent" fit tactics as described above produce the numerical
result $\sigma_{\mu^x} =
\sigma_{\mu^y}$ in the course of the least-squares estimation, essentially
circularizing the beam in all epochs.

\section{Kinematic model results}
\label{app:veltab}

Table~\ref{table:proper} lists the results of the kinematic modeling for all
779 components from sources participating in the proper-motion
analysis.  Column (1) lists the IAU name, column (2) gives the optical
classification (Q: quasar, B: BL Lac object, G: galaxy, U: unclassified
object), column (3) lists the redshift, column (4) gives the jet
component identification, column (5) the quality class as defined in
section~\ref{sec:modelf}.  The following three columns (6), (7), and
(8) give the number of times this component has been detected, the time
span from the first to the last detection, and the reference epoch.
Column (9) and (10) list the component's position at the reference
epoch in rectangular coordinates. Columns (11) and (12) list the proper
motion components in $x$ and $y$ respectively, and the next pair of
columns (13) and (14) show the radial and orthogonal proper motion
components.  Column (15) gives the total proper motion value,
calculated from $\mu^x, \mu^y$.  The last three columns
(16)--(18) list the radial, orthogonal, and total apparent
velocities. Parameters in columns (9)--(18) are listed with their
uncertainties.  The uncertainties for the ``splitters'' are typically
~2-3 times the uncertainties for a ``normal'' component, driven mostly
by scaling the high reduced chi-square of their fit.  Components that
have observations at only two epochs do not have an associated
uncertainty; with no degrees of freedom in the fit for
the kinematic model (four constraints and four unknowns), the scaling to
$\tilde\chi=1$ is ill-defined.  In Table~\ref{table:proper} we indicate
this missing uncertainty by a bar.\\

\begin{figure*}[htb]
\includegraphics[clip,width=19.45cm]{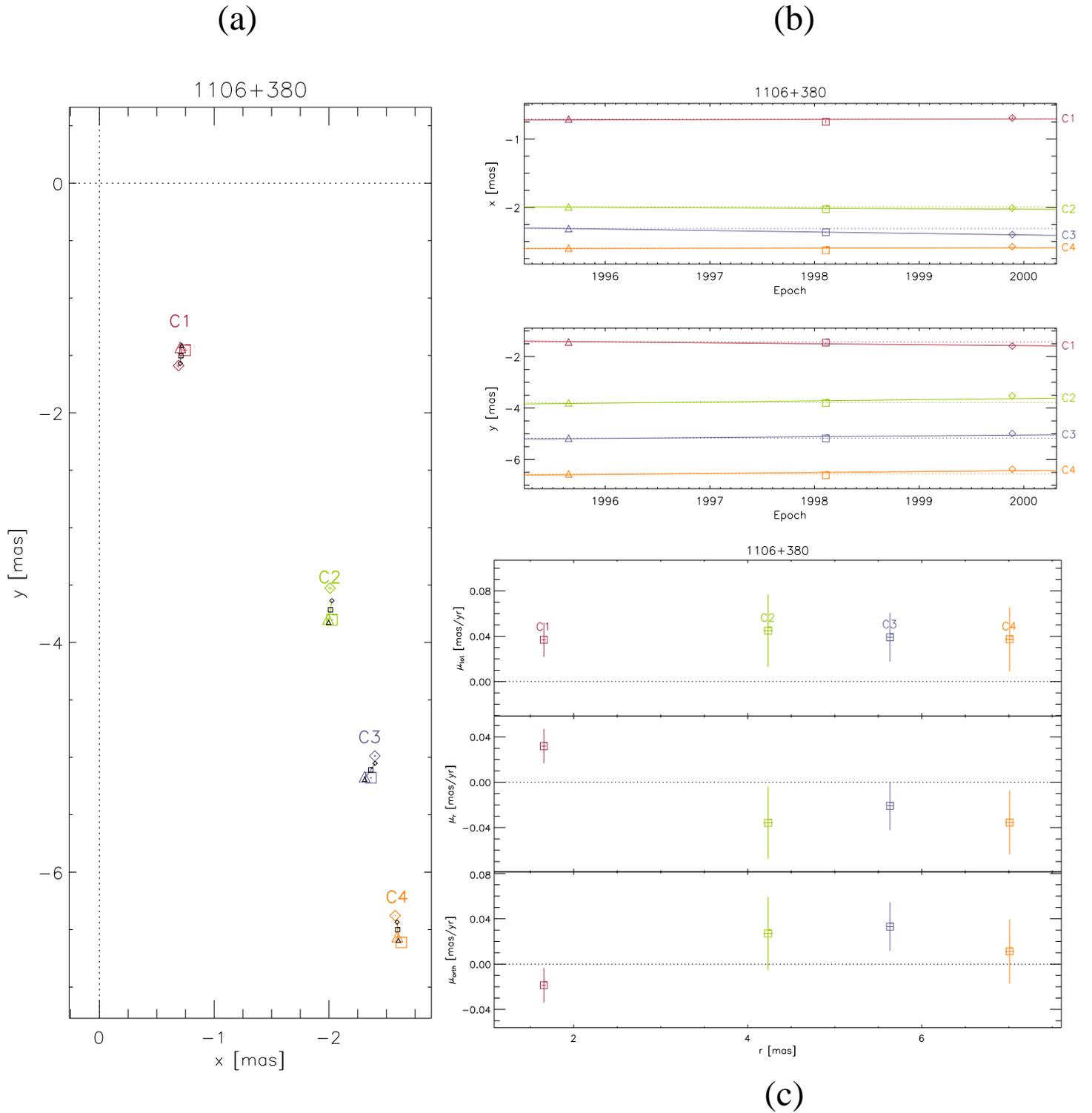}
\caption{Example, for 1106+380, of the kinematic modeling figures.
Panel (a) shows the positions of all components at all epochs.  The components
are color-coded and the epochs are represented by different plotting symbols.
The modeled positions at each epoch are shown in small black symbols.
Panel (b) plots $x$ and $y$ of each component as a function of time.  For
each component, a ``stationary" dotted line is plotted for reference.
Panel (c) plots $\mu^{\rm tot}$, $\mu^r$, and $\mu^o$ as a function of radial
distance from the core.  The component color-coding of panel (a) is carried
forward into the other two panels.}
\label{fig:bplots}
\end{figure*}

Additionally, we display the kinematic modeling results graphically
. Fig.~\ref{fig:bplots} shows the source 1106+380 as an example.
We show three panels:\\
Panel (a) shows a plot of the positions of all components at all epochs on a
plane tangent to the core.  The components are color-coded, and the
epochs are represented by symbols (first = triangle, second = square,
third = diamond, fourth = $\times$, fifth = asterisk).  The scaled
uncertainties for each component from Paper~I are also overplotted on
top of the symbols.  In addition to the observed positions plotted in
color, the modeled positions at each epoch are shown in small black
symbols.  Each of these plots is drawn so that one mas on the sky has
the same scale in both the $x$- and $y$-axes on the plot (the scale of
course can vary from source to source).\\
Panel (b) shows a plot of the $x$ and $y$ positions of the components as a function
of time.  The component color-coding and epoch symbols remain the same as in
the X-Y plot.  To guide the eye for small motions, a horizontal dotted line
is plotted for each component.\\
Panel (c) shows a plot of the total proper motion as well as the radial and orthogonal
proper-motion components, for each jet component plotted as a function of
radial distance from the core.  The jet-component color-coding remains the
same as in the other plots. The scaled errors in both $r$ and the 
proper-motion component are plotted (unscaled
errors used for components with only two epochs of observations).
Plots for all 237 CJF sources are available in electronic form at the CDS via anonymous ftp to cdsarc.u-strasbg.fr (130.79.128.5) or via http://cdsweb.u-strasbg.fr/cgi-bin/qcat?J/A+A/. In addition, the plots can be downloaded from a CJF-archive page at the MPIfR in Bonn (http://www.mpifr-bonn.mpg.de/staff/sbritzen/cjf.html).

\onecolumn
\landscape
\tabcolsep0.6mm
{\scriptsize
\hspace*{-6cm}
\tablecaption{The proper motion and $\beta_{\rm app}$-values of a subsample of the CJF sources (18 sources). The complete information for 237 sources is available in electronic form at the CDS via anonymous ftp to cdsarc.u-strasbg.fr (130.79.128.5) or via http://cdsweb.u-strasbg.fr/cgi-bin/qcat?J/A+A/} {\smallskip}
\hspace*{-6cm}
\tablehead{\noalign{\smallskip} \hline \noalign{\smallskip}
(1)   &(2)&  (3) &(4) & (5) &(6)&(7)&(8)&(9)&(10)&(11)&(12)&(13)&(14)&(15)&(16)&(17)&(18)\\
\hline
Source& Cl&z&Id&q&n.ep.&span&ref.ep.&X$_{0}\pm\Delta$X$_{0}$ & Y$_{0}\pm\Delta$Y$_{0}$ &$\mu^x\pm\Delta$$\mu^x$ &$\mu^y\pm\Delta\mu^y$&$\mu^r\pm\Delta\mu^r$ &$\mu^{\o}\pm\Delta\mu^{\o}$ &$\mu^{\rm tot}\pm\Delta\mu^{\rm tot}$&$\beta_{\rm app}^r$&$\beta_{\rm app}^{\o}$&$\beta_{\rm app}^{\rm tot}$  \\
\multicolumn{1}{c}{} & 
\multicolumn{1}{c}{} &
\multicolumn{1}{c}{} &
\multicolumn{1}{c}{} &
\multicolumn{1}{c}{} &
\multicolumn{1}{c}{} &
\multicolumn{1}{c}{\scriptsize [year]} &
\multicolumn{1}{c}{} &
\multicolumn{1}{c}{\scriptsize [mas] } & 
\multicolumn{1}{c}{\scriptsize [mas]} &
\multicolumn{1}{c}{\scriptsize [mas/year]}&
\multicolumn{1}{c}{\scriptsize [mas/year]} & 
\multicolumn{1}{c}{\scriptsize [mas/year]} &
\multicolumn{1}{c}{\scriptsize [mas/year]} &
\multicolumn{1}{c}{\scriptsize [mas/year]}&
\multicolumn{1}{c}{\scriptsize [$c$]}&
\multicolumn{1}{c}{\scriptsize [$c$]}&
\multicolumn{1}{c}{\scriptsize [$c$]}\\
\noalign{\smallskip} \hline \hline  \noalign{\smallskip}}
\tabletail{\hline\multicolumn{10}{r}{\small continued on next  page}\\}
\tablelasttail{\hline}
\begin{supertabular}{ccccccccccccccccccccccccccc}
\label{table:proper}
0003+380&G, {\it Sy}&   0.229&C1&2&3&3.21& 49806.3333&    0.605$\pm$         0.052&   -0.266$\pm$         0.052&   -0.003$\pm$         0.038&    0.008$\pm$         0.038&   -0.006$\pm$         0.038&   -0.006$\pm$         0.038&    0.009$\pm$         0.038&   -0.087$\pm$         0.554&   -0.087$\pm$         0.554&    0.131$\pm$         0.554\\
0003+380&G&   0.229&C2&2&3&3.21& 49806.3333&    1.351$\pm$         0.048&   -0.964$\pm$         0.048&    0.124$\pm$         0.037&   -0.053$\pm$         0.037&    0.132$\pm$         0.037&   -0.029$\pm$         0.035&    0.135$\pm$         0.037&    1.925$\pm$         0.540&   -0.423$\pm$         0.510&    1.968$\pm$         0.540\\
0003+380&G&   0.229&C3&3&2&0.99& 49806.3333&    3.221$\pm$ ------             &   -1.857$\pm$ ------             &    0.085$\pm$  ------            &    0.118$\pm$   ------           &    0.014$\pm$  ------            &   -0.144$\pm$     ------         &    0.145$\pm$  ------            &    0.204$\pm$    ------          &   -2.100$\pm$    ------          &    2.114$\pm$    ------          \\
0003+380&G&   0.229&C4&3&2&2.22& 49806.3333&    4.452$\pm$  ------            &   -2.600$\pm$ ------             &    0.012$\pm$ ------             &   -0.336$\pm$------              &    0.180$\pm$------              &    0.284$\pm$ ------             &    0.336$\pm$------              &    2.625$\pm$ ------             &    4.141$\pm$ ------             &    4.899$\pm$    ------          \\
\hline
0010+405&G, {\it Sy 1.9}&   0.255&C1&2&3&6.46& 49142.3333&   -0.389$\pm$         0.051&    0.660$\pm$         0.051&   -0.013$\pm$         0.022&    0.037$\pm$         0.022&    0.039$\pm$         0.022&    0.008$\pm$         0.021&    0.039$\pm$         0.022&    0.630$\pm$         0.355&    0.129$\pm$         0.339&    0.630$\pm$         0.355\\
0010+405&G&   0.255&C2&2&3&6.46& 49142.3333&   -1.179$\pm$         0.093&    2.103$\pm$         0.093&    0.003$\pm$         0.040&   -0.065$\pm$         0.040&   -0.058$\pm$         0.040&   -0.029$\pm$         0.041&    0.065$\pm$         0.040&   -0.936$\pm$         0.646&   -0.468$\pm$         0.662&    1.049$\pm$         0.646\\
\hline
0014+813&Q&   3.366&C1&1&3&3.91& 49604.3333&   -0.042$\pm$         0.018&   -0.665$\pm$         0.018&    0.002$\pm$         0.012&    0.003$\pm$         0.012&   -0.003$\pm$         0.012&   -0.001$\pm$         0.012&    0.004$\pm$         0.012&   -0.323$\pm$         1.293&   -0.108$\pm$         1.293&    0.431$\pm$         1.293\\
0014+813&Q&   3.366&C2&1&3&3.91& 49604.3333&   -1.332$\pm$         0.024&   -4.893$\pm$         0.024&    0.034$\pm$         0.015&   -0.079$\pm$         0.015&    0.067$\pm$         0.015&   -0.054$\pm$         0.015&    0.086$\pm$         0.015&    7.221$\pm$         1.617&   -5.820$\pm$         1.617&    9.269$\pm$         1.617\\
0014+813&Q&   3.366&C3&2&3&3.91& 49604.3333&   -1.443$\pm$         0.030&   -8.864$\pm$         0.030&    0.065$\pm$         0.018&   -0.090$\pm$         0.018&    0.078$\pm$         0.018&   -0.079$\pm$         0.018&    0.111$\pm$         0.018&    8.407$\pm$         1.940&   -8.514$\pm$         1.940&   11.963$\pm$         1.940\\
\hline
0016+731&Q&   1.781&C1&1&2&1.07& 51239.0000&   -0.643$\pm$------              &    0.361$\pm$  ------            &   -0.086$\pm$ ------             &    0.185$\pm$------              &    0.166$\pm$------              &    0.119$\pm$ ------             &    0.204$\pm$ ------             &   12.933$\pm$ ------             &    9.271$\pm$ ------             &   15.893$\pm$    ------          \\
\hline
0018+729&G&        &C1&2&3&5.19& 50478.6667&   -1.117$\pm$         0.093&    0.376$\pm$         0.093&   -0.017$\pm$         0.044&   -0.027$\pm$         0.044&    0.008$\pm$         0.044&   -0.031$\pm$         0.044&    0.032$\pm$         0.044&                  &                  &                      \\
0018+729&G&        &C2&1&3&5.19& 50478.6667&   -3.485$\pm$         0.171&   -0.330$\pm$         0.171&   -0.043$\pm$         0.092&   -0.135$\pm$         0.092&    0.055$\pm$         0.092&   -0.131$\pm$         0.091&    0.142$\pm$         0.092&                  &                  &               \\
0018+729&G& &C3&3&3&5.19& 50478.6667&   -6.265$\pm$         0.247&    0.526$\pm$         0.247&    0.165$\pm$         0.111&    0.019$\pm$         0.111&   -0.163$\pm$         0.111&    0.033$\pm$         0.108&    0.166$\pm$         0.111&    &    &    \\
\hline
0022+390&Q&   1.946&C1&2&3&6.46& 49142.3333&    0.608$\pm$         0.126&   -1.137$\pm$         0.126&    0.077$\pm$         0.043&   -0.082$\pm$         0.043&    0.109$\pm$         0.043&   -0.029$\pm$         0.036&    0.113$\pm$         0.043&    8.943$\pm$         3.528&   -2.379$\pm$         2.954&    9.272$\pm$         3.528\\
0022+390&Q&   1.946&C2&2&3&6.46& 49142.3333&    0.484$\pm$         0.174&   -3.982$\pm$         0.174&    0.023$\pm$         0.057&   -0.067$\pm$         0.057&    0.070$\pm$         0.057&   -0.015$\pm$         0.059&    0.071$\pm$         0.057&    5.743$\pm$         4.677&   -1.231$\pm$         4.841&    5.826$\pm$         4.677\\
\hline
0035+367&Q&   0.366&C1&2&3&5.30& 50916.7500&    7.058$\pm$         0.027&    7.981$\pm$         0.027&   -0.074$\pm$         0.012&    0.097$\pm$         0.012&    0.024$\pm$         0.012&   -0.120$\pm$         0.012&    0.122$\pm$         0.012&    0.542$\pm$         0.271&   -2.710$\pm$         0.271&    2.756$\pm$         0.271\\
\hline
0035+413&Q&   1.353&C1&1&3&4.23& 50770.0000&    1.317$\pm$         0.033&   -0.396$\pm$         0.033&    0.069$\pm$         0.017&   -0.047$\pm$         0.017&    0.080$\pm$         0.017&    0.025$\pm$         0.016&    0.083$\pm$         0.017&    5.234$\pm$         1.112&    1.636$\pm$         1.047&    5.430$\pm$         1.112\\
0035+413&Q&   1.353&C2&3&2&1.78& 50770.0000&    2.288$\pm$ ------             &   -0.955$\pm$------              &    0.136$\pm$------              &    0.105$\pm$ ------             &    0.085$\pm$ ------             &   -0.149$\pm$ ------             &    0.171$\pm$ ------             &    5.561$\pm$ ------             &   -9.749$\pm$      ------        &   11.188$\pm$   ------           \\
0035+413&Q&   1.353&C3&1&3&4.23& 50770.0000&    6.534$\pm$         0.058&   -1.946$\pm$         0.058&    0.061$\pm$         0.032&   -0.018$\pm$         0.032&    0.063$\pm$         0.032&    0.000$\pm$         0.032&    0.063$\pm$         0.032&    4.122$\pm$         2.094&    0.000$\pm$         2.094&    4.122$\pm$         2.094\\
0035+413&Q&   1.353&C4&2&3&4.23& 50770.0000&   11.573$\pm$         0.177&   -4.879$\pm$         0.177&    0.064$\pm$         0.106&   -0.164$\pm$         0.106&    0.123$\pm$         0.106&    0.126$\pm$         0.105&    0.176$\pm$         0.106&    8.048$\pm$         6.935&    8.244$\pm$         6.870&   11.515$\pm$         6.935\\
\hline
0102+480&U&        &C1&2&3&6.46& 49142.3333&   -0.170$\pm$         0.019&   -0.695$\pm$         0.019&   -0.003$\pm$         0.007&   -0.017$\pm$         0.007&    0.017$\pm$         0.007&   -0.002$\pm$         0.007&    0.017$\pm$         0.007&                  &                  &                \\
0102+480&U&        &CC1&1&3&6.46& 49142.3333&    0.426$\pm$         0.132&    1.441$\pm$         0.132&    0.082$\pm$         0.053&    0.030$\pm$         0.053&    0.052$\pm$         0.054&    0.070$\pm$         0.053&    0.088$\pm$         0.053&                  &                  &                  \\
\hline
0108+388&G&   0.669&C1&2&3&4.31& 51239.0000&   -1.162$\pm$         0.033&   -0.275$\pm$         0.033&   -0.016$\pm$         0.017&    0.018$\pm$         0.017&    0.011$\pm$         0.017&    0.021$\pm$         0.017&    0.024$\pm$         0.017&    0.422$\pm$         0.652&    0.805$\pm$         0.652&    0.920$\pm$         0.652\\
0108+388&G&   0.669&C2&1&3&4.31& 51239.0000&   -4.870$\pm$         0.031&   -2.732$\pm$         0.031&    0.000$\pm$         0.016&    0.012$\pm$         0.016&   -0.006$\pm$         0.016&    0.011$\pm$         0.016&    0.012$\pm$         0.016&   -0.230$\pm$         0.614&    0.422$\pm$         0.614&    0.460$\pm$         0.614\\
\hline
0109+351&Q&   0.450&C1&2&3&4.66& 49986.3333&   -0.308$\pm$         0.049&   -0.694$\pm$         0.049&   -0.020$\pm$         0.024&   -0.043$\pm$         0.024&    0.047$\pm$         0.024&    0.001$\pm$         0.022&    0.047$\pm$         0.024&    1.279$\pm$         0.653&    0.027$\pm$         0.599&    1.279$\pm$         0.653\\
0109+351&Q&   0.450&C2&3&2&2.45& 49986.3333&   -0.965$\pm$ ------             &   -2.133$\pm$ ------             &   -0.060$\pm$  ------            &   -0.091$\pm$ ------             &    0.107$\pm$------              &    0.017$\pm$------              &    0.108$\pm$ ------             &    2.913$\pm$ ------             &    0.463$\pm$        ------      &    2.940$\pm$   ------           \\
\hline
0110+495&Q&   0.389&C1&2&3&4.66& 49986.3333&   -0.328$\pm$         0.097&    0.608$\pm$         0.097&    0.022$\pm$         0.044&   -0.001$\pm$         0.044&   -0.011$\pm$         0.045&    0.019$\pm$         0.045&    0.022$\pm$         0.044&   -0.263$\pm$         1.074&    0.454$\pm$         1.074&    0.525$\pm$         1.051\\
0110+495&Q&   0.389&C2&1&3&4.66& 49986.3333&   -1.463$\pm$         0.025&    2.138$\pm$         0.025&   -0.008$\pm$         0.012&    0.032$\pm$         0.012&    0.031$\pm$         0.012&    0.011$\pm$         0.012&    0.033$\pm$         0.012&    0.740$\pm$         0.287&    0.263$\pm$         0.287&    0.788$\pm$         0.287\\
0110+495&Q&   0.389&C3&3&3&4.66& 49986.3333&   -4.324$\pm$         0.498&    7.261$\pm$         0.498&    0.042$\pm$         0.249&    0.007$\pm$         0.249&   -0.015$\pm$         0.249&    0.040$\pm$         0.249&    0.042$\pm$         0.249&   -0.358$\pm$         5.945&    0.955$\pm$         5.945&    1.003$\pm$         5.945\\
\hline
0133+476&Q&   0.859&C1&2&3&4.31& 51239.0000&   -0.431$\pm$         0.055&    0.839$\pm$         0.055&   -0.188$\pm$         0.060&    0.139$\pm$         0.060&    0.210$\pm$         0.061&   -0.104$\pm$         0.072&    0.234$\pm$         0.060&    9.860$\pm$         2.864&   -4.883$\pm$         3.380&   10.986$\pm$         2.817\\
0133+476&Q&   0.859&C2&1&3&4.31& 51239.0000&   -1.472$\pm$         0.003&    2.023$\pm$         0.003&    0.044$\pm$         0.001&    0.054$\pm$         0.001&    0.018$\pm$         0.001&    0.067$\pm$         0.001&    0.070$\pm$         0.001&    0.845$\pm$         0.047&    3.146$\pm$         0.047&    3.287$\pm$         0.047\\
\hline
0145+386&Q&   1.442&C1&2&3&4.66& 49985.3333&   -0.576$\pm$         0.028&    0.687$\pm$         0.028&    0.062$\pm$         0.016&    0.084$\pm$         0.016&    0.025$\pm$         0.017&    0.102$\pm$         0.016&    0.105$\pm$         0.016&    1.706$\pm$         1.160&    6.962$\pm$         1.092&    7.167$\pm$         1.092\\
\hline
0151+474&Q&   1.026&C1&3&3&4.66& 49985.3333&   -0.093$\pm$         0.068&   -0.496$\pm$         0.068&    0.022$\pm$         0.052&    0.006$\pm$         0.052&   -0.009$\pm$         0.052&   -0.021$\pm$         0.052&    0.023$\pm$         0.052&   -0.484$\pm$         2.796&   -1.129$\pm$         2.796&    1.237$\pm$         2.796\\
0151+474&Q&   1.026&C2&3&3&4.66& 49985.3333&   -0.016$\pm$         0.028&   -2.151$\pm$         0.028&   -0.010$\pm$         0.018&   -0.087$\pm$         0.018&    0.087$\pm$         0.018&    0.009$\pm$         0.017&    0.087$\pm$         0.018&    4.679$\pm$         0.968&    0.484$\pm$         0.914&    4.679$\pm$         0.968\\
\hline
0153+744&Q&   2.338&C1&3&2&3.24& 51239.0000&    4.738$\pm$ ------             &   -2.089$\pm$  ------            &    0.060$\pm$ ------             &    0.028$\pm$ ------             &    0.043$\pm$  ------            &   -0.050$\pm$  ------            &    0.066$\pm$  ------            &    3.901$\pm$  ------            &   -4.535$\pm$  ------            &    5.987$\pm$  ------            \\
0153+744&Q&   2.338&C2&3&3&4.31& 51239.0000&    6.054$\pm$         0.293&   -4.845$\pm$         0.293&   -0.017$\pm$         0.154&    0.019$\pm$         0.154&   -0.025$\pm$         0.154&   -0.004$\pm$         0.154&    0.025$\pm$         0.154&   -2.268$\pm$        13.969&   -0.363$\pm$        13.969&    2.268$\pm$        13.969\\
0153+744&Q&   2.338&C3&3&3&4.31& 51239.0000&    4.605$\pm$         0.598&   -7.987$\pm$         0.598&    0.395$\pm$         0.290&    0.242$\pm$         0.290&   -0.012$\pm$         0.292&   -0.463$\pm$         0.291&    0.463$\pm$         0.290&   -1.089$\pm$        26.487&  -41.999$\pm$        26.397&   41.999$\pm$        26.306\\
0153+744&Q&   2.338&C4&3&3&4.31& 51239.0000&    4.738$\pm$         0.344&   -8.355$\pm$         0.344&   -0.555$\pm$         0.189&   -0.424$\pm$         0.189&    0.095$\pm$         0.190&    0.691$\pm$         0.192&    0.698$\pm$         0.189&    8.617$\pm$        17.235&   62.680$\pm$        17.416&   63.315$\pm$        17.144\\
0153+744&Q&   2.338&C5&3&3&4.31& 51239.0000&    4.558$\pm$         0.122&   -9.310$\pm$         0.122&    0.171$\pm$         0.089&   -0.023$\pm$         0.089&    0.096$\pm$         0.089&   -0.143$\pm$         0.089&    0.173$\pm$         0.089&    8.708$\pm$         8.073&  -12.972$\pm$         8.073&   15.693$\pm$         8.073\\
0153+744&Q&   2.338&C6&3&3&4.31& 51239.0000&    5.045$\pm$         0.064&   -9.654$\pm$         0.064&    0.015$\pm$         0.027&   -0.017$\pm$         0.027&    0.022$\pm$         0.027&   -0.006$\pm$         0.027&    0.022$\pm$         0.027&    1.996$\pm$         2.449&   -0.544$\pm$         2.449&    1.996$\pm$         2.449\\
\hline
0212+735&Q&   2.367&C1&1&3&4.31& 51239.0000&    0.479$\pm$         0.038&   -0.220$\pm$         0.038&    0.047$\pm$         0.021&    0.001$\pm$         0.021&    0.042$\pm$         0.021&   -0.021$\pm$         0.022&    0.047$\pm$         0.021&    3.834$\pm$         1.917&   -1.917$\pm$         2.008&    4.291$\pm$         1.917\\
0212+735&Q&   2.367&C2&3&3&4.31& 51239.0000&    1.490$\pm$         0.143&   -0.222$\pm$         0.143&   -0.172$\pm$         0.078&    0.038$\pm$         0.078&   -0.176$\pm$         0.078&   -0.012$\pm$         0.078&    0.176$\pm$         0.078&  -16.068$\pm$         7.121&   -1.096$\pm$         7.121&   16.068$\pm$         7.121\\
0212+735&Q&   2.367&C3&3&3&4.31& 51239.0000&    2.823$\pm$         0.113&   -0.825$\pm$         0.113&   -0.005$\pm$         0.063&    0.036$\pm$         0.063&   -0.015$\pm$         0.063&   -0.033$\pm$         0.062&    0.036$\pm$         0.063&   -1.369$\pm$         5.751&   -3.013$\pm$         5.660&    3.287$\pm$         5.751\\
0212+735&Q&   2.367&C4&2&3&4.31& 51239.0000&    6.350$\pm$         0.316&   -1.298$\pm$         0.316&   -0.022$\pm$         0.183&   -0.010$\pm$         0.183&   -0.019$\pm$         0.183&    0.015$\pm$         0.183&    0.024$\pm$         0.183&   -1.735$\pm$        16.707&    1.369$\pm$        16.707&    2.191$\pm$        16.707\\
0212+735&Q&   2.367&C5&1&3&4.31& 51239.0000&   14.138$\pm$         0.111&   -0.383$\pm$         0.111&    0.034$\pm$         0.059&    0.032$\pm$         0.059&    0.033$\pm$         0.059&   -0.033$\pm$         0.059&    0.047$\pm$         0.059&    3.013$\pm$         5.386&   -3.013$\pm$         5.386&    4.291$\pm$         5.386\\
\hline
0219+428&B&   0.444&C1&1&3&2.45& 50372.6667&   -0.232$\pm$         0.015&   -1.242$\pm$         0.015&   -0.012$\pm$         0.015&   -0.216$\pm$         0.015&    0.215$\pm$         0.015&   -0.027$\pm$         0.015&    0.216$\pm$         0.015&    5.783$\pm$         0.403&   -0.726$\pm$         0.403&    5.810$\pm$         0.403\\
0219+428&B&   0.444&C2&1&3&2.45& 50372.6667&    0.079$\pm$         0.025&   -2.530$\pm$         0.025&    0.007$\pm$         0.024&   -0.041$\pm$         0.024&    0.041$\pm$         0.024&   -0.006$\pm$         0.024&    0.042$\pm$         0.024&    1.103$\pm$         0.646&   -0.161$\pm$         0.646&    1.130$\pm$         0.646\\
0219+428&B&   0.444&C3&1&3&2.45& 50372.6667&    1.222$\pm$         0.086&   -6.162$\pm$         0.086&    0.111$\pm$         0.093&   -0.189$\pm$         0.093&    0.207$\pm$         0.093&   -0.072$\pm$         0.092&    0.219$\pm$         0.093&    5.568$\pm$         2.501&   -1.937$\pm$         2.475&    5.890$\pm$         2.501\\
0219+428&B&   0.444&C4&3&3&2.45& 50372.6667&    3.381$\pm$         0.193&  -13.370$\pm$         0.193&    0.005$\pm$         0.187&    0.522$\pm$         0.187&   -0.505$\pm$         0.187&   -0.133$\pm$         0.188&    0.522$\pm$         0.187&  -13.583$\pm$         5.030&   -3.577$\pm$         5.057&   14.040$\pm$         5.030\\
\hline
\end{supertabular}
}
\endlandscape
\end{appendix}

\twocolumn

\end{document}